\documentclass[preprint2]{aastex}
\usepackage{natbib}
\usepackage{lscape}
\usepackage{subfig}

\usepackage{graphicx}
\usepackage{enumerate}

\providecommand{\e}[1]{\ensuremath{\times 10^{#1}}}	

\newcommand{\superscript}[1]{\ensuremath{^{\textrm{#1}}}}

\newcommand{\mf}{$\langle$F$_{\nu}\rangle$}
\newcommand{\mfg}{$\langle$F$_{\gamma}\rangle$}

\newcommand{\osi}{$\alpha_{o}$}
\newcommand{\xsi}{$\alpha_{X}$}
\newcommand{\gsi}{$\alpha_{\gamma}$}
\newcommand{\oxsi}{$\alpha_{ox}$}
\newcommand{\xgsi}{$\alpha_{xg}$}

\usepackage{datatool}
\usepackage{hyperref}
\usepackage[acronym,nosuper,nonumberlist]{glossaries}
\usepackage{wasysym}	

\newacronym[plural=AGNs, firstplural=active galactic nuclei (AGNs)]{agn}{AGN}{active galactic nucleus}
\newacronym{blr}{BLR}{broad-line region}
\newacronym[plural=FSRQs, firstplural=flat spectrum radio quasars (FSRQs)]{fsrq}{FSRQ}{flat spectrum radio quasar}
\newacronym{grb}{GRB}{gamma ray burst}
\newacronym{hbl}{HBL}{high-frequency BL Lacs}
\newacronym{lat}{LAT}{Large Area Telescope}
\newacronym{lbl}{LBL}{low-frequency BL Lacs}
\newacronym{nlr}{NLR}{narrow-line region}
\newacronym{ovv}{OVV}{optically violent variable}
\newacronym{pc}{PC}{photon counting}
\newacronym{psf}{PSF}{point-spread function}
\newacronym{sed}{SED}{spectral energy distribution}
\newacronym{ssc}{SSC}{synchrotron self-Compton}
\newacronym{uvot}{UVOT}{UV/Optical Telescope}
\newacronym{wt}{WT}{windowed timing}
\newacronym{xrt}{XRT}{X-ray Telescope}

\makeglossaries

\newcommand{\noprint}[1]{}

\begin{document}
\title{Comprehensive Monitoring of Gamma-ray Bright Blazars.\\ I. Statistical Study of Optical, X-ray, and Gamma-ray Spectral Slopes
}
\author{Karen E. Williamson\altaffilmark{1}, Svetlana G. Jorstad\altaffilmark{1,2}, Alan P. Marscher\altaffilmark{1}, Valeri M. Larionov\altaffilmark{2,5,6}, Paul S. Smith\altaffilmark{3}, Iv\'an Agudo \altaffilmark{4,1}, Arkady A. Arkharov\altaffilmark{5}, Dmitry A. Blinov\altaffilmark{7,2}, Carolina Casadio\altaffilmark{4},  Natalia V. Efimova\altaffilmark{2,5}, Jos\'e L. G\'omez\altaffilmark{4}, Vladimir A. Hagen-Thorn\altaffilmark{2,6}, Manasvita Joshi\altaffilmark{1}, Tatiana S. Konstantinova\altaffilmark{2}, Evgenia N. Kopatskaya\altaffilmark{2}, 
Elena G. Larionova\altaffilmark{2}, Liudmilla V. Larionova\altaffilmark{2}, Michael P. Malmrose\altaffilmark{1}, Ian M. McHardy\altaffilmark{8}, 
Sol N. Molina\altaffilmark{4}, Daria A. Morozova\altaffilmark{2}, Gary D. Schmidt\altaffilmark{9}, Brian W. Taylor\altaffilmark{1,10}, and Ivan S. Troitsky\altaffilmark{2}}
\altaffiltext{1}{Institute for Astrophysical Research, Boston University, 725 Commonwealth Avenue, Boston, MA 02215}
\email{kwilliam@bu.edu}
\altaffiltext{2}{Astronomical Institute, St. Petersburg State University, Universitetskij Pr. 28, Petrodvorets, 
198504 St. Petersburg, Russia}
\altaffiltext{3}{Steward Observatory, University of Arizona, Tucson, AZ 85721-0065}
\altaffiltext{4}{Instituto de Astrof\'{\i}sica de Andaluc\'{\i}a, CSIC, Apartado 3004, 18080,
Granada, Spain}
\altaffiltext{5}{Main (Pulkovo) Astronomical Observatory of RAS, Pulkovskoye shosse, 60, 196140,
St. Petersburg, Russia}
\altaffiltext{6}{Isaac Newton Institute of Chile, St. Petersburg Branch, St. Petersburg, Russia}
\altaffiltext{7}{Department  of Physics, University of Crete, 71003, Heraklion, Greece}
\altaffiltext{8}{Department of Physics and Astronomy, University of Southampton, Southampton, SO17 1BJ,
United Kingdom}
\altaffiltext{9}{National Science Foundation, 4201 Wilson Ave., Arlington, VA, 22230 USA}
\altaffiltext{10}{Lowell Observatory, Flagstaff, AZ 86001}
\shorttitle{MULTI-FREQUENCY EMISSION IN BLAZARS}
\shortauthors{Williamson et al.}

\begin{abstract}
We present $\gamma$-ray, X-ray, ultraviolet, optical, and near-infrared light curves of 33 $\gamma$-ray bright blazars over four years that we have been monitoring since 2008 August with multiple optical, ground-based telescopes and the {\it Swift} satellite, and augmented by data from the {\it Fermi Gamma-ray Space Telescope} and other publicly available data from {\it Swift}. The sample consists of 21 flat-spectrum radio quasars (FSRQs) and 12 BL Lac objects (BL Lacs). We identify quiescent and active states of the sources based on their $\gamma$-ray behavior. We derive $\gamma$-ray, X-ray, and optical spectral indices, $\alpha_\gamma$, $\alpha_X$, and $\alpha_o$, respectively ($F_\nu\propto\nu^\alpha$), and construct spectral energy distributions (SEDs) during quiescent and active states. We analyze the relationships between different spectral indices, blazar classes, and activity states. We find (i) significantly steeper $\gamma$-ray spectra of FSRQs than for BL Lacs during quiescent states, but a flattening of the spectra for FSRQs during active states while the BL Lacs show no significant change; (ii) a small difference of $\alpha_X$ within each class between states, with BL Lac X-ray spectra significantly steeper than in FSRQs; (iii) a highly peaked distribution of X-ray spectral slopes of FSRQs at $\sim-$0.60, but a very broad distribution of $\alpha_X$ of BL Lacs during active states; (iv) flattening of the optical spectra of FSRQs during quiescent states, but no statistically significant change of $\alpha_o$ of BL Lacs between states; and (v) a positive correlation between optical and $\gamma$-ray spectral slopes of BL Lacs, with similar values of the slopes. We discuss the findings with respect to the relative prominence of different components of high-energy and optical emission as the flux state changes.

\end{abstract}


\section{Introduction}

\label{Introduction}
Blazars are active galactic nuclei characterized by ultra-luminous, broad-band, non-thermal radio to $\gamma$-ray continuum radiation, and by irregular, rapid flux variability across wavebands. They are divided into two classes, flat spectrum radio quasars (FSRQs) and BL Lac objects (BL Lacs). A primary method employed to probe our understanding of these objects is to study their spectral energy distributions (SEDs). Until recently, however, studies of blazar SEDs have been hindered by an insufficient number of simultaneous observations across the spectrum for a large enough sample of objects to allow a statistical analysis of their behavior in varying states of activity. A significant advance occurred with the launch of the \textit{Fermi Gamma-ray Space Telescope}. With its sensitivity and its ability to scan the entire sky every three hours, the \textit{Fermi} \gls{lat} \citep{2009ApJ...697.1071A} provides continuous coverage of blazars in the $\gamma$-ray regime. One year prior to the onset of the science mission of \textit{Fermi}, we began international, collaborative, multiwavelength monitoring of 33 blazars at radio to optical bands. These observations, combined with the $\gamma$-ray data from \textit{Fermi} and the X-ray, ultraviolet (UV), and optical data from the \textit{Swift} space observatory \citep{2004ApJ...611.1005G},
as well as measurements with several ground-based instruments, provide a rich dataset to study the behavior of these objects. We focus on measurements across the electromagnetic spectrum, made within 24 hours of each other, at multiple epochs
when the objects are in different $\gamma$-ray activity states.

Long-term monitoring of blazars reveals variability of emission best described by a ``red noise" power spectrum, where the amplitude of variations is greater on longer time-scales \citep[e.g.,][]{2009ApJ...691.1021D,2012ApJ...749..191C}. The light curves contain periods of relative quiescence interrupted by sometimes sudden, prominent outbursts with durations of weeks to several months in one or more energy bands, as well as more rapid lower-level fluctuations.  These outbursts can vary dramatically in both time profile and amplitude. Critical to unraveling the physics of blazars is to study how the SED changes between such quiescent and active periods.  Many studies have examined a small number of objects in an active state, sometimes contrasting activity at different flux levels \citep[e.g.,][]{2008A&A...491..755R, 2012A&A...545A..48R, 2012ApJ...754..114H, 2013ApJ...773..147J}.  Fewer \citep[e.g.,][]{2007A&A...473..819R,2008A&A...491..755R,2008ApJ...679.1029T,Q0528} have studied objects in a quiescent or low $\gamma$-ray state. With the increased sensitivity of instruments and the time coverage of \textit{Fermi}, studies of larger samples are beginning to unveil trends in the behavior of blazars at different $\gamma$-ray activity states \citep[e.g., ][]{2009MNRAS.396L.105G, 2010ApJ...710.1271A}. A statistical analysis of SEDs from optical to $\gamma$-ray wavelengths based on simultaneous observations at different activity states for a sample of blazars should, therefore, be instructive. 
  
A distinctive characteristic of a blazar's SED is its two-peaked shape, with one maximum at infrared (IR) to X-ray frequencies and the other at $\gamma$-ray frequencies. The shape of the SED, combined with polarization characteristics, provides considerable evidence that the emission produced from radio to optical wavelengths is dominated by synchrotron radiation. 
If the accretion disk luminosity is important, it will be seen in the UV portion of the spectrum. Commonly seen in other classes of \glspl{agn}, the ``big blue bump" (BBB) is often less prominent, or even undetectable in blazars, owing to the strong, relativistically beamed non-thermal radiation. \citet{2003MNRAS.339.1081D} found that the non-thermal component of the optical/UV emission of FSRQs accounts for an average of $\sim$ 85\% of the total power. Only in about 9\% of the objects they studied did the thermal component dominate.  Signatures of the BBB in FSRQs include a decrease of the degree of polarization with frequency \citep[e.g.,][]{Smith1986} and a redder color index at brighter flux states \citep[e.g.,][]{Bregman1986, 2012A&A...545A..48R}. A number of observations have indicated that the accretion disk is less prominent in BL Lacs \citep[e.g.,][]{2009MNRAS.396L.105G,2012MNRAS.420.2899G}. An alternative possibility for flatter-spectrum emission in the UV region of some blazars was suggested by \citet{Raiteri0235}. Studying the spectrum of 0235+164, these authors see  the signature of a second synchrotron component. 

The higher-energy SED is consistent with inverse Compton (IC) scattering off photons either from inside the jet (synchrotron self-Compton mechanism, SSC) or external to the jet (external Compton mechanism, EC) by relativistic electrons in the jet \citep[e.g.,][]{2010ApJ...710L.126M}. Other mechanisms, e.g., proton synchrotron emission 
\citep{BOTTCHER13}, might play a role as well. In IC models, we expect the spectral slope of high-energy emission to be similar to the slope of the synchrotron radiation emitted by the electrons responsible for scattering seed photons up to high energies.

The locations of radiative dissipation zones within the jet and the physical processes involved are still under debate. Polarization and timing of flares relative to changes in images of parsec-scale jets of blazars indicate that near-infrared (NIR) to optical synchrotron flares often take place near the end of the jet's acceleration zone  \citep{J07,2008Natur.452..966M}.  Using Very Long Baseline Array (VLBA) images, \citet{sj2012} conclude that enhanced $\gamma$-ray emission is produced downstream of the broad emission line clouds, while others \citep[e.g.,][]{2010MNRAS.405L..94T} argue for a sub-parsec origin, based on short timescales of $\gamma$-ray variability. The outbursts, which occur across the electromagnetic spectrum, can be caused by shock formations in the jet or other processes that increase the particle density, magnetic field strength, or seed photon field, change the
magnetic field orientation, and/or enhance the Doppler boosting. The characteristics of the SED represented by spectral indices at different wavebands can provide insights into the interplay between different factors responsible for the outbursts, as well as between different emission components (e.g., the accretion disk and jet) and processes (synchrotron, inverse Compton, and thermal) during active and quiescent states. These insights will improve our understanding of the physics and location of energetic phenomena in blazars. 

Here we statistically study how the spectral indices at $\gamma$-ray, X-ray, and optical frequencies change as the flux state varies, as well as whether the behavior depends on the type of blazar. 
We  present over four years of data (from early 2008 to late 2012) in 13 frequency bands from NIR to $\gamma$-rays. From this compilation, we select epochs of quasi-simultaneous data at both active and quiescent states, compute spectral indices, and examine the trends and correlations between them.
The sample of blazars and the data reduction are described in $\S$2. In  $\S$3, we define \textit{quiescent} and \textit{active} states and describe the selection of epochs for our statistical analysis. We describe the computation of spectral indices in $\S$4 and present the trends and correlations of those indices and in the relationships between them in $\S$5. Using these statistical trends, we describe  a ``typical'' quiescent and active BL Lac object and FSRQ in $\S$6 and discuss the implications of our results for physical models. We summarize our findings in $\S$7. An expanded version of this paper with a complete set of light curves and SEDs for all sources can be found at \url{www.bu.edu/blazars/VLBAproject.html}.

\section{Observations and Data Reduction}
\label{Observations}

\subsection{The Sample}
Since 2007, we have been collecting multi-waveband fluxes, polarization measurements, and radio images of blazars to provide the data for understanding the physics of the jets \citep[see, e.g.,][]{2012arXiv1201.5402M}.  This study includes 28 of the original 30 objects selected for the monitoring campaign, confirmed as $\gamma$-ray sources by EGRET (Energetic $\gamma$-Ray Experiment Telescope) on the {\it Compton Gamma Ray Observatory}, have an \textit{R}-band brightness exceeding 18 mag (bright enough for optical polarization measurements at a 1 $-$ 2 meter class optical telescope without needing excessive amounts of telescope time), exceed 0.5 Jy at 43 GHz, and have a declination accessible to the collaboration's observatories ($> -30^\circ$). Three additional BL Lacs  (1055+018, 1308+326, and 1749+096) and two FSRQs (3C345 and 3C446) included in this analysis were among those added when they were detected as $\gamma$-ray sources by the {\it Fermi} LAT \citep{LATsources2009ApJ...700..597A}.

Table \ref{tab:sources} presents general information about these 33 blazars. Column 1 is an object reference number that will be used in plots to identify each source, column 2 is the object name as used in this writing, column 3 is an alternate, commonly used name,  column 4 is the object's name as listed in the 2FGL catalog \citep{ackermann2fgl}, column 5 gives the redshift as reported in the NASA/IPAC Extragalactic Database (NED)\footnote{\url{http://nedwww.ipac.caltech.edu}}, and columns 6 and 7 are the right ascension and declination of the object as retrieved by Simbad and reported in \url{http://heasarc.gsfc.nasa.gov}. From \citet{ackermann2fgl}, we include  the object's optical classification and the SED classification in columns 8 and 9, respectively. Of the 33 blazars, 12 have optical classifications  as  BL Lacs and 21 as FSRQs. Of the 12 BL Lacs, 5 have an SED classification \citep{2010ApJ...716...30A} of low synchrotron peak frequency (LSP, $\lesssim 10^{14}$ Hz), 6 as  intermediate synchrotron peak frequency (ISP, between $10^{14}$ and $10^{15}$ Hz), and 1 as high synchrotron peak frequency (HSP, $\gtrsim 10^{15}$ Hz) blazar. All of the  FSRQs have an SED classification of LSP.

\begin{deluxetable}{llllrrrll}
\tabletypesize{\small}
\tablewidth{0pt}
\tablecaption{Sources Analyzed}
\tablecolumns{9}
\tablehead{ 
			\multicolumn{1}{l}{Ref}&
			\multicolumn{1}{c}{Object}&
			\multicolumn{1}{c}{Alternate}&
			\multicolumn{1}{c}{2FGL Catalog}&
			\multicolumn{1}{c}{}&
			\multicolumn{1}{c}{}&
			\multicolumn{1}{c}{}&
			\multicolumn{1}{c}{Optical}&
			\multicolumn{1}{c}{SED}
			\cr
			\multicolumn{1}{l}{Num}&
			\multicolumn{1}{c}{Name}&
			\multicolumn{1}{c}{Name}&
			\multicolumn{1}{c}{Name\tablenotemark{a}}&
			\multicolumn{1}{c}{\textit{z}\tablenotemark{b}}&
			\multicolumn{1}{c}{R.A. 2000\tablenotemark{c}}&
			\multicolumn{1}{c}{Dec. 2000\tablenotemark{c}}&
			\multicolumn{1}{c}{Class\tablenotemark{a}}&
			\multicolumn{1}{c}{Class\tablenotemark{a}}
					  \cr
		  	\multicolumn{1}{l}{(1)}&
		  	\multicolumn{1}{c}{(2)}&
		 	\multicolumn{1}{c}{(3)}&
		 	\multicolumn{1}{c}{(4)}&
		 	\multicolumn{1}{c}{(5)}&
		  	\multicolumn{1}{c}{(6)}&
		  	\multicolumn{1}{c}{(7)}&
		 	\multicolumn{1}{c}{(8)}&
		 	\multicolumn{1}{c}{(9)}
			}    

\startdata
    1     & 3C66A & 0219+428 & J0222.6+4302 & 0.444? & 02 22 39.61 & +43 02 07.8 & BL Lac & ISP \\
    2     & 0235+164 &       & J0238.7+1637 & 0.940 & 02 38 38.93 & +16 36 59.3 & BL Lac & LSP \\
    3     & 0336-019 & CTA26 & J0339.4-0144 & 0.852 & 03 39 30.94 & -01 46 35.8  & FSRQ  & LSP \\
    4     & 0420-014 & OA129 & J0423.2-0120 & 0.916 & 04 23 15.80 & -01 20 33.1  & FSRQ  & LSP \\
    5     & 0528+134 &       & J0530.8+1333 & 2.060 & 05 30 56.42 & +13 31 55.1 & FSRQ  & LSP \\
    6     & 0716+714 &       & J0721.9+7120 & 0.300\tablenotemark{d} & 07 21 53.45 & +71 20 36.4 & BL Lac & ISP \\
    7     & 0735+178 &       & J0738.0+1742 & 0.424 & 07 38 07.39 & +17 42 19.0 & BL Lac & LSP \\
    8     & 0827+243 & OJ248 & J0830.5+2407 & 0.940 & 08 30 52.09 & +24 10 59.8 & FSRQ  & LSP \\
    9     & 0829+046 &       & J0831.9+0429 & 0.174 & 08 31 48.88 & +04 29 39.1 & BL Lac & LSP \\
    10    & 0836+710 &       & J0841.6+7052 & 2.172 & 08 41 24.37 & +70 53 42.2 & FSRQ  & LSP \\
    11    & OJ287 & 0851+202 & J0854.8+2005 & 0.306 & 08 54 48.87 & +20 06 30.6 & BL Lac & ISP \\
    12    & 0954+658 &       & J0958.6+6533 & 0.368 & 09 58 47.25 & +65 33 54.8  & BL Lac & ISP \\
    13    & 1055+018 & 4C+01.28 & J1058.4+0133 & 0.890 & 10 58 29.61 & +01 33 58.8  & BL Lac & LSP \\
    14    & Mkn421 & 1101+384 & J1104.4+3812 & 0.030 & 11 04 27.31 & +38 12 31.8 & BL Lac & HSP \\
    15    & 1127-145 &       & J1130.3-1448 & 1.184 & 11 30 07.05 & -14 49 27.4 & FSRQ  & LSP \\
    16    & 1156+295 & 4C+29.45 & J1159.5+2914 & 0.724 & 11 59 31.83 & +29 14 43.8 & FSRQ  & LSP \\
    17    & 1219+285 & WCom  & J1221.4+2814 & 0.102 & 12 21 31.69 & +28 13 58.5 & BL Lac & ISP \\
    18    & 1222+216 & 4C+21.35 & J1224.9+2122 & 0.432 & 12 24 54.45 & +21 22 46.5 & FSRQ  & LSP \\
    19    & 3C273 & 1226+023 & J1229.1+0202 & 0.158 & 12 29 06.70 & +02 03 08.7  & FSRQ  & LSP \\
    20    & 3C279 & 1253-055 & J1256.1-0547 & 0.536 & 12 56 11.17 & -05 47 21.5 & FSRQ  & LSP \\
    21    & 1308+326 &       & J1310.6+3222 & 0.996 & 13 10 28.66 & +32 20 43.8 & FSRQ  & LSP \\
    22    & 1406-076 &       & J1408.8-0751 & 1.494 & 14 08 56.48 & -07 52 26.7  & FSRQ  & LSP \\
    23    & 1510-089 &       & J1512.8-0906 & 0.360 & 15 12 50.53 & -09 05 59.8  & FSRQ  & LSP \\
    24    & 1611+343 & DA406 & J1613.4+3409 & 1.397 & 16 13 41.06 & +34 12 47.9 & FSRQ  & LSP \\
    25    & 1622-297 &       & J1626.1-2948 & 0.815 & 16 26 06.02 & -29 51 27.0 & FSRQ  & LSP \\
    26    & 1633+382 & 4C+38.41 & J1635.2+3810 & 1.814 & 16 35 15.49 & +38 08 04.5 & FSRQ  & LSP \\
    27    & 3C345 & 1641+399 & J1642.9+3949 & 0.593 & 16 42 58.81 & +39 48 37.0  & FSRQ  & LSP \\
    28    & 1730-130 & NRAO 530 & J1733.1-1307 & 0.902 & 17 33 02.71 & -13 04 49.5  & FSRQ  & LSP \\
    29    & 1749+096 & OT081 & J1751.5+0938 & 0.322 & 17 51 32.82 & +09 39 00.7  & BL Lac & LSP \\
    30    & BL Lacertae & 2200+420 & J2202.8+4216 & 0.069 & 22 02 43.29 & +42 16 40.0 & BL Lac & ISP \\
    31    & 3C446 & 2223-052 & J2225.6-0454 & 1.404 & 22 25 47.26 & -04 57 01.4 & FSRQ  & LSP \\
    32    & CTA102 & 2230+114 & J2232.4+1143 & 1.037 & 22 32 36.42 & +11 43 50.8  & FSRQ  & LSP \\
    33    & 3C454.3 & 2251+158 & J2253.9+1609 & 0.859 & 22 53 57.75 & +16 08 53.6  & FSRQ  & LSP \\

\enddata 
\tablenotetext{a}{\citet{ackermann2fgl}.}
\tablenotetext{b}{Information taken from the NASA/IPAC Extragalactic Database  (\url{http://nedwww.ipac.caltech.edu/}).}
\tablenotetext{c}{Simbad resolver as reported in \url{http://heasarc.gsfc.nasa.gov}.}
\tablenotetext{d}{\citet{Danforth} set  0.2315 $<$ z  $<$ 0.372 (99.7\%).}
   
  \label{tab:sources}%
\end{deluxetable}
\subsection{Gamma-ray Data}
\label{sec:gammareduction}
The $\gamma$-ray data were obtained by the \gls{lat} on board the \textit{Fermi Gamma Ray Space Telescope}. To construct the $\gamma$-ray light curves, we reduced the \textit{Fermi} data  using Pass 7 photon and spacecraft data, the V9r23p1 version of the Fermi Science Tools, and the instrument responses for the gal$\_$2yearp7v6$\_$v0 and iso$\_$p7v6clean.txt diffuse source models. All of these are available on the \textit{Fermi} website.\footnote{\url{http://fermi.gsfc.nasa.gov/ssc/}} We modeled the $\gamma$-ray emission between 0.1 and 200 GeV  from a given target and other point sources within a 15-degree radius of the target. Comprehensive reduction of the data was first performed with spectral models corresponding to those listed in the 2FGL catalog, typically with a seven-day bin size. However, because the power-law photon index in the 2FGL catalog was computed from the flux collected by \textit{Fermi} over two years \citep{Nolan2FGL}, and because a typical blazar spends less than 5\% of its time in a $\gamma$-ray active state \citep{Abdovariability}, this index best represents the object in a quiescent state. To obtain a spectral index for each object while in an active state (to be defined in $\S$\ref{sec:quiactdefinition}), we re-reduced the data during active states, typically with a 1-3 day bin size, using a simple power law model while allowing the photon index to vary. To obtain a spectral index during long periods of quiescence (defined in $\S$\ref{sec:quiactdefinition}) when only upper limits were obtained with 7-day binning, we re-reduced the data using extended bin sizes.
  


\subsection{X-Ray Data}
\label{sec:X-Ray}
The X-ray data, including the photon index  and its uncertainty, were obtained at a photon energy range of 0.3$-$10 keV by the \gls{xrt} \citep{2005SSRv..120..165B} on board the \textit{Swift} satellite. We reduced the data using the standard HEAsoft package (version 6.11). The standard \texttt{xrtpipeline} task was used to calibrate and clean the events.
We selected events with grades 0$-$12 in \gls{pc} mode and 0$-$2 in \gls{wt} mode. An ancillary response file was created with PSF correction  using the \texttt{xrtmkarf} task, and the the data were rebinned with the  \texttt{grppha} task to ensure a minimum of 10 photons in every newly defined channel.  We fit the spectra with the spectral analysis tool \texttt{xspec}, using a power-law model with minimum $\chi^2$  value, and, except for 0235+164, fixing the hydrogen column density (N$_{H}$) according to the measurements of \cite{Dickey}. For 0235+164, a value of N$_{H}$ of 2.8\e{21} cm\superscript{-2} was used to include an intervening \textit{z} = 0.524 absorber \citep{Madejski, Ackermann0235}. A Monte-Carlo method was used to test the goodness of fit. 

The photon counts of the sources were checked for pileup. The threshold for pileup is 0.5~counts~s\superscript{-1} and 100~counts~s\superscript{-1} for \gls{pc} mode and \gls{wt} mode, respectively.  
Each event with pileup was individually re-examined to remove the center of the point-spread function (PSF), following the process outlined on the \textit{Swift} website.\footnote{\url{http://www.swift.ac.uk/analysis/xrt/pileup.php}.} We created a new annular source region, determining  the inner radius by modeling the PSF as a King function.  None of the WT mode events exceeded the threshold for pileup.

\subsection{Swift Optical and Ultraviolet Data}
\label{sec:uvot}
 
\gls{uvot} \citep{RomingUVOT} data were reduced by using the standard HEAsoft package (version 6.11) and the calibration files released in 2011 July.  For each object, we defined a selection region centered on the source with a standard radius of $5''$, except  for  very faint objects (e.g., 0528+134, 0827+243), for which we chose a $3''$ radius and performed aperture correction according to \citet{Poole}.  The background region was defined in a source-free region with a circular aperture of $20''$. Unaligned exposures were individually aligned.  All extensions within an image were summed with \texttt{uvotimsum} and processed with \texttt{uvotsource} using a sigma value of five. Only epochs with  a summed exposure time exceeding 40 seconds were retained.

\subsection{Ground-Based Optical and Near-Infrared Data}
\label{sec:Ground-Based Optical}
In addition to UVOT data, we used optical data from eight ground-based observatories.
Table \ref{table:obs} provides the symbol we use to identify each observatory in light curves and SEDs (column 1), the identifying color of the observatory in light curves (column 2), the location of the observatory (column 3),  the diameter of the telescope (column 4), and the wavebands of the data used in this study (column 5). References to the data reduction procedures are listed in the footnotes of the table.


\begin{deluxetable}{cllll}
\rotate
\tablewidth{0pt}
\tabletypesize{\scriptsize }
\tablecolumns{5}
\tablecaption{List of Observatories Providing Measurements for this Study}
\tablehead{	
			\multicolumn{2}{l}{\hspace{40pt}Symbol}&
			\colhead{}&
			\colhead{Telescope}&
		  	\colhead{}
		  	\cr
		  	\colhead{Shape}& 
		  	\colhead{Color (Light curves)}&  
			\colhead{Observatory (Telescope or Monitoring Program) and Location}&
			\colhead{Diameter}&
		  	\colhead{Wavebands}
		  	
		  	}
\startdata
Space-based&&&\\
$\Diamond$					&black 	&\textit{Fermi} Gamma Ray Space Telescope (LAT) && Gamma-ray (0.1 GeV $-$ 300 GeV) 	\\
\scalebox{.8}{$\triangle$}	&black&\textit{Swift} Space Satellite (XRT) 	&		&  X-ray (0.3 $-$ 10 keV) \\
\scalebox{.8}{$\triangle$}, $\Leftcircle$, \underline{$\,\sqcap\,$}		&black, green, orange&\textit{Swift} Space Satellite (UVOT)	&& \textit{UVW1, UVM2, UVW2}  \\
\scalebox{.8}{$\triangle$}	&black&\textit{Swift} Space Satellite (UVOT)	&		& \textit{U, B, V}  \\
Ground-based&&&\\
$\times$					&indigo	&Lowell Observatory (Perkins Telescope), Flagstaff, Arizona\tablenotemark{a} 		&1.83 m				& \textit{B, V, R, I} 			 \\
$\triangleleft$				&light blue&Crimean Astrophysical Observatory (AZT-8), Nauchnij, Ukraine\tablenotemark{b}	&0.70 m		& \textit{B, V, R, I}			 \\
$\triangledown$				&green	&Observatorio del Roque de los Muchachos 	&2.00 m				& \textit{R}			\\
 & & \hspace{25 pt}(Liverpool Telescope), La Palma, Spain\tablenotemark{a}	&			& 			\\
$\Rightcircle$				&dark orange	&Calar Alto Observatory (MAPCAT), Andaluc\'{\i}a, Spain\tablenotemark{c}	&2.20 m				& \textit{R}			\\
\scalebox{.8}{$\square$}	&blue	&Cerro Tololo Inter-American Observatory (SMARTS), 	&0.90 $-$ 1.50 m		& \textit{B, V, R, J, K}		 \\
 & 	& \hspace{25 pt}Cerro Tololo, Chile\tablenotemark{d}		& &  		 \\
$\triangleright$			&red	&St. Petersburg University (LX-200), St. Petersburg, Russia\tablenotemark{b}	&0.40 m 	& \textit{B, V, R, I}			 \\
\scalebox{.75}{$\bigcirc$}	&yellow	&Steward Observatory (Kuiper and Bok Telescopes), 	&1.54, 2.30 m			& \textit{V}	 \\
 	& & \hspace{25 pt}Mt. Bigelow and Kitt Peak, Arizona\tablenotemark{e}	&   & 	 \\
\rotatebox{90}{$\bowtie$}	&red	&Istituto Nazionale di Astrofisica  (AZT-24),&1.10 m				&	\textit{J, H, K}		 \\	
	&	&  \hspace{25 pt}Campo Imperatore, Italy\tablenotemark{f}		&	&		 \\				
\enddata
\tablenotetext{a}{Data reduction is performed with the ESO software package MIDAS; refer to  \citet{2010ApJ...715..362J}.}
\tablenotetext{b}{Data reduction details provided in \citet{2008A&A...492..389L}.}
\tablenotetext{c}{Monitoring AGN with Polarimetry at the Calar Alto Telescopes (MAPCAT); data reduction details provided in  \citet{mapcat}.}
\tablenotetext{d}{The Small and Moderate Aperture Research Telescope System (SMARTS) daily monitoring program; refer to \url{http://www.astro.yale.edu/smarts/}.}
\tablenotetext{e}{Data reduction details provided in \citet{Smith2009}.}
\tablenotetext{f}{AZT-24 observations are made within an agreement between Pulkovo Astronomical Observatory, Rome Astronomical Observatory, and Collurania-Teramo Observatory. Data reduction details provided in \citet{2008ApJ...672...40H}.}

\label{table:obs}
\end{deluxetable}
\subsection{Dereddening and Flux Conversion}
\label{dereddening}
For the UV observations, we dereddened the fluxes using the \citet{Fitzpatrick99} interstellar extinction curve with an $R_{v}$ of 3.1 and  $A_{\lambda}$ values \citep{Schlafly} as retrieved from NED in 2012 November.  Optical and NIR magnitudes were dereddened using the \citet{Schlafly} values. Dereddening of 0235+164 is complicated by  intervening sources of dust and optical emission. We followed the procedure of \citet{Raiteri2008} to remove the additional flux from a foreground galaxy and applied the extinction values from  \citet{Raiteri0235}  and   \citet{Ackermann0235}. 
We converted the dereddened magnitudes to fluxes using the zero points and Pickles star spectra conversion factors from \citet{Poole} for \textit{Swift} observations and \citet{Mead} for ground-based observations.
For most objects in our sample, the host galaxy contribution is negligible in the UV. However, host  galaxy contamination was subtracted for two nearby objects, BL Lacertae and Mkn 421.  The host contribution in the UV is expected to be negligible for these two sources. We used the \textit{R}-Band host galaxy flux values derived by \citet{Nilsson2007} and average effective colors for elliptical galaxies determined by \citet{2001MNRAS.326..745M}. Converting these values as above, we obtained the dereddened host galaxy flux values, reported in Table \ref{table:hostgalaxy}. We subtracted these constant values from the dereddened measured flux.

\begin{deluxetable}{rccrrrrrrr}

\tablewidth{0pt}
\tabletypesize{\small}
\tablecolumns{10}
\tablecaption{Host Galaxy Contaminating Flux}
\tablehead{
         	\multicolumn{1}{l}{Object and} &     
			\multicolumn{1}{l}{Uncorrected}&
			\multicolumn{8}{c}{Dereddened Host Galaxy Flux}
					  \cr 
			
			\multicolumn{1}{r}{\hspace{10pt}Measurement}&   
			\multicolumn{1}{l}{\textit{R}-Band Flux\tablenotemark{a}}&
			\multicolumn{8}{c}{[mJy]}
					  \cr 
			\multicolumn{1}{r}{Source}&             
			\multicolumn{1}{l}{(uncertainty)}&
			\multicolumn{1}{c}{\textit{U}}&
			\multicolumn{1}{c}{\textit{B}}&
			\multicolumn{1}{c}{\textit{V}}&
			\multicolumn{1}{c}{\textit{R}}&
			\multicolumn{1}{c}{\textit{I}}&
			\multicolumn{1}{c}{\textit{J}}&
			\multicolumn{1}{c}{\textit{H}}&
			\multicolumn{1}{c}{\textit{K}}
					  \cr
		  	\multicolumn{1}{c}{(1)}&
		  	\multicolumn{1}{c}{(2)}&
		 	\multicolumn{1}{c}{(3)}&
		 	\multicolumn{1}{c}{(4)}&
		 	\multicolumn{1}{c}{(5)}&
		  	\multicolumn{1}{c}{(6)}&
		  	\multicolumn{1}{c}{(7)}&
		  	\multicolumn{1}{c}{(8)}&
		  	\multicolumn{1}{c}{(9)}&
		  	\multicolumn{1}{c}{(10)}
			}

\startdata
    \multicolumn{10}{l}{BL Lacertae}\\
    Ground-based & 1.35 (0.03) & 0.23  & 0.84  & 1.79  & 2.61  & 3.85  & 7.38  & 9.08  & 6.84 \\
    Swift   & 1.13 (0.03) & 0.19  & 0.70  & 1.50  &       &       &       &       &  \\
    \cr
    \multicolumn{10}{l}{Mkn 421}\\ 
    Ground-based & 7.8 (0.4) & 0.7   & 2.6   & 5.5   & 8.0   & 11.9  &       &       &  \\
    Swift   & 6.2 (0.4) & 0.6   & 2.1   & 4.4   &       &       &       &       &  \\
\enddata
\tablenotetext{a}{\citet{Nilsson2007}.}
\tablecomments{Ground-based values are for a typical aperture radius of 7 arcsec, and for Swift, the typical 5 arcsec radius.}
\label{table:hostgalaxy}
\end{deluxetable}

\subsection{Calibration of Near-Infrared through Ultraviolet Spectra}
\label{sec:calibrate}

To determine if any observatory has magnitudes for a band that are consistently higher or lower than other observatories, we examined all measurements for all objects, selecting sets of measurements when a minimum of two observatories observed an object in the same band within the same day. We restricted the observations to days when the source was not active in any NIR through \textit{U} bands. If an observatory had multiple observations within a given day, we computed a weighted mean for each such day and band. We then analyzed the differences between the fluxes from different observatories for a given band based on different epochs and sources.

Overall, no systematic discrepancies appear to be present in any band for any observatory, with the exception of the SMARTS \textit{K} band; hence, these data  are used with caution. All light curves were checked for outliers, which were deleted in the final analysis.


\section{Quiescent and Active Epochs}
\subsection{Properties of Quiescent and Active States}
\label{sec:quiactdefinition}
Our monitoring program has resulted in a sufficient number of quasi-simultaneous measurements of each object at different frequencies to compute and compare the spectral indices when objects were in an active versus a quiescent state. Barring a few definitions of ``bright"  or ``flares'' \citep[see, e.g.,][]{Abdovariability,Nalewajko}, there is no standard definition for ``quiescent'' and ``active.''    We  define these states based on the weighted mean flux, \mf, and its weighted standard deviation, 
\begin{equation}
\sigma_{w_{\nu}} = \sqrt{\frac{\sum \limits_{i=1}^{N} w_{i}(x_{i}-\langle F_{\nu}\rangle)^2} {\frac{(N-1)\sum \limits_{i=1}^{N}  w_{i}}{N}}},
\end{equation}
where $x_{i}$ is a measurement with uncertainty $\sigma_{i}$, $w_i = 1/\sigma_{i}^{2}$ is the weight of the individual measurement,  and $N$ is the number of observations of the source within a given energy band $\nu$.  All measurements from all observatories, subject to restrictions stated in Section \ref{sec:calibrate} and with a self-imposed minimum of ten measurements within a band, were used to compute these values. We set as an upper limit to a \textit{quiescent} flux level  \mf, and as a lower limit to an  \textit{active} flux level   \mf\, +  1$\sigma_{w_{\nu}}$. Between these levels, we consider the source to be in a transitional state. Additionally, we further define a \textit{flaring} flux level to be when the flux exceeds \mf\, +  3$\sigma_{w_{\nu}}$. For \textit{Fermi} data, we include upper limits in the computation of \mfg, replacing both the flux and its error with the value of the upper limit. Table \ref{table:statedata} presents  \mf, its weighted standard deviation, and the number of data points used in its computation for each of the selected frequency bands for each object.


We restrict our analysis to epochs when $\gamma$-ray emission was in a sustained period of either quiescence or active flux levels.  A flaring flux level is, by definition, part of an active state.  Based on our typical 7-day binning of $\gamma$-ray data, we require a quiescent period to extend a minimum of 21 consecutive days, with upper limits considered quiescent, and an active period to extend a minimum of 14 consecutive days. As an example, for 7-day binning of \textit{Fermi} data, a minimum of two consecutive data points at least 1$\sigma_{w_{\gamma}}$ above \mfg\, are required before we consider the source to be in an active state. Active $\gamma$-ray periods thus determined initially have been reevaluated using \textit{Fermi} light curves computed with the photon index allowed to vary. A minor exception is made for the active epochs of 1749+096. We allow two active periods to include epochs when the $\gamma$-ray measurement fell marginally below, and well within the uncertainties, of the lower limit for active periods.  To evaluate the state of a source at the other bands during a given $\gamma$-ray state, no minimum duration is imposed.

Table \ref{tab:fermiperiodsq} presents a summary of the $\gamma$-ray periods of quiescence for BL Lacs and FSRQs. The columns of Table \ref{tab:fermiperiodsq} are as follows: 1 - the object name, 2 - the number of quiescent periods, 3 - the total number of days during which the object was in a quiescent period (note that this excludes any days for which the object had a low flux value but for less than an uninterrupted 21-day period), and 4$-$6 - the number of days in the longest uninterrupted period of quiescence and the dates of the beginning and end of the longest period, respectively. 
Similarly, Table \ref{tab:fermiperiodsa} presents a summary of the $\gamma$-ray active periods (all active periods are identified from the data computed with a fixed photon index): column 1 is the object name, column 2 is the number of active periods identified for the object, column 3 is the number of active periods that have a flux value considered to be in a flaring state, and column 4 is the total number of days during which the object was in an active period. Columns 5$-$7 list the number of days in the longest uninterrupted active period and the dates of the beginning and end of the longest active period, respectively. Columns 8$-$10 give the maximum flux observed, its uncertainty, and the central date of the bin, respectively. The spectral index at the time of measurement of the maximum flux is listed in column 11. If the maximum flux was computed using the fixed photon index in the 2FGL catalog, an ``F" is inserted in column 12; otherwise, if the photon index was allowed to vary, a ``V" is inserted in column 12 (see Section \ref{sec:xandgammasi}). Columns 13 and 14 present the ratio of the maximum flux to \mfg\, and the uncertainty that characterizes an amplitude of $\gamma$-ray variability.


\begin{deluxetable}{lrrrrrrrrrrrrrrr}

\rotate
\tablewidth{0pt}
\tabletypesize{\tiny}
\tablecolumns{16}
\tablecaption{Weighted Mean Flux and Weighted Standard Deviations (Part 1 of 3): Gamma-ray through UVW1 Bands}
\tablehead{ 
			\multicolumn{1}{l}{Object}&
			\multicolumn{3}{c}{\textit{Fermi} $\gamma$-ray [phot cm\superscript{-2} s\superscript{-1}]}&
			\multicolumn{3}{c}{\textit{Swift} XRT [erg cm\superscript{-2} s\superscript{-1}]}&
			\multicolumn{3}{c}{\textit{Swift UVW2} [mJy]}&
			\multicolumn{3}{c}{\textit{Swift UVM2} [mJy]}&
			\multicolumn{3}{c}{\textit{Swift UVW1} [mJy]}			
			\cr
			\multicolumn{1}{l}{Name}&
			\multicolumn{1}{r}{$\langle$F$_{\gamma}\rangle$}&
			\multicolumn{1}{r}{1-$\sigma_{w_{\gamma}}$}&
			\multicolumn{1}{r}{\# Items}&
			\multicolumn{1}{r}{$\langle$F$_{X}\rangle$}&
			\multicolumn{1}{r}{1-$\sigma_{w_{X}}$}&
			\multicolumn{1}{r}{\# Items}&
			\multicolumn{1}{r}{$\langle$F$_{W2}\rangle$}&
			\multicolumn{1}{r}{1-$\sigma_{w_{W2}}$}&
			\multicolumn{1}{r}{\# Items}&
			\multicolumn{1}{r}{$\langle$F$_{M2}\rangle$}&
			\multicolumn{1}{r}{1-$\sigma_{w_{M2}}$}&
			\multicolumn{1}{r}{\# Items}&
			\multicolumn{1}{r}{$\langle$F$_{W1}\rangle$}&
			\multicolumn{1}{r}{1-$\sigma_{w_{W1}}$}&
			\multicolumn{1}{r}{\# Items}
					  \cr
		  	\multicolumn{1}{l}{(1)}&
		  	\multicolumn{1}{r}{(2)}&
		 	\multicolumn{1}{r}{(3)}&
		 	\multicolumn{1}{r}{(4)}&
		 	\multicolumn{1}{r}{(5)}&
		  	\multicolumn{1}{r}{(6)}&
		  	\multicolumn{1}{r}{(7)}&
		  	\multicolumn{1}{r}{(8)}&
		  	\multicolumn{1}{r}{(9)}&
		  	\multicolumn{1}{r}{(10)}&
		  	\multicolumn{1}{r}{(11)}&
		  	\multicolumn{1}{r}{(12)}&
		  	\multicolumn{1}{r}{(13)}&
		  	\multicolumn{1}{r}{(14)}&
		  	\multicolumn{1}{r}{(15)}&
		 	\multicolumn{1}{r}{(16)}
			}

\startdata
    3C66A & 1.25E-07 & 6.99E-08 & 213   & 4.09E-12 & 2.15E-12 & 19    & 2.313 & 0.832 & 16    & 2.368 & 0.883 & 13    & 3.549 & 1.320 & 15 \\
    0235+164 & 1.88E-07 & 2.74E-07 & 206   & 2.28E-12 & 1.96E-12 & 91    & 0.186 & 0.277 & 99    & 0.250 & 0.384 & 95    & 0.292 & 0.460 & 99 \\
    0336-019 & 1.23E-07 & 8.20E-08 & 213   &       &       & 5     &       &       & 7     &       &       & 6     &       &       & 7 \\
    0420-014 & 1.33E-07 & 6.81E-08 & 213   & 2.62E-12 & 6.87E-13 & 16    & 0.136 & 0.038 & 12    & 0.203 & 0.078 & 12    & 0.243 & 0.080 & 13 \\
    0528+134 & 1.10E-07 & 8.28E-08 & 216   & 2.36E-12 & 1.41E-12 & 73    & 0.476 & 0.177 & 10    &       &       & 9     & 0.328 & 0.195 & 14 \\
    0716+714 & 2.17E-07 & 1.38E-07 & 501   & 7.60E-12 & 4.31E-12 & 103   & 3.975 & 2.034 & 76    & 4.316 & 2.107 & 72    & 6.080 & 3.015 & 83 \\
    0735+178 & 7.08E-08 & 3.02E-08 & 212   & 8.12E-13 & 3.75E-13 & 14    & 0.285 & 0.127 & 11    & 0.339 & 0.137 & 10    &       &       & 9 \\
    0827+243 & 1.51E-07 & 1.30E-07 & 228   & 2.76E-12 & 1.85E-12 & 63    & 0.241 & 0.045 & 44    & 0.280 & 0.052 & 50    & 0.376 & 0.071 & 47 \\
    0829+046 & 6.41E-08 & 3.78E-08 & 209   & 1.13E-12 & 5.74E-13 & 16    &       &       & 7     &       &       & 7     &       &       & 7 \\
    0836+710 & 1.58E-07 & 1.42E-07 & 208   & 1.64E-11 & 4.38E-12 & 28    & 0.057 & 0.008 & 12    & 0.077 & 0.009 & 11    & 0.172 & 0.022 & 13 \\
    OJ287 & 1.05E-07 & 8.11E-08 & 209   & 4.81E-12 & 2.31E-12 & 127   & 1.088 & 0.464 & 100   & 1.170 & 0.504 & 90    & 1.773 & 0.757 & 119 \\
    0954+658 & 6.86E-08 & 4.71E-08 & 209   & 2.15E-12 & 1.17E-12 & 14    & 0.097 & 0.046 & 11    &       &       & 9     & 0.164 & 0.067 & 14 \\
    1055+018 & 1.17E-07 & 6.69E-08 & 214   & 2.74E-12 & 9.51E-13 & 13    &       &       & 8     &       &       & 5     &       &       & 5 \\
    Mkn421 & 1.79E-07 & 7.29E-08 & 219   & 6.09E-10 & 1.72E-10 & 288   & 11.771 & 3.668 & 415   & 11.924 & 3.817 & 402   & 15.449 & 5.086 & 388 \\
    1127-145 & 1.03E-07 & 6.01E-08 & 219   & 5.50E-12 & 1.74E-12 & 23    & 0.217 & 0.058 & 19    & 0.274 & 0.069 & 19    & 0.434 & 0.100 & 20 \\
    1156+295 & 1.43E-07 & 1.04E-07 & 214   & 1.34E-12 & 4.73E-13 & 21    &       &       & 8     &       &       & 8     & 0.454 & 0.392 & 10 \\
    1219+285 & 6.03E-08 & 2.62E-08 & 219   & 2.14E-12 & 1.97E-12 & 74    & 1.054 & 0.510 & 70    & 1.119 & 0.540 & 68    & 1.660 & 0.747 & 75 \\
    1222+216 & 2.02E-07 & 3.37E-07 & 213   & 3.21E-12 & 7.65E-13 & 66    & 1.671 & 0.417 & 38    & 1.554 & 0.380 & 41    & 1.983 & 0.502 & 41 \\
    3C273 & 3.14E-07 & 3.34E-07 & 214   & 1.20E-10 & 4.73E-11 & 148   & 24.924 & 3.002 & 29    & 24.202 & 2.688 & 23    & 30.848 & 2.869 & 27 \\
    3C279 & 3.43E-07 & 2.71E-07 & 208   & 1.03E-11 & 2.85E-12 & 284   & 0.353 & 0.454 & 142   & 0.400 & 0.512 & 136   & 0.643 & 0.739 & 170 \\
    1308+326 & 7.38E-08 & 3.64E-08 & 219   & 1.43E-12 & 9.03E-13 & 14    & 0.078 & 0.055 & 10    & 0.120 & 0.064 & 11    &       &       & 7 \\
    1406-076 & 8.50E-08 & 4.02E-08 & 219   & 7.14E-13 & 1.75E-13 & 20    & 0.007 & 0.003 & 42    & 0.015 & 0.007 & 42    & 0.036 & 0.014 & 46 \\
    1510-089 & 6.11E-07 & 6.29E-07 & 214   & 6.78E-12 & 1.85E-12 & 157   & 0.537 & 0.244 & 154   & 0.615 & 0.242 & 141   & 0.731 & 0.335 & 153 \\
    1611+343 & 2.02E-08 & 1.92E-08 & 219   &       &       & 7     &       &       & 4     &       &       & 4     &       &       & 6 \\
    1622-297 & 9.98E-08 & 5.61E-08 & 156   & 2.23E-12 & 9.47E-13 & 39    & 0.256 & 0.060 & 43    & 0.302 & 0.092 & 41    & 0.282 & 0.108 & 47 \\
    1633+382 & 2.44E-07 & 1.87E-07 & 216   & 2.29E-12 & 1.59E-12 & 72    & 0.022 & 0.009 & 54    & 0.044 & 0.018 & 59    & 0.160 & 0.066 & 60 \\
    3C345 & 1.26E-07 & 6.56E-08 & 222   & 4.76E-12 & 1.05E-12 & 27    & 0.234 & 0.079 & 21    & 0.241 & 0.090 & 19    & 0.327 & 0.121 & 23 \\
    1730-130 & 1.82E-07 & 1.25E-07 & 213   & 1.65E-12 & 6.61E-13 & 46    & 0.114 & 0.045 & 37    & 0.182 & 0.063 & 25    & 0.197 & 0.084 & 48 \\
    1749+096 & 8.39E-08 & 5.17E-08 & 233   & 4.17E-12 & 3.05E-12 & 25    & 0.694 & 0.994 & 11    & 0.548 & 0.843 & 11    & 1.104 & 1.642 & 13 \\
    BL Lacertae & 2.40E-07 & 1.63E-07 & 268   & 1.05E-11 & 6.78E-12 & 196   & 1.117 & 0.755 & 182   & 1.443 & 1.015 & 175   & 2.052 & 1.334 & 189 \\
    3C446 & 7.87E-08 & 4.33E-08 & 225   &       &       & 9     &       &       & 2     &       &       & 3     &       &       & 2 \\
    CTA102 & 2.05E-07 & 2.16E-07 & 324   & 4.43E-12 & 2.44E-12 & 53    & 0.437 & 0.306 & 36    & 0.514 & 0.351 & 32    & 0.692 & 0.493 & 39 \\
    3C454.3 & 7.79E-07 & 1.58E-06 & 1214  & 3.41E-11 & 3.31E-11 & 331   & 0.987 & 0.465 & 255   & 1.222 & 0.573 & 251   & 1.682 & 0.865 & 280 \\
 

\enddata
\label{table:statedata}
\end{deluxetable}

\addtocounter{table}{-1}
\begin{deluxetable}{lrrrrrrrrrrrr}
\rotate
\tablewidth{0pt}
\tabletypesize{\tiny}
\tablecolumns{13}
\tablecaption{Weighted Mean Flux and Weighted Standard Deviations (Part 2 of 3): \textit{U - R} Bands}
\tablehead{ 
			\multicolumn{1}{l}{Object}&
			\multicolumn{3}{c}{\textit{U}-BAND [mJy]}&
			\multicolumn{3}{c}{\textit{B}-BAND [mJy]}&
			\multicolumn{3}{c}{\textit{V}-BAND [mJy]}&
			\multicolumn{3}{c}{\textit{R}-BAND [mJy]}			
			\cr
			\multicolumn{1}{l}{Name}&
			\multicolumn{1}{r}{$\langle$F$_{U}\rangle$}&
			\multicolumn{1}{r}{1-$\sigma_{w_{U}}$}&
			\multicolumn{1}{r}{\# Items}&
			\multicolumn{1}{r}{$\langle$F$_{B}\rangle$}&
			\multicolumn{1}{r}{1-$\sigma_{w_{B}}$}&
			\multicolumn{1}{r}{\# Items}&
			\multicolumn{1}{r}{$\langle$F$_{V}\rangle$}&
			\multicolumn{1}{r}{1-$\sigma_{w_{V}}$}&
			\multicolumn{1}{r}{\# Items}&
			\multicolumn{1}{r}{$\langle$F$_{R}\rangle$}&
			\multicolumn{1}{r}{1-$\sigma_{w_{R}}$}&
			\multicolumn{1}{r}{\# Items}
					  \cr
		  	\multicolumn{1}{l}{(1)}&
		  	\multicolumn{1}{r}{(2)}&
		 	\multicolumn{1}{r}{(3)}&
		 	\multicolumn{1}{r}{(4)}&
		 	\multicolumn{1}{r}{(5)}&
		  	\multicolumn{1}{r}{(6)}&
		  	\multicolumn{1}{r}{(7)}&
		  	\multicolumn{1}{r}{(8)}&
		  	\multicolumn{1}{r}{(9)}&
		  	\multicolumn{1}{r}{(10)}&
		  	\multicolumn{1}{r}{(11)}&
		  	\multicolumn{1}{r}{(12)}&
		  	\multicolumn{1}{r}{(13)}
			}

\startdata
    3C66A & 4.492 & 1.733 & 15    & 5.525 & 2.170 & 412   & 7.090 & 2.783 & 590   & 6.795 & 2.941 & 786 \\
    0235+164 & 0.507 & 0.817 & 82    & 0.207 & 0.312 & 427   & 0.497 & 0.750 & 690   & 0.523 & 0.814 & 1249 \\
    0336-019 &       &       & 6     & 0.412 & 0.189 & 13    & 0.377 & 0.129 & 29    & 0.600 & 0.242 & 211 \\
    0420-014 & 0.247 & 0.072 & 10    & 0.441 & 0.330 & 78    & 0.579 & 0.343 & 154   & 0.686 & 0.653 & 365 \\
    0528+134 & 0.262 & 0.122 & 41    & 0.334 & 0.093 & 173   & 0.334 & 0.127 & 203   & 0.296 & 0.049 & 356 \\
    0716+714 & 8.111 & 3.735 & 79    & 10.626 & 4.916 & 960   & 14.759 & 6.199 & 1112  & 21.304 & 9.848 & 1908 \\
    0735+178 & 0.658 & 0.229 & 10    & 0.865 & 0.240 & 21    & 1.218 & 0.402 & 94    & 1.350 & 0.346 & 240 \\
    0827+243 & 0.415 & 0.115 & 45    & 0.460 & 0.088 & 42    & 0.488 & 0.105 & 192   & 0.481 & 0.080 & 276 \\
    0829+046 &       &       & 7     & 0.708 & 0.373 & 29    & 1.491 & 0.754 & 46    & 1.726 & 0.672 & 149 \\
    0836+710 & 0.474 & 0.039 & 11    & 0.582 & 0.050 & 67    & 0.633 & 0.060 & 104   & 0.677 & 0.094 & 313 \\
    OJ287 & 2.400 & 0.960 & 110   & 3.257 & 1.381 & 826   & 4.669 & 1.905 & 1051  & 5.258 & 2.260 & 1215 \\
    0954+658 & 0.261 & 0.099 & 12    & 0.491 & 0.295 & 182   & 0.689 & 0.385 & 229   & 0.950 & 0.552 & 933 \\
    1055+018 &       &       & 7     & 0.355 & 0.165 & 11    & 0.657 & 0.368 & 13    & 0.756 & 0.238 & 100 \\
    Mkn421 & 8.914 & 3.656 & 10    & 18.382 & 6.930 & 94    & 19.096 & 7.779 & 304   & 24.475 & 12.253 & 263 \\
    1127-145 &       &       & 4     & 0.599 & 0.153 & 31    &       &       & 9     & 0.781 & 0.076 & 112 \\
    1156+295 & 0.430 & 0.441 & 13    & 0.824 & 0.932 & 118   & 0.397 & 0.483 & 94    & 0.582 & 0.751 & 565 \\
    1219+285 & 2.102 & 0.887 & 73    & 2.662 & 1.001 & 170   & 3.749 & 1.394 & 334   & 4.729 & 1.369 & 304 \\
    1222+216 & 1.927 & 0.448 & 48    & 2.074 & 0.494 & 50    & 2.162 & 0.596 & 234   & 2.406 & 0.853 & 228 \\
    3C273 &       &       & 4     & 27.088 & 1.877 & 423   & 31.260 & 2.166 & 574   & 31.750 & 2.366 & 248 \\
    3C279 & 0.916 & 1.063 & 171   & 0.779 & 1.040 & 636   & 1.632 & 2.093 & 817   & 0.928 & 0.857 & 914 \\
    1308+326 &       &       & 7     & 0.143 & 0.115 & 20    & 0.174 & 0.158 & 22    & 0.264 & 0.147 & 173 \\
    1406-076 & 0.076 & 0.024 & 47    & 0.100 & 0.033 & 230   & 0.119 & 0.036 & 222   & 0.144 & 0.046 & 286 \\
    1510-089 & 0.963 & 0.431 & 147   & 1.096 & 0.522 & 660   & 1.236 & 0.759 & 850   & 1.347 & 0.642 & 1129 \\
    1611+343 &       &       & 6     & 0.373 & 0.028 & 50    & 0.358 & 0.057 & 54    & 0.433 & 0.055 & 258 \\
    1622-297 & 0.333 & 0.067 & 42    & 0.441 & 0.100 & 223   & 0.756 & 0.222 & 232   & 0.409 & 0.150 & 239 \\
    1633+382 & 0.332 & 0.142 & 56    & 0.439 & 0.228 & 241   & 0.395 & 0.235 & 429   & 0.470 & 0.211 & 717 \\
    3C345 & 0.322 & 0.162 & 22    & 0.474 & 0.332 & 220   & 0.540 & 0.367 & 287   & 0.450 & 0.207 & 699 \\
    1730-130 & 0.183 & 0.083 & 46    & 0.331 & 0.126 & 272   & 0.473 & 0.209 & 271   & 0.874 & 0.470 & 368 \\
    1749+096 & 2.531 & 2.991 & 11    & 2.798 & 3.591 & 55    & 3.213 & 4.443 & 79    & 0.907 & 0.490 & 459 \\
    BL Lacertae & 3.144 & 2.065 & 184   & 4.844 & 3.021 & 818   & 7.710 & 4.571 & 1152  & 13.385 & 7.517 & 1318 \\
    3C446 &       &       & 5     & 0.030 & 0.053 & 13    & 0.152 & 0.035 & 25    & 0.216 & 0.079 & 188 \\
    CTA102 & 0.806 & 0.777 & 33    & 0.752 & 0.890 & 175   & 1.162 & 1.446 & 286   & 1.046 & 1.344 & 794 \\
    3C454.3 & 2.144 & 1.406 & 225   & 2.549 & 1.728 & 988   & 4.061 & 2.780 & 1304  & 3.221 & 3.090 & 1578 \\

\enddata

\label{table:statedata2}
\end{deluxetable}

\addtocounter{table}{-1}
\begin{deluxetable}{lrrrrrrrrrrrr}
\rotate
\centering
\tablewidth{0pt}
\tabletypesize{\tiny}
\tablecolumns{13}
\tablecaption{Weighted Mean Flux and Weighted Standard Deviations (Part 3 of 3): \textit{I - K} Bands}
\tablehead{ 
			\multicolumn{1}{l}{Object}&
			\multicolumn{3}{c}{\textit{I}-BAND [mJy]}&
			\multicolumn{3}{c}{\textit{J}-BAND [mJy]}&
			\multicolumn{3}{c}{\textit{H}-BAND [mJy]}&
			\multicolumn{3}{c}{\textit{K}-BAND [mJy]}			
			\cr
			\multicolumn{1}{l}{Name}&
			\multicolumn{1}{r}{$\langle$F$_{I}\rangle$}&
			\multicolumn{1}{r}{1-$\sigma_{w_{I}}$}&
			\multicolumn{1}{r}{\# Items}&
			\multicolumn{1}{r}{$\langle$F$_{J}\rangle$}&
			\multicolumn{1}{r}{1-$\sigma_{w_{J}}$}&
			\multicolumn{1}{r}{\# Items}&
			\multicolumn{1}{r}{$\langle$F$_{H}\rangle$}&
			\multicolumn{1}{r}{1-$\sigma_{w_{H}}$}&
			\multicolumn{1}{r}{\# Items}&
			\multicolumn{1}{r}{$\langle$F$_{K}\rangle$}&
			\multicolumn{1}{r}{1-$\sigma_{w_{K}}$}&
			\multicolumn{1}{r}{\# Items}
					  \cr
		  	\multicolumn{1}{l}{(1)}&
		  	\multicolumn{1}{r}{(2)}&
		 	\multicolumn{1}{r}{(3)}&
		 	\multicolumn{1}{r}{(4)}&
		 	\multicolumn{1}{r}{(5)}&
		  	\multicolumn{1}{r}{(6)}&
		  	\multicolumn{1}{r}{(7)}&
		  	\multicolumn{1}{r}{(8)}&
		  	\multicolumn{1}{r}{(9)}&
		  	\multicolumn{1}{r}{(10)}&
		  	\multicolumn{1}{r}{(11)}&
		  	\multicolumn{1}{r}{(12)}&
		  	\multicolumn{1}{r}{(13)}
			}   
\startdata
    3C66A & 9.323 & 3.788 & 498   & 12.355 & 4.022 & 245   & 18.191 & 5.699 & 238   & 24.548 & 7.169 & 252 \\
    0235+164 & 0.933 & 2.219 & 469   & 1.835 & 2.465 & 646   & 2.530 & 2.837 & 376   & 5.122 & 6.002 & 547 \\
    0336-019 & 0.773 & 0.323 & 33    &       &       &       &       &       &       &       &       &  \\
    0420-014 & 1.456 & 1.238 & 112   & 1.850 & 0.933 & 55    & 2.833 & 1.632 & 55    & 2.821 & 2.430 & 55 \\
    0528+134 & 0.312 & 0.037 & 16    & 0.627 & 0.303 & 233   & 0.952 & 0.691 & 36    & 1.201 & 0.873 & 36 \\
    0716+714 & 26.156 & 12.430 & 990   & 22.779 & 8.237 & 346   & 45.092 & 13.925 & 137   & 57.176 & 21.710 & 134 \\
    0735+178 & 2.121 & 0.546 & 75    & 3.550 & 0.489 & 19    & 5.131 & 0.630 & 16    & 6.830 & 0.673 & 16 \\
    0827+243 & 0.496 & 0.146 & 36    & 0.602 & 0.119 & 18    & 0.635 & 0.170 & 15    & 0.807 & 0.302 & 15 \\
    0829+046 & 2.315 & 0.926 & 45    & 3.963 & 1.110 & 20    & 5.867 & 1.724 & 16    & 8.277 & 1.849 & 16 \\
    0836+710 & 0.770 & 0.048 & 22    & 1.046 & 0.483 & 35    & 1.106 & 0.612 & 23    & 1.502 & 1.016 & 21 \\
    OJ287 & 7.363 & 3.002 & 460   & 13.000 & 5.704 & 407   & 19.070 & 6.749 & 62    & 35.597 & 17.029 & 313 \\
    0954+658 & 1.345 & 0.538 & 98    & 2.473 & 1.193 & 60    & 3.917 & 1.761 & 54    & 5.756 & 3.064 & 49 \\
    1055+018 &       &       & 9     &       &       &       &       &       &       &       &       &  \\
    Mkn421 & 32.674 & 7.614 & 11    &       &       &       &       &       &       &       &       &  \\
    1127-145 &       &       & 5     & 0.568 & 0.126 & 31    &       &       &       &       &       & 9 \\
    1156+295 & 1.954 & 1.868 & 181   & 1.288 & 0.797 & 33    & 2.176 & 1.339 & 23    & 4.115 & 2.980 & 16 \\
    1219+285 & 5.861 & 1.635 & 109   & 10.653 & 1.798 & 29    & 15.241 & 2.342 & 27    & 17.913 & 2.886 & 23 \\
    1222+216 &       &       & 9     &       &       &       &       &       &       &       &       &  \\
    3C273 & 40.046 & 2.983 & 214   & 40.485 & 2.299 & 310   & 54.576 & 3.834 & 15    & 96.516 & 3.114 & 16 \\
    3C279 & 1.801 & 1.276 & 164   & 2.721 & 2.646 & 416   & 3.693 & 2.067 & 37    & 8.811 & 9.013 & 362 \\
    1308+326 & 0.206 & 0.098 & 31    &       &       &       &       &       &       &       &       &  \\
    1406-076 &       &       & 4     & 0.272 & 0.146 & 181   &       &       &       & 0.379 & 0.144 & 45 \\
    1510-089 & 1.879 & 0.938 & 250   & 2.393 & 1.676 & 444   & 3.323 & 1.101 & 76    & 8.011 & 6.008 & 382 \\
    1611+343 & 0.520 & 0.068 & 57    & 0.418 & 0.062 & 26    & 0.631 & 0.061 & 25    & 0.586 & 0.086 & 24 \\
    1622-297 &       &       & 1     & 0.887 & 0.430 & 189   &       &       &       & 1.669 & 1.146 & 182 \\
    1633+382 & 0.707 & 0.521 & 261   & 0.656 & 0.211 & 62    & 0.923 & 0.338 & 56    & 1.292 & 0.573 & 49 \\
    3C345 & 1.008 & 0.806 & 296   & 1.312 & 0.461 & 67    & 2.249 & 0.768 & 64    & 3.372 & 1.233 & 63 \\
    1730-130 &       &       & 4     & 1.331 & 0.523 & 242   &       &       &       & 3.102 & 1.635 & 213 \\
    1749+096 & 2.275 & 1.197 & 97    &       &       & 8     &       &       &       &       &       &  \\
    BL Lacertae & 18.983 & 8.823 & 750   & 47.059 & 10.521 & 62    & 71.696 & 16.745 & 58    & 92.027 & 21.234 & 59 \\
    3C446 &       &       & 8     & 0.378 & 0.096 & 53    &       &       &       &       &       & 2 \\
    CTA102 & 0.795 & 0.850 & 176   & 0.915 & 0.233 & 26    & 0.929 & 0.285 & 66    & 1.372 & 0.638 & 63 \\
    3C454.3 & 5.153 & 4.962 & 448   & 4.053 & 4.680 & 574   & 2.924 & 3.419 & 141   & 9.302 & 12.282 & 506 \\
\enddata

\label{table:statedata3}
\end{deluxetable}


\begin{deluxetable}{lrrrrr}

\tabletypesize{\small}
\tablewidth{0pt}
\tablecaption{Gamma-Ray Periods of Quiescence}
\tablecolumns{6}
\tablehead{
& Number & Total Days &\multicolumn{3}{c}{\hspace{20pt}Longest Quiescent Period} \\
Object &of & in All & No. of &Start &End \\
Name & Periods & Periods & Days &Date &Date \\
\multicolumn{1}{l}{(1)}&\multicolumn{1}{r}{(2)}&\multicolumn{1}{r}{(3)}&
\multicolumn{1}{r}{(4)}&\multicolumn{1}{r}{(5)}&\multicolumn{1}{r}{(6)}
}
\startdata
    3C66A 		& 11    & 517   & 133   & 5931.55 & 6064.55 \\
    0235+164 	& 3     & 1260  & 1162  & 4958.52 & 6120.55 \\
    0716+714 	& 10    & 431   & 97    & 4810.50 & 4907.52 \\
    0735+178 	& 17    & 701   & 97    & 5413.55 & 5511.51 \\
    0829+046 	& 8     & 889   & 637   & 5504.51 & 6141.54 \\
    OJ287 		& 14    & 1051  & 288   & 4810.50 & 5098.54 \\
    0954+658 	& 11    & 1506  & 267   & 4684.50 & 4951.54 \\
    1055+018 	& 10    & 854   & 439   & 5022.54 & 5462.51 \\
    Mkn421 		& 10    & 433   & 84    & 4858.16 & 4942.16 \\
    1219+285 	& 20    & 715   & 195   & 5638.53 & 5833.53 \\
    1749+096 	& 15    & 1216  & 175   & 5194.16 & 5369.16 \\
    BL Lacertae & 12    & 723   & 189   & 5439.16 & 5628.16 \\
    \\
\textbf{FSRQs} &       &       &       &       &  \\
    0336-019 	& 10    & 869   & 491   & 4691.50 & 5182.51 \\
    0420-014	 & 11   & 784   & 370   & 5344.55 & 5714.55 \\
    0528+134	 & 13   & 1247  & 441   & 4880.50 & 5321.50 \\
    0827+243 	& 5     & 1399  & 780   & 4684.50 & 5464.50 \\
    0836+710 	& 7     & 1072  & 419   & 4895.56 & 5315.52 \\
    1127-145 	& 2     & 1063  & 1042  & 5035.54 & 6078.53 \\
    1156+295 	& 17    & 805   & 168   & 5595.51 & 5763.53 \\
    1222+216 	& 11    & 715   & 253   & 4684.50 & 4937.52 \\
    3C273 		& 12    & 792   & 378   & 5669.50 & 6047.50 \\
    3C279 		& 6     & 650   & 350   & 5845.00 & 6195.00 \\
    1308+326 	& 12    & 1080  & 336   & 5434.51 & 5770.53 \\
    1406-076 	& 19    & 1157  & 210   & 5980.53 & 6190.53 \\
    1510-089 	& 16    & 762   & 112   & 6050.53 & 6162.53 \\
    1611+343 	& 4     & 1213  & 940   & 4683.16 & 5623.53 \\
    1622-297	& 9     & 884   & 401   & 4683.16 & 5084.54 \\
    1633+382 	& 8     & 770   & 238   & 5860.50 & 6098.50 \\
    3C345 		& 13    & 1167  & 399   & 5791.53 & 6190.53 \\
    1730-130 	& 12    & 1251  & 476   & 4790.52 & 5266.54 \\
    3C446 		& 19    & 1260  & 187   & 6062.16 & 6249.16 \\
    CTA102 		& 12    & 1064  & 490   & 4963.16 & 5453.16 \\
    3C454.3 	& 3     & 614   & 477   & 5708.16 & 6185.16 \\
\enddata
\tablecomments{Time has not been adjusted for redshift.\\
\hspace{33pt} --- If two or more quiescent periods have the same longest duration, only the first is shown.}
\label{tab:fermiperiodsq}%
\end{deluxetable}

\begin{deluxetable}{lrrrrrrrrrrrrr}
\rotate
\tabletypesize{\scriptsize}
\tablewidth{0pt}
\tablecaption{Gamma-Ray Active Periods}
\tablecolumns{14}
\centering
\tablehead{
          & Number	    & Number      &Total Days  &\multicolumn{3}{c}{Longest Active Period} &\multicolumn{3}{c}{Overall Highest Flux Measured} &Spec-       &       &       \\
 Object   &of 	        &Flaring      & in All     & No. of  &Start &End  &     &       &       &tral &            & $\langle$F$_{max}\rangle$ / \\
    Name  & Periods     &Periods      & Periods    & Days &Date  &Date &$\langle$F$_{max}\rangle$ & 1-$\sigma$  & Date  &Index&Source  & \mfg   & 1-$\sigma$  \\
    					 
		  	\multicolumn{1}{l}{(1)}&
		  	\multicolumn{1}{r}{(2)}&
		 	\multicolumn{1}{r}{(3)}&
		 	\multicolumn{1}{r}{(4)}&
		 	\multicolumn{1}{r}{(5)}&
		  	\multicolumn{1}{r}{(6)}&
		  	\multicolumn{1}{r}{(7)}&
		  	\multicolumn{1}{r}{(8)}&
		  	\multicolumn{1}{r}{(9)}&
		  	\multicolumn{1}{r}{(10)}&
		  	\multicolumn{1}{r}{(11)}&
		  	\multicolumn{1}{r}{(12)}&
		  	\multicolumn{1}{r}{(13)}&
		  	\multicolumn{1}{r}{(14)}
		  	 }

\startdata
\textbf{BL Lacs}& &       &       &       &       &       &       &       &       &       &       &       &     \\
    3C66A & 10    & 2     & 238   & 70    & 4916.52 & 4986.52 & 7.43E-07 & 6.54E-08 & 4969.02 & -0.85 & F     & 6.0   & 3.4 \\
    0235+164 & 1     & 1     & 84    & 84    & 4691.50 & 4775.50 & 1.39E-06 & 1.27E-07 & 4728.00 & -0.96 & V     & 7.4   & 10.9 \\
    0716+714 & 6     & 3     & 105   & 21    & 6101.50 & 6122.50 & 2.18E-06 & 1.66E-07 & 5857.00 & -0.91 & V     & 10.0  & 6.4 \\
    0735+178 & 5     & 3     & 133   & 77    & 6085.55 & 6162.55 & 2.50E-07 & 4.44E-08 & 6138.05 & -1.05 & F     & 3.5   & 1.6 \\
    0829+046 & 5     & 2     & 119   & 42    & 5182.51 & 5224.51 & 4.43E-07 & 7.93E-08 & 5130.04 & -1.09 & V     & 6.9   & 4.3 \\
    OJ287 & 4     & 3     & 147   & 84    & 5819.54 & 5903.54 & 7.40E-07 & 8.81E-08 & 5872.04 & -1.14 & V     & 7.1   & 5.5 \\
    0954+658 & 3     & 3     & 49    & 21    & 5665.53 & 5686.53 & 3.02E-07 & 4.40E-08 & 5641.03 & -1.42 & F     & 4.4   & 3.1 \\
    1055+018 & 6     & 3     & 161   & 42    & 5623.53 & 5665.53 & 5.11E-07 & 1.54E-07 & 5648.01 & -1.33 & V     & 4.4   & 2.8 \\
    Mkn421 & 5     & 1     & 168   & 98    & 6099.53 & 6197.53 & 8.45E-07 & 6.53E-08 & 6124.03 & -0.75 & V     & 4.7   & 2.0 \\
    1219+285 & 3     & 1     & 63    & 28    & 4683.16 & 4711.16 & 1.88E-07 & 4.08E-08 & 4686.66 & -1.02 & F     & 3.1   & 1.5 \\
    1749+096 & 4     & 3     & 63    & 21    & 4683.16 & 4704.16 & 3.79E-07 & 5.41E-08 & 4686.66 & -1.10 & F     & 4.5   & 2.9 \\
    BL Lacertae & 9     & 5     & 247   & 91    & 5691.16 & 5782.16 & 9.89E-07 & 7.41E-08 & 5708.66 & -1.11 & F     & 4.1   & 2.8 \\
\\
\textbf{FSRQs}& &       &       &       &       &       &       &       &       &       &       &       &     \\
    0336-019 & 8     & 4     & 133   & 21    & 5532.50 & 5553.50 & 4.37E-07 & 6.96E-08 & 5550.00 & -1.48 & F     & 3.6   & 2.4 \\
    0420-014 & 7     & 1     & 168   & 56    & 5175.51 & 5231.51 & 4.92E-07 & 5.70E-08 & 5221.01 & -1.30 & F     & 3.7   & 1.9 \\
    0528+134 & 3     & 2     & 49    & 21    & 5805.55 & 5826.55 & 5.71E-07 & 7.19E-08 & 4723.00 & -1.55 & V     & 5.2   & 4.0 \\
    0827+243 & 2     & 2     & 119   & 77    & 6183.54 & 6260.54 & 7.11E-07 & 8.34E-08 & 6285.04 & -1.30 & V     & 4.7   & 4.1 \\
    0836+710 & 3     & 2     & 91    & 42    & 5894.01 & 5936.01 & 1.61E-06 & 1.32E-07 & 5870.05 & -1.61 & V     & 10.2  & 9.2 \\
    1127-145 & 2     & 1     & 63    & 35    & 4809.16 & 4844.16 & 2.99E-07 & 5.90E-08 & 4777.66 & -1.61 & V     & 2.9   & 1.8 \\
    1156+295 & 4     & 3     & 189   & 84    & 5420.51 & 5504.51 & 9.55E-07 & 7.31E-08 & 5431.01 & -1.13 & V     & 6.7   & 4.9 \\
    1222+216 & 9     & 6     & 350   & 105   & 5343.55 & 5448.55 & 5.82E-06 & 1.78E-07 & 5368.01 & -1.08 & V     & 28.8  & 48.0 \\
    3C273 & 8     & 4     & 273   & 84    & 5049.50 & 5133.50 & 5.30E-06 & 3.94E-07 & 5094.00 & -1.40 & V     & 16.9  & 18.0 \\
    3C279 & 9     & 4     & 287   & 63    & 4839.56 & 4902.56 & 1.91E-06 & 7.74E-08 & 4800.00 & -1.20 & V     & 5.6   & 4.4 \\
    1308+326 & 4     & 3     & 84    & 42    & 4683.16 & 4725.16 & 3.59E-07 & 4.16E-08 & 4714.66 & -1.10 & F     & 4.9   & 2.5 \\
    1406-076 & 4     & 1     & 56    & 14    & 4802.16 & 4816.16 & 2.08E-07 & 5.85E-08 & 5424.01 & -1.43 & F     & 2.5   & 1.4 \\
    1510-089 & 8     & 6     & 315   & 84    & 4951.54 & 5035.54 & 6.37E-06 & 2.01E-07 & 5872.03 & -1.29 & F     & 10.4  & 10.7 \\
    1611+343 & 0     & 0     & 0     &      & &  &  &  &       &  &      &   &  \\
    1622-297 & 2     & 0     & 28    & 14    & 5294.51 & 5308.51 & 2.39E-07 & 7.62E-08 & 5368.01 & -1.34 & F     & 2.4   & 1.6 \\
    1633+382 & 10    & 3     & 357   & 91    & 5678.50 & 5769.50 & 1.50E-06 & 1.08E-07 & 6193.00 & -1.25 & F     & 6.2   & 4.7 \\
    3C345 & 3     & 2     & 91    & 42    & 4951.54 & 4993.54 & 4.32E-07 & 9.28E-08 & 4976.04 & -1.45 & V     & 3.4   & 1.9 \\
    1730-130 & 2     & 1     & 49    & 35    & 5490.50 & 5525.50 & 9.15E-07 & 7.37E-08 & 5501.05 & -0.97 & V     & 5.0   & 3.5 \\
    3C446 & 1     & 0     & 21    & 21    & 4970.16 & 4991.16 & 1.64E-07 & 5.04E-08 & 4987.66 & -1.44 & F     & 2.1   & 1.3 \\
    CTA102 & 7     & 4     & 205   & 84    & 6178.53 & 6262.53 & 4.11E-06 & 1.95E-07 & 6194.03 & -1.01 & V     & 20.1  & 21.2 \\
    3C454.3 & 6     & 5     & 307   & 110   & 5480.16 & 5590.16 & 4.16E-05 & 4.86E-07 & 5522.04 & -1.26 & V     & 53.4  & 108.0 \\
\enddata 
\tablecomments{All flux values are in photon cm$^{-2}$ s$^{-1}$. Time is in the observer's frame, not adjusted for redshift.\\
\hspace{26pt} --- If two or more active periods have the same longest duration, only the first is shown.}   
\label{tab:fermiperiodsa}%
\end{deluxetable}

\begin{figure}[t]%
\centering
\subfloat{
	\label{fig:numper-a}
	\includegraphics[trim=0cm 0cm 0cm 0cm, clip=true,height=0.75\linewidth, angle=0]{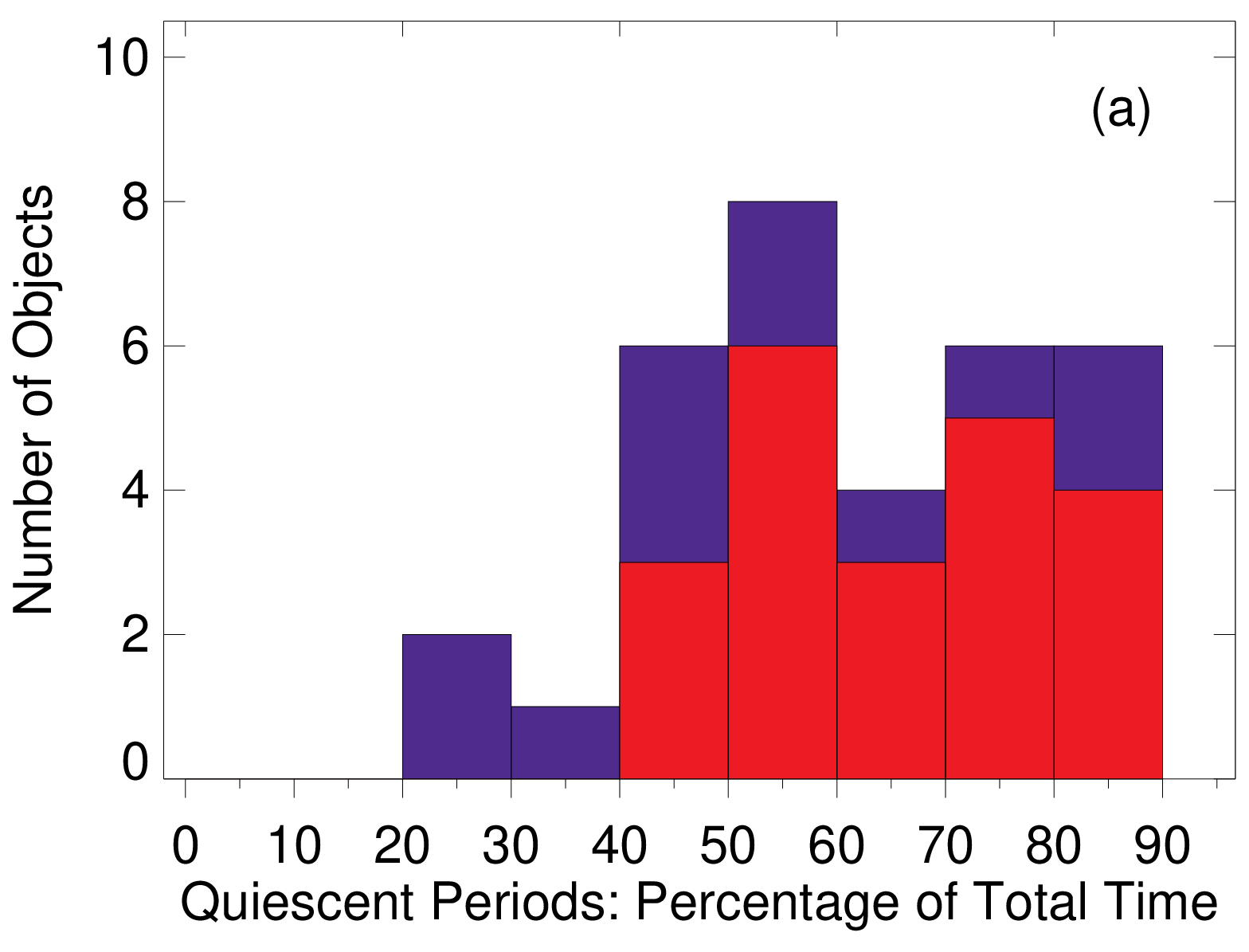}}

\subfloat{
	\label{fig:numper-b}%
	\includegraphics[trim=0cm 0cm 0cm 0cm, clip=true,height=0.75\linewidth, angle=0]{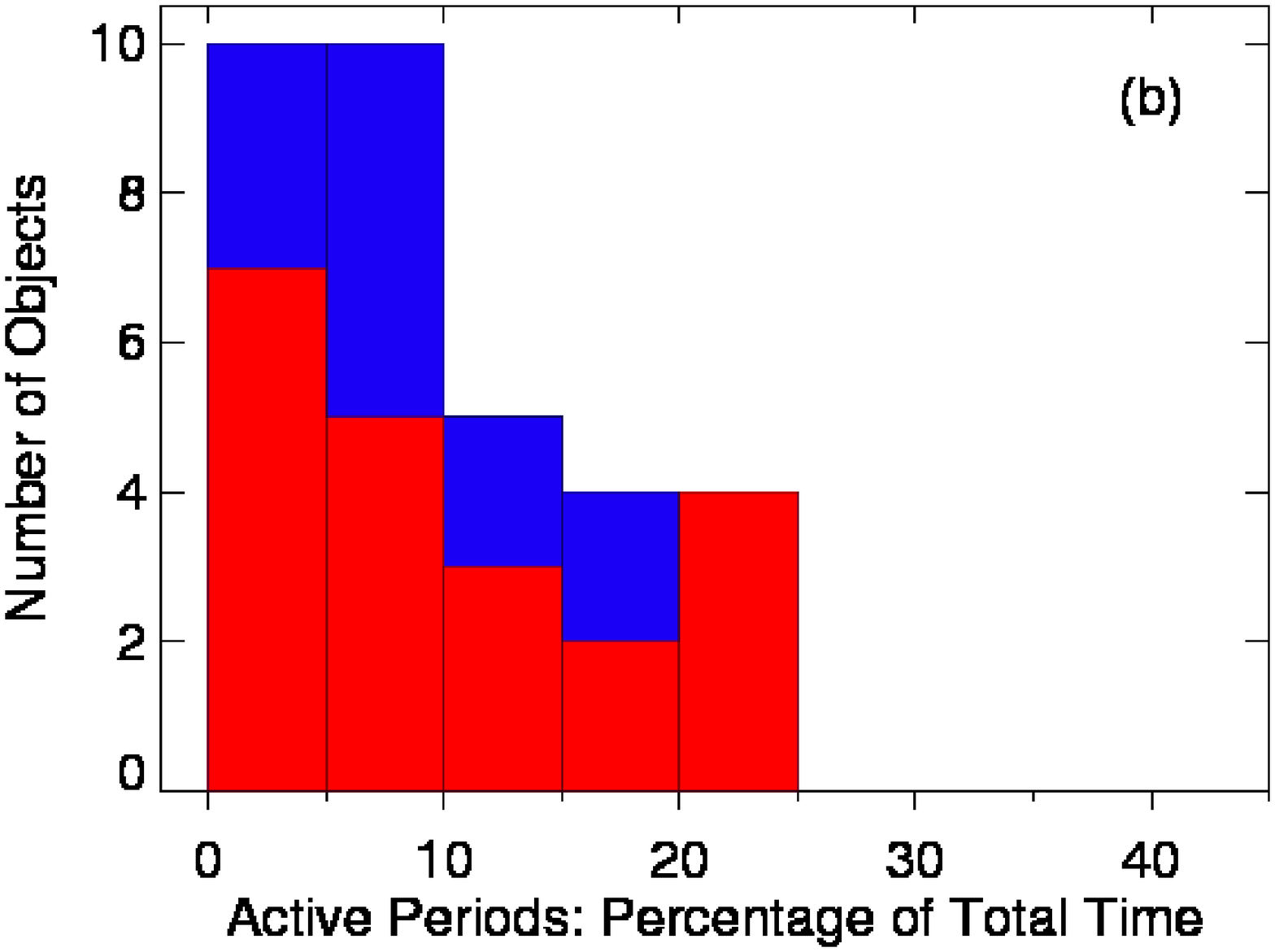}}
	\\

\caption{
Histograms of the percent of time that sources were in a $\gamma$-ray quiescent \protect\subref{fig:numper-a} or active
\protect\subref{fig:numper-b} period. (See text for definitions of periods.) FSRQs are red-filled  and BL Lac objects, blue-filled.}%
\label{fig:numperiods}%
\end{figure}

To identify trends based on the class of objects, we generate a series of plots using the values in Tables \ref{tab:fermiperiodsq} and \ref{tab:fermiperiodsa}. Histograms of the percentage of time that each source was in a quiescent or active period are presented in Figure \ref{fig:numperiods}. Note that measurements in a transitory state or in isolated quiescent/active states are included in the total time. BL Lacs and FSRQs show similar behavior. BL Lacs spend an average of 55 $\pm$ 20\% of their time in quiescent periods, while FSRQs spend 65 $\pm$ 15\% of their time in quiescent periods. Time spent in active periods for BL Lacs is 9 $\pm$ 4\% and for FSRQs, 10 $\pm$ 8\%. Both averaged 5 $\pm$ 3 active periods over the 4.2 years of Fermi measurements included in this study, and BL Lacs averaged 12 $\pm$ 4 quiescent periods and FSRQs 11 $\pm$ 5.

\begin{figure}[t]%
\centering
\subfloat{%
	\label{fig:lengthqui}%
	\includegraphics[trim=0cm 0cm 0cm 0cm, clip=true,height=0.75\linewidth, angle=0]{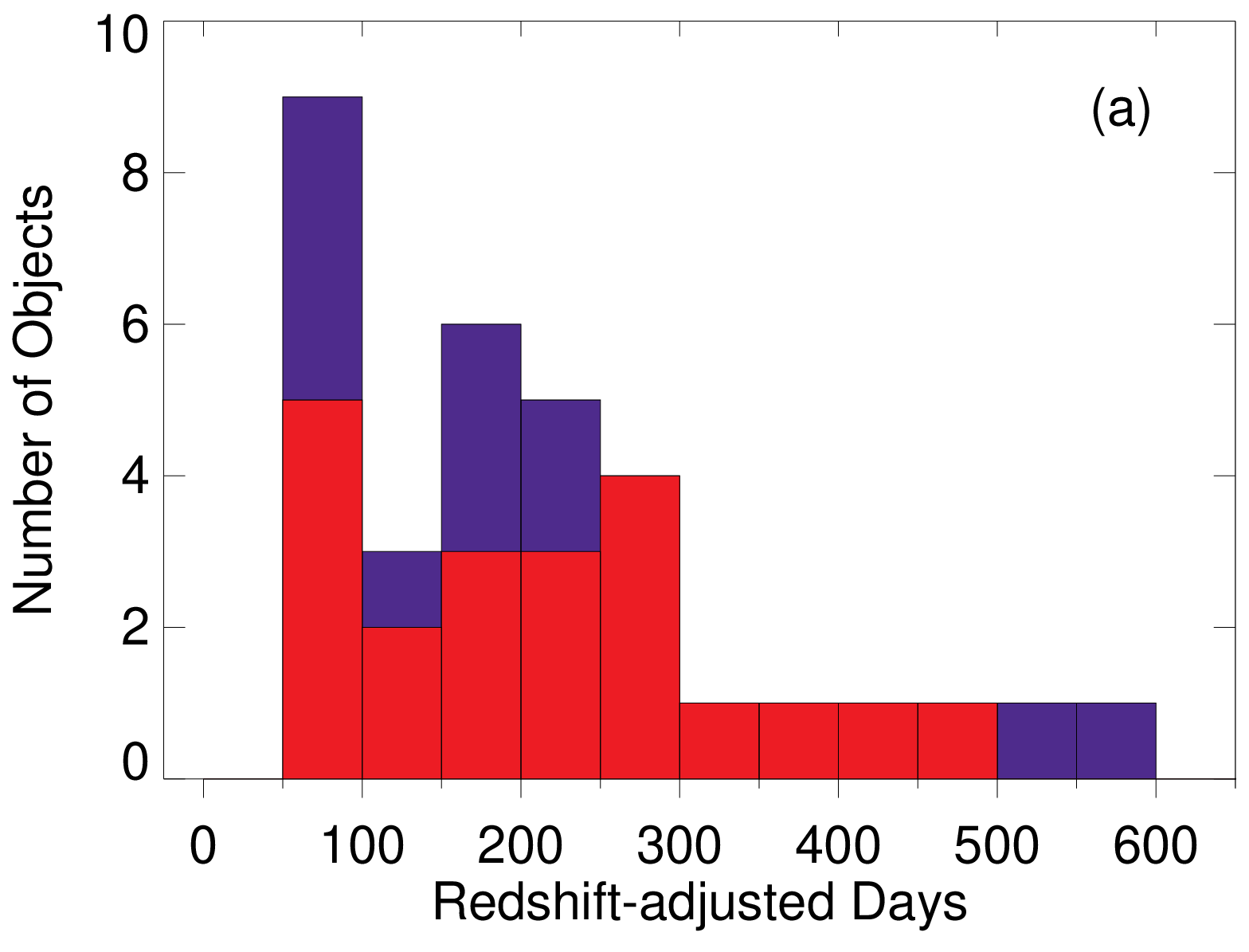}}%
	
\subfloat{%
	\label{fig:lengthact}%
	\includegraphics[trim=0cm 0cm 0cm 0cm, clip=true,height=0.75\linewidth, angle=0]{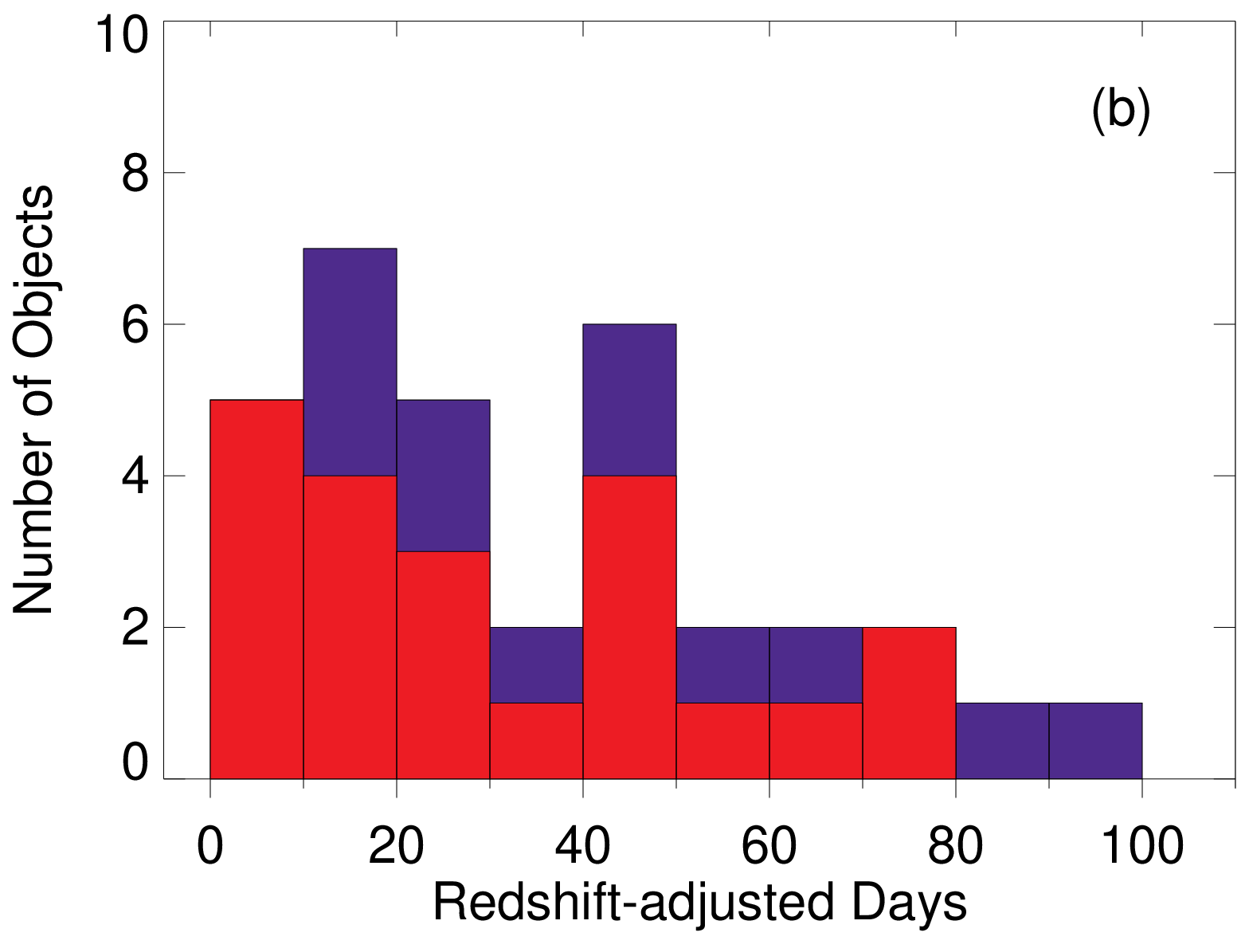}}%
\caption{Histograms of the durations of the longest uninterrupted periods of $\gamma$-ray  \protect\subref{fig:lengthqui} quiescent   or 
\protect\subref{fig:lengthact} active activity, adjusted for redshift. FSRQs are red-filled  and BL Lacs, blue-filled.}%
\label{fig:perlength}%
\end{figure}

Histograms of the longest uninterrupted quiescent and active periods for each of the sources are displayed in Figure \ref{fig:perlength}.  Time is in the host galaxy frame, adjusted for redshift. We checked the ends of the light curves for the longest uninterrupted periods. Four of our objects (0827+243, 0954+658, 1222+216, and 1622-297), were within their longest uninterrupted quiescent period at the start of the \textit{Fermi} mission and the longest uninterrupted active period was in progress for 1308+326. Additionally, 4 of our objects (3C279, 3C345, 3C446, and 3C454.3) were within their longest uninterrupted quiescent period at the end of the monitoring period for this paper. Thus, for these objects, our longest uninterrupted periods represent lower limits. No obvious trends exist for either subclass while in a quiescent state, with both having wide dispersions. The longest uninterrupted quiescent period for most BL Lacs ran from 68 days (0735+178) to 232 days (1055+018), but 0235+164 and 0829+046 had 599 and 543 days, respectively. All but four FSRQs (3C273, 1611+343, 0827+243, and 1127-145) had fewer than 265 days in their longest uninterrupted quiescent period, with the length generally equally dispersed from a minimum of 78 days (3C446). The longest uninterrupted active periods were also highly dispersed for both  subclasses, with BL Lac objects generally having a longer  uninterrupted period (ranging from 15 to 95 days and averaging 43 $\pm$ 27 days) than FSRQs (ranging from 6 to 73 days and averaging 30 $\pm$ 23 days) when converted to the respective galaxies' restframes.  

\begin{figure}[t]%
\centering
	\includegraphics[trim=1.8cm 0cm 0cm 0cm, clip=true,height=0.75\linewidth, angle=0]{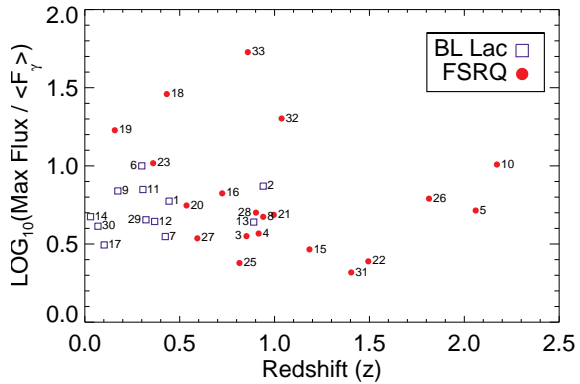}%

\caption{Maximum amplitude of $\gamma$-ray variations  achieved by each object (values listed in Table \ref{tab:fermiperiodsa}) vs.\ redshift. The labels refer to the object reference number (see Table \ref{tab:sources}). The highest amplitudes correspond to 3C454.3 (\#33), 1222+216 (\#18), CTA102 (\#32), and 3C273 (\#19). 
 }%
\label{fig:pernorm}%
\end{figure}

We plot the normalized amplitude of flux variations vs.\ redshift in Figure \ref{fig:pernorm}. Noticeable is the lack of BL Lacs displaying large amplitudes. The average normalized amplitudes are $5.5 \pm 2.0$ for the BL Lacs and a highly dispersed 10 with a standard deviation of 12 for the FSRQs. However, without the four quasars exhibiting the largest values of  maximum to mean fluxes (3C454.3, 1222+216, CTA102, and 3C273), the normalized maximum flux  average for FSRQs drops to  5.0 $\pm$ 2.5. If the BL Lacs displayed such large amplitudes of $\gamma$-ray outbursts at the same rate as the FSRQs, we could expect, at least, 2 BL Lac objects with large outbursts. This implies that the process responsible for activity in the BL Lacs is more uniform, while the FSRQs appear to have different levels of activity.

\subsection{Selection of Representative Epochs}
\label{sec:composition}
To form a well-sampled, representative selection of data for a statistical study of spectral indices, we establish minimum requirements for epochs of data to be extracted for analysis.  Because many objects have multiple epochs that can be classified as quiescent or active, and in order to avoid skewing the analysis towards any particular object, four epochs per object are selected for analysis for the majority of the sources, two within $\gamma$-ray quiescent periods and two within $\gamma$-ray active periods. Fewer than four epochs are used for ten sources because of either weak $\gamma$-ray emission and insufficient optical-UV data, or lack of simultaneity  of observations across bands. An ideal epoch would include a sufficient number of observations to construct a complete SED and compute spectral indices for the $\gamma$-ray, X-ray, and UV-optical-NIR regions, although some epochs are accepted without X-ray measurements. Epochs are carefully selected to include a minimum separation of time between earliest and latest NIR through X-ray observations, never to exceed 24 hours, resulting in an average elapsed time of measurements for all selected epochs of 9.0 hours. Preference is given to epochs that include a wide range of NIR to UV wavebands and to epochs containing observations obtained from the greatest number of observatories to mitigate potential bias introduced by the use of data from a single observatory.

\subsection{Light Curves and SEDs}

\begin{figure*}
\centering
\includegraphics[width=0.65\paperwidth, angle=0]{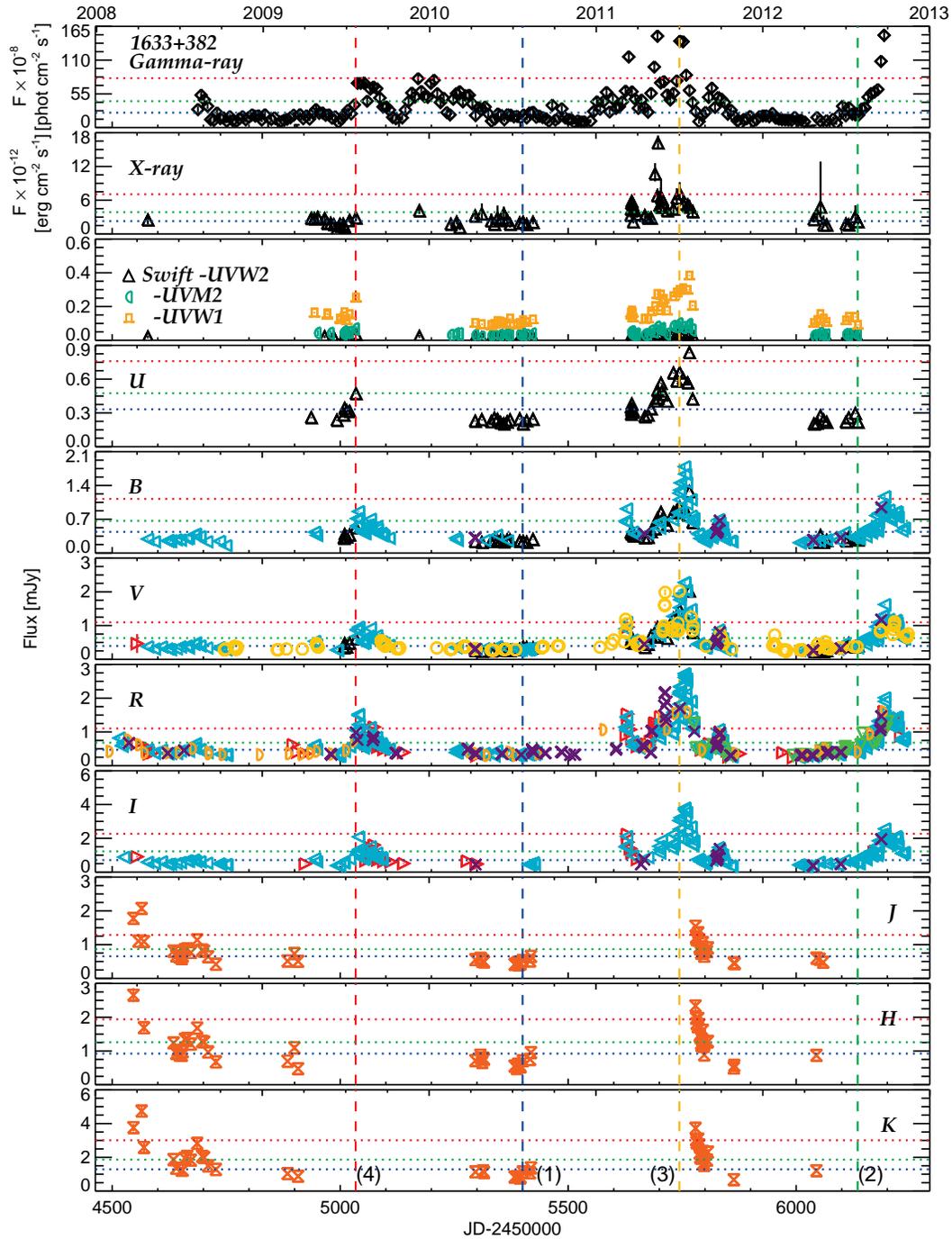}
\caption{Light curves at different wavebands from NIR to $\gamma$-ray frequencies, with 1633+382 presented as an example. Energy range of the $\gamma$-ray flux is 0.1$-$200 GeV and for X-ray flux, 0.3$-$10 keV. Symbols identify telescopes used in measurements (see Table~\ref{table:obs}). Horizontal dotted lines on the light curves indicate the upper limit for quiescent states (blue) and lower limits for active states (green) and flaring states (red). Vertical dashed lines indicate specific epochs of interest, each designated with an identifying number located in the lowest panel. [Light curves collected for all objects can be found in an expanded version of this paper at \url{www.bu.edu/blazars/VLBAproject.html}.]}
\label{fig:1633lc}
\end{figure*}

Figure~\ref{fig:1633lc} presents the light curves of the quasar 1633+382 as an example of the data used in the analysis. (Light curves collected for all objects can be found in an expanded version of this paper at \url{www.bu.edu/blazars/VLBAproject.html}.) The light curves are presented in a series of sub-panels, with the  highest frequency in the top panel and the lowest in the bottom panel. The energy range of the $\gamma$-ray flux is 0.1$-$200 GeV and of the X-ray flux, 0.3$-$10 keV.  The observatory making the measurement is identified by the color and shape of the symbol. Table~\ref{table:obs} presents the legend for the observatories. As explained in $\S$\ref{sec:composition}, up to four epochs per source were selected for analysis. These are indicated on the light curve plots by vertical dashed lines, with each epoch identified by a number and a color.  Quiescent epochs are colored blue and green, active epochs are yellow and red. Horizontal dotted lines indicate upper limits of quiescent states and lower limits of active and flaring states.

\begin{figure*}[t]
\centering
\subfloat{%
	\label{fig:SED0716}%
	\includegraphics[trim=0cm 0cm 0cm 0cm, clip=true,height=0.37\linewidth, angle=0]{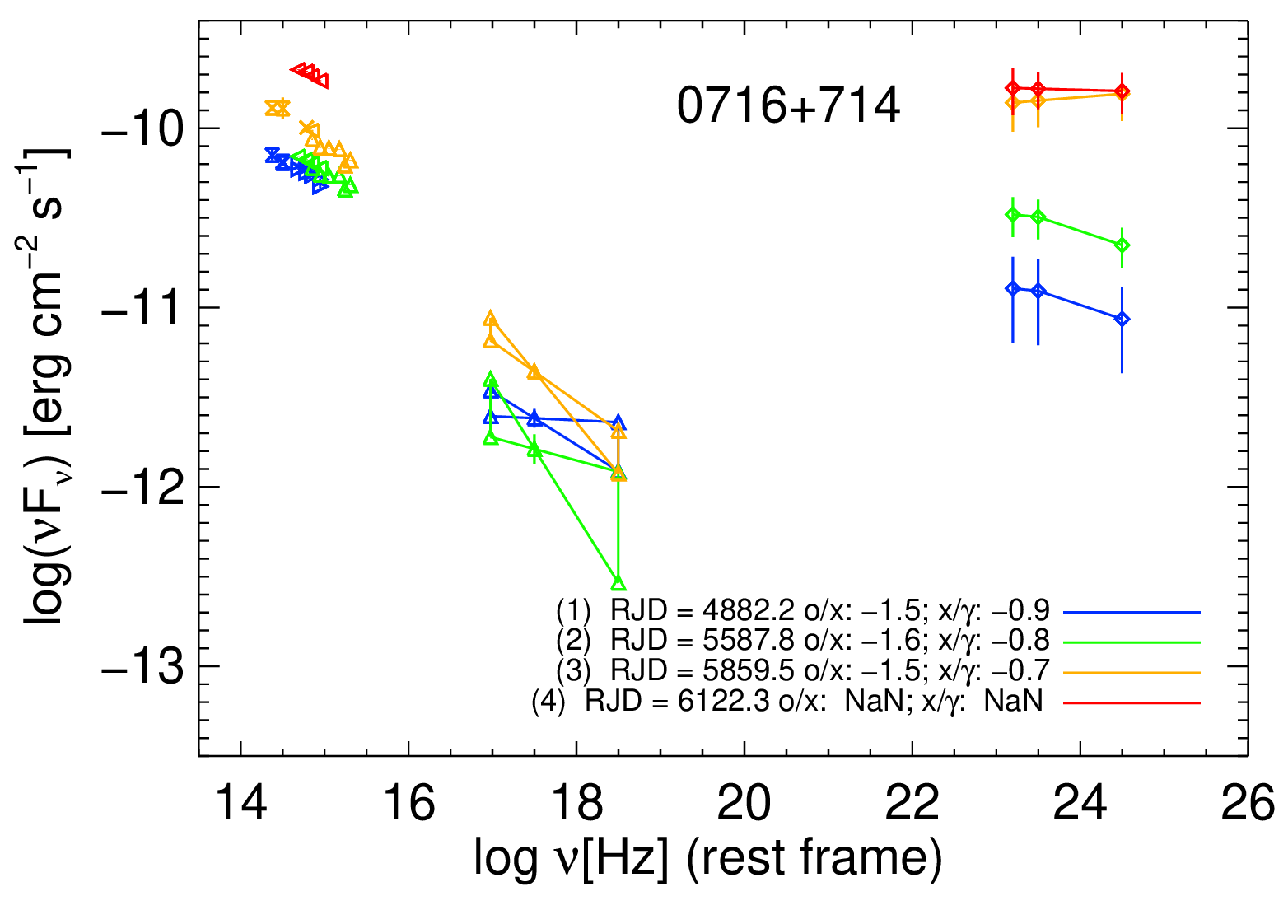}}%
\subfloat{%
	\label{fig:SED1633}%
	\includegraphics[trim=1.4cm 0cm 0cm 0cm, clip=true,height=0.37\linewidth, angle=0]{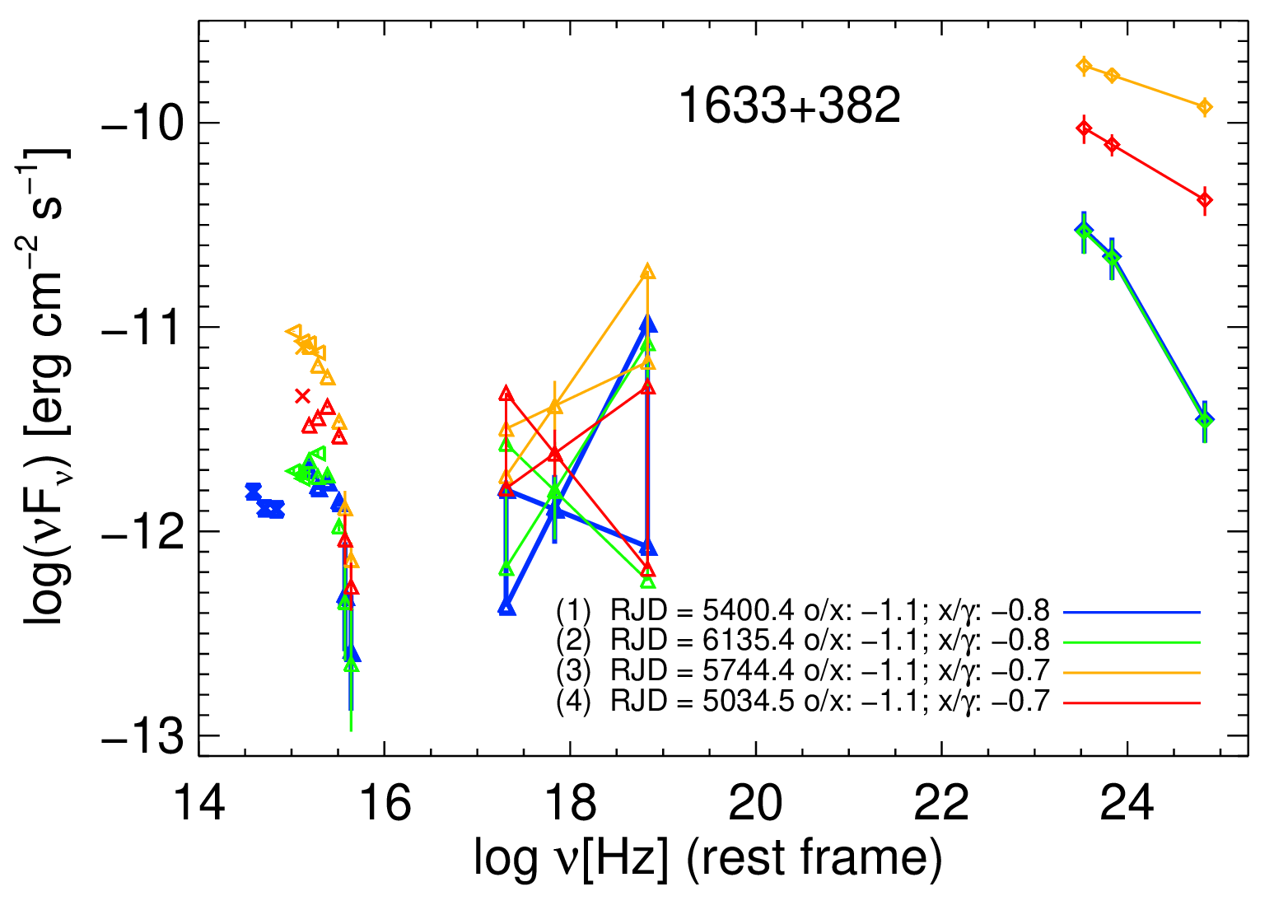}}\\
	
\caption{SEDs for 0716+714 and 1633+382, shown as examples.  Each epoch retains the identifying color and epoch number as displayed with vertical dashed lines on the light curves. The symbols (but not the color) refer to the observatory making the measurement (see Table \ref{table:obs}). Frequency is adjusted to the object's rest frame. For convenience, \oxsi\, and \xgsi\, are shown if \textit{Swift} X-ray data are available at the epoch. [SEDs for all objects can be found in an expanded version of this paper at \url{www.bu.edu/blazars/VLBAproject.html}.]
}%
\label{fig:SED}%

\centering
\subfloat{%
	\label{fig:SI0716}%
	\includegraphics[trim=0cm 0cm 0cm 0cm, clip=true,height=0.35\linewidth, angle=0]{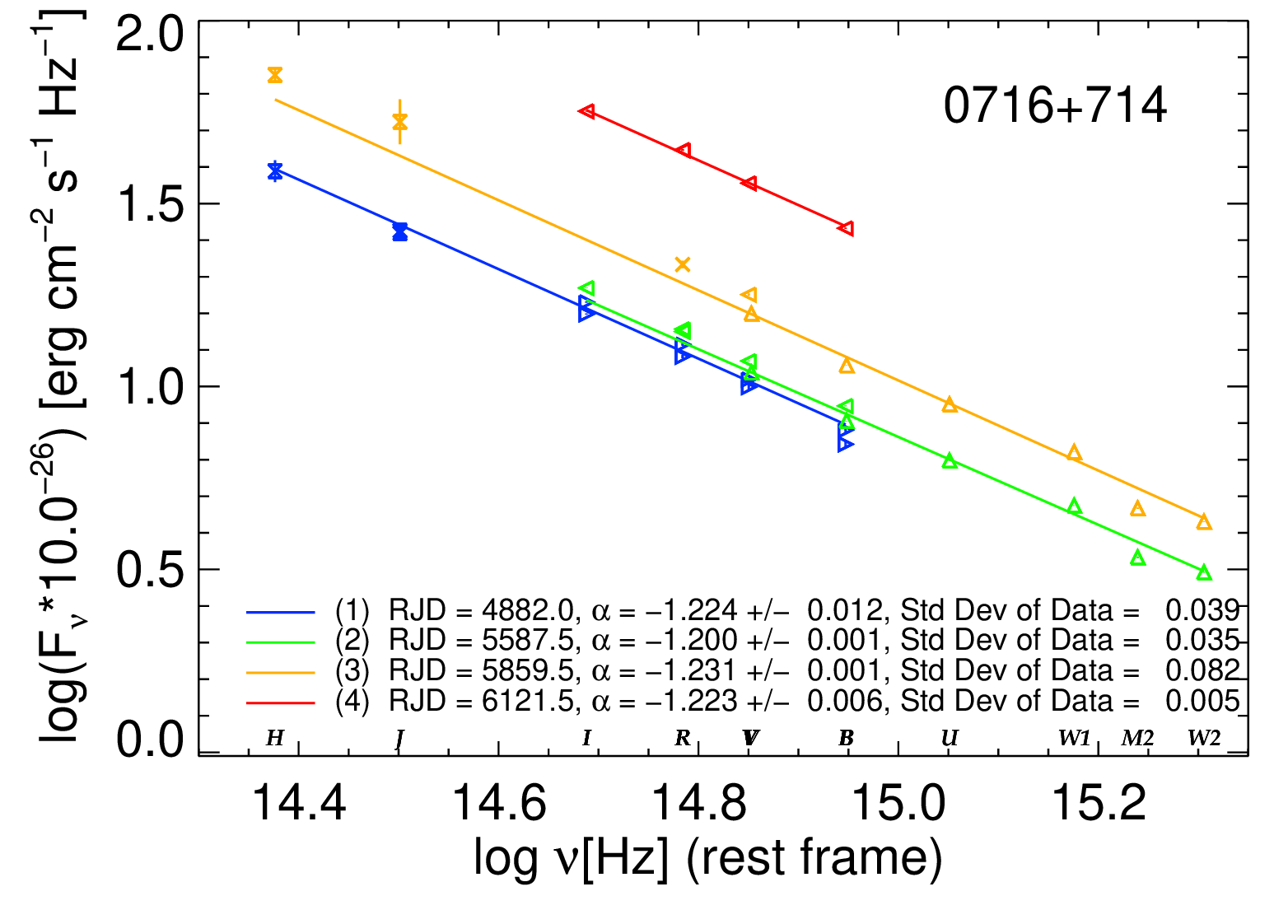}}%
\hspace{8pt}%
\subfloat{%
	\label{fig:SI1633}%
	\includegraphics[trim=1.8cm 0cm 0cm 0cm, clip=true,height=0.35\linewidth, angle=0]{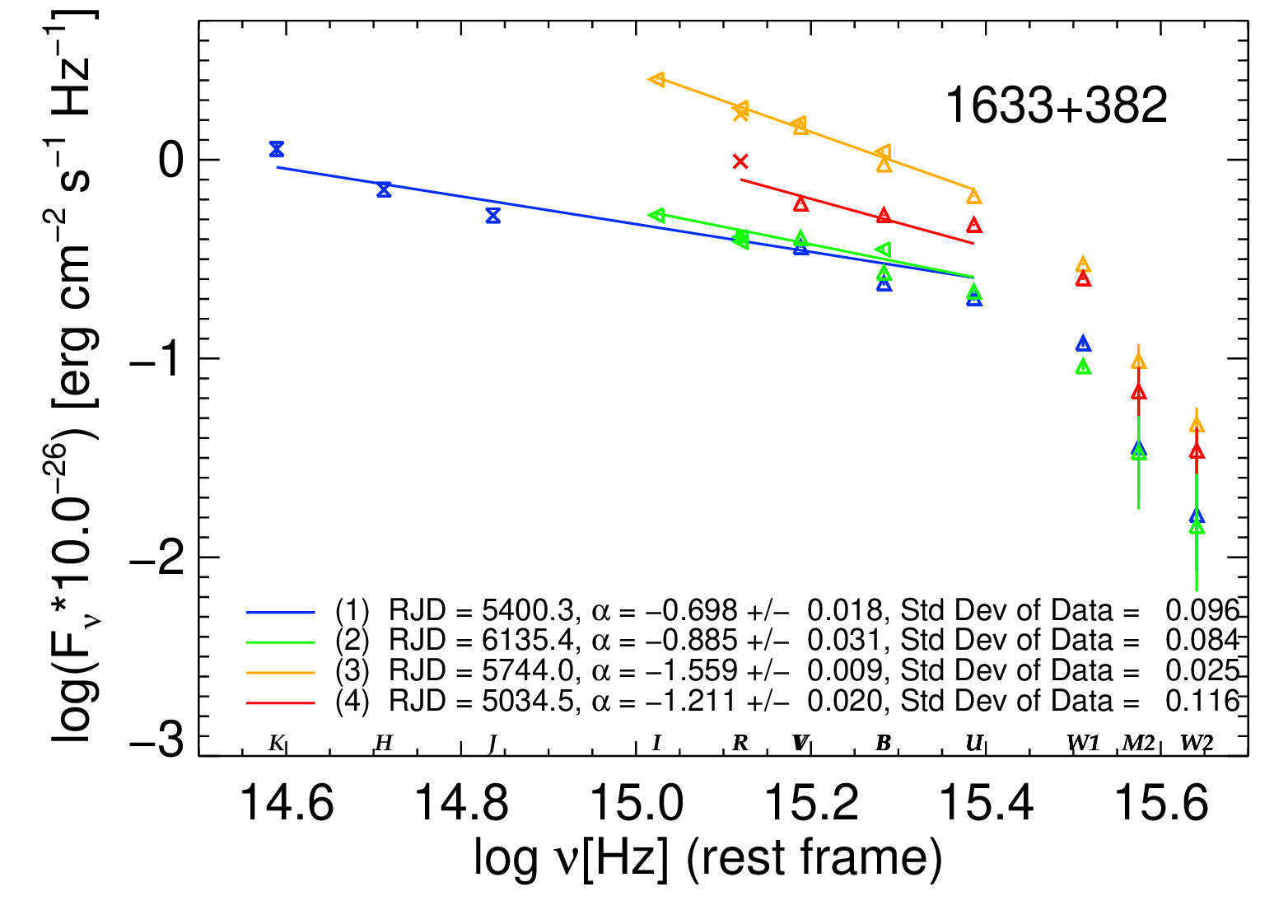}}\\
	
\caption{Examples of optical spectral index computation. Each epoch retains the identifying color and epoch number as displayed with vertical dashed lines on the light curves. The symbols indicate the observatory (see Table \ref{table:obs}). The frequency band of the observation is denoted immediately above the \textsl{X}-axis. Frequencies are adjusted for redshift.
}%
\label{fig:SI}%
\end{figure*}

Figure \ref{fig:SED} presents SEDs for 0716+714 and 1633+382 as examples. (SEDs for all objects can be found in an expanded version of this paper at \url{www.bu.edu/blazars/VLBAproject.html}.) The SEDs display the flux data for each selected epoch, with the frequency adjusted to the rest frame of the host galaxy. Information about the selected epochs is given in Table \ref{tab:selecteddata}, where column 1 is the object name, column 2 is the identifying epoch number (corresponding to the number displayed on the light curve plot), column 3 is the date of the earliest NIR $-$ X-ray observation within the epoch, column 4 is the elapsed time in days between the earliest and the latest NIR $-$ X-ray  observations of the epoch, column 5 is the date of the center of the \textit{Fermi} binned record, and column 6 is the bin size for that record. Columns 8-20 indicate the activity state of the  object at different bands during the epoch: ``Q'' is quiescent, ``A'' is active, ``F'' is flaring, and ``T'' is transient. A dash indicates that although we had some data available for the band, there were fewer than 10 measurements and we did not compute \mf. If there are multiple observations at a particular waveband, the activity state is determined based on the weighted mean of the observations.

\clearpage
\begin{deluxetable}{lcrrrrccccccccccccc}

\rotate
\tablewidth{0pt}
\tabletypesize{\tiny}
\tablecolumns{16}
\tablecaption{Epochs Selected For Study}
\tablehead{
			\multicolumn{2}{l}{}&
			\multicolumn{2}{c}{Non-$\gamma$ Observations}&
			\multicolumn{2}{c}{$Fermi$ Obsvs.}&
			\multicolumn{10}{l}{}&						
			\cr
			\multicolumn{1}{l}{Object}&
			\multicolumn{1}{c}{Epoch}&
			\multicolumn{1}{r}{Earliest Date }&
			\multicolumn{1}{r}{Elapsed}&
			\multicolumn{1}{r}{Mid-Bin}&
			\multicolumn{1}{r}{Bin}&
			\multicolumn{13}{c}{Activity State of Frequency Band}

			\cr
			\multicolumn{1}{l}{Name}&
			\multicolumn{1}{c}{Number}&
			\multicolumn{1}{r}{Within Epoch}&
			\multicolumn{1}{r}{Timespan}&
			\multicolumn{1}{r}{Date}&
			\multicolumn{1}{r}{Size}&
			\multicolumn{1}{c}{$G$}&
			\multicolumn{1}{c}{$X$}&
			\multicolumn{1}{c}{$W2$}&
			\multicolumn{1}{c}{$M2$}&
			\multicolumn{1}{c}{$W1$}&
			\multicolumn{1}{c}{$U$}&
			\multicolumn{1}{c}{$B$}&
			\multicolumn{1}{c}{$V$}&
			\multicolumn{1}{c}{$R$}&
			\multicolumn{1}{c}{$I$}&
			\multicolumn{1}{c}{$J$}&
			\multicolumn{1}{c}{$H$}&
			\multicolumn{1}{c}{$K$}
					  \cr
		  	\multicolumn{1}{l}{(1)}&
		  	\multicolumn{1}{c}{(2)}&
		 	\multicolumn{1}{r}{(3)}&
		 	\multicolumn{1}{r}{(4)}&
		 	\multicolumn{1}{r}{(5)}&
		  	\multicolumn{1}{r}{(6)}&
		  	\multicolumn{1}{c}{(7)}&
		  	\multicolumn{1}{c}{(8)}&
		  	\multicolumn{1}{c}{(9)}&
		  	\multicolumn{1}{c}{(10)}&
		  	\multicolumn{1}{c}{(11)}&
		  	\multicolumn{1}{c}{(12)}&
		  	\multicolumn{1}{c}{(13)}&
		  	\multicolumn{1}{c}{(14)}&
		  	\multicolumn{1}{c}{(15)}&
		  	\multicolumn{1}{c}{(16)}&
		  	\multicolumn{1}{c}{(17)}&
		  	\multicolumn{1}{c}{(18)}&
		 	\multicolumn{1}{c}{(19)}
		 	}    

\startdata
    3C66A & 1     & 5784.410 & 0.129 & 5781.048 & 7.0   & Q     & Q     & Q     & Q     & Q     & Q     & Q     & Q     & Q     & Q     & Q     & Q     & Q \\
          & 2     & 6185.907 & 0.965 & 6187.048 & 7.0   & Q     &       &       &       &       &       & Q     & Q     & Q     & Q     &       &       &  \\
          & 3     & 4744.658 & 0.763 & 4744.000 & 7.0   & F     & A     & A     & A     & T     & T     & T     & T     & A     & T     &       &       &  \\
          & 4     & 5390.561 & 0.977 & 5479.999 & 7.0   & A     &       &       &       &       &       & A     & A     & A     & A     & F     & F     & A \\
    0235+164 & 1     & 5087.774 & 0.052 & 5087.996 & 7.0   & Q     & Q     &       &       & Q     &       & T     & T     & T     &       & Q     &       & Q \\
          & 2     & 5128.683 & 0.192 & 5130.009 & 7.0   & Q     &       &       &       &       &       & T     & Q     & Q     & Q     & Q     &       & Q \\
          & 3     & 4729.762 & 0.393 & 4731.000 & 3.0   & F     & T     & A     & A     & A     & A     & F     & F     & F     &       & F     &       &  \\
          & 4     & 4758.502 & 0.970 & 4758.000 & 3.0   & F     & F     & A     & A     & F     & A     & F     & F     & F     &       & F     & F     & F \\
    0336-019 & 1     & 4711.555 & 0.032 & 4714.500 & 5.5   & Q     &       &       &       &       &       &       & T     & Q     & Q     &       &       &  \\
          & 2     & 4917.231 & 0.109 & 4894.500 & 60.0  & Q     & -     & -     & -     & -     & -     & Q     & Q     &       &       &       &       &  \\
          & 3     & 5832.461 & 0.466 & 5831.644 & 7.0   & A     &       &       &       &       &       & T     & A     & T     & T     &       &       &  \\
          & 4     & 5858.888 & 0.016 & 5859.644 & 7.0   & F     &       &       &       &       &       & A     & F     & A     & A     &       &       &  \\
    0420-014 & 1     & 5124.447 & 0.469 & 5123.009 & 7.0   & Q     &       &       &       &       &       & T     & T     & T     & Q     &       &       &  \\
          & 2     & 5508.897 & 0.964 & 5512.509 & 32.5  & Q     & Q     & Q     & Q     & Q     & Q     & Q     & Q     & Q     &       &       &       &  \\
          & 3     & 5217.245 & 0.082 & 5214.009 & 7.0   & A     &       &       &       &       &       & F     & F     & F     & F     &       &       &  \\
          & 4     & 5899.334 & 0.015 & 5900.048 & 7.0   & A     &       &       &       &       &       & T     & T     & T     & Q     &       &       &  \\
    0528+134 & 1     & 5120.762 & 0.201 & 5122.000 & 7.0   & Q     &       &       &       &       &       & Q     & Q     & Q     & T     & Q     &       &  \\
          & 3     & 5825.600 & 0.407 & 5823.048 & 7.0   & A     &       &       &       &       &       & T     & T     & A     & A     &       &       &  \\
    0716+714 & 1     & 4882.182 & 0.371 & 4882.000 & 3.0   & Q     & Q     &       &       &       &       & Q     & Q     & Q     & Q     & T     & Q     &  \\
          & 2     & 5587.747 & 0.631 & 5588.000 & 3.0   & Q     & Q     & Q     & Q     & Q     & Q     & Q     & Q     & Q     & Q     &       &       &  \\
          & 3     & 5859.502 & 0.529 & 5860.000 & 3.0   & A     & T     & T     & T     & T     & T     & T     & T     & T     &       & F     & A     &  \\
          & 4     & 6122.279 & 0.007 & 6183.000 & 3.0   & A     &       &       &       &       &       & F     & F     & A     & A     &       &       &  \\
    0735+178 & 1     & 5503.001 & 0.628 & 5501.009 & 7.0   & Q     & Q     & Q     & Q     & -     & Q     & Q     &       & Q     &       & T     & T     & T \\
          & 2     & 6011.277 & 0.347 & 6012.048 & 7.0   & Q     &       &       &       &       &       &       & T     & A     & T     &       &       &  \\
          & 3     & 6070.398 & 0.244 & 6068.048 & 7.0   & A     & F     & A     & A     & -     & A     & A     & A     & A     &       &       &       &  \\
    0827+243 & 1     & 4767.528 & 0.501 & 4759.500 & 30.0  & Q     &       &       &       &       &       & Q     & Q     & Q     &       &       &       &  \\
          & 2     & 5503.630 & 0.365 & 5509.500 & 30.0  & Q     & A     & Q     & T     & Q     & Q     &       & T     & T     &       & A     & A     & A \\
          & 3     & 6198.701 & 0.585 & 6201.041 & 7.0   & A     & A     & A     & A     & A     & A     &       & A     & F     &       &       &       &  \\
          & 4     & 6284.264 & 0.039 & 6285.041 & 7.0   & F     & F     & A     & A     & A     & A     &       &       &       &       &       &       &  \\
    0829+046 & 1     & 5663.736 & 0.006 & 5655.510 & 30.0  & Q     &       &       &       &       &       & T     & Q     & Q     & Q     &       &       &  \\
          & 2     & 6089.713 & 0.010 & 6090.541 & 30.0  & Q     & A     & -     &       &       & -     &       & Q     &       &       &       &       &  \\
          & 3     & 5234.339 & 0.020 & 5235.044 & 7.0   & A     &       &       &       &       &       & F     & A     & A     & A     &       &       &  \\
    0836+710 & 1     & 5624.396 & 0.024 & 5627.028 & 14.0  & Q     &       &       &       &       &       & Q     & T     & T     &       &       &       &  \\
          & 2     & 6020.937 & 0.011 & 6023.506 & 7.0   & Q     & T     & T     & T     & T     & A     & T     & Q     &       &       &       &       &  \\
          & 3     & 5869.336 & 0.348 & 5870.047 & 7.0   & F     & T     & T     &       &       &       &       &       &       &       & Q     & Q     & Q \\
          & 4     & 5923.485 & 0.731 & 5926.047 & 7.0   & F     & F     & T     & T     & T     & T     & T     & Q     & A     &       &       &       &  \\
    OJ287 & 1     & 5296.542 & 0.837 & 5298.024 & 7.0   & Q     & Q     & T     & T     & T     & T     & T     & T     & T     & T     & Q     &       & Q \\
          & 2     & 5340.494 & 0.852 & 5340.024 & 7.0   & Q     & Q     & Q     & Q     & Q     & Q     & Q     & Q     & Q     & Q     & Q     &       & Q \\
          & 3     & 5129.847 & 0.567 & 5130.044 & 7.0   & A     & A     & T     & T     & T     & T     & T     & T     & T     & T     & A     &       & A \\
          & 4     & 6038.508 & 0.833 & 6040.041 & 7.0   & A     &       &       &       &       &       & A     & A     & A     & A     & A     &       & A \\
    0954+658 & 1     & 4766.685 & 0.918 & 4774.500 & 60.0  & Q     &       &       &       &       &       & Q     & Q     & Q     &       & Q     & Q     & Q \\
          & 2     & 4781.697 & 0.006 & 4774.500 & 60.0  & Q     &       &       &       &       &       &       &       &       &       & Q     & Q     & Q \\
          & 3     & 5636.232 & 0.384 & 5634.028 & 7.0   & A     &       &       &       &       &       & A     & F     & A     &       &       &       &  \\
          & 4     & 5667.827 & 0.539 & 5669.028 & 7.0   & F     &       &       &       &       &       & A     & A     & A     & A     &       &       &  \\
    1055+018 & 1     & 5305.329 & 0.571 & 5306.542 & 30.0  & Q     & Q     & -     & -     & -     & -     & T     & Q     & Q     &       &       &       &  \\
          & 2     & 6046.729 & 0.022 & 6047.028 & 7.0   & Q     &       &       &       &       &       & Q     & Q     & Q     & -     &       &       &  \\
          & 3     & 5664.801 & 0.007 & 5662.010 & 7.0   & A     &       &       &       &       &       & A     & A     & A     & -     &       &       &  \\
          & 4     & 5709.314 & 0.041 & 5711.010 & 7.0   & A     & A     & -     & -     & -     & -     & F     & A     &       &       &       &       &  \\
    Mkn421 & 1     & 5306.337 & 0.010 & 5305.011 & 7.0   & Q     &       &       &       &       &       & Q     & Q     & Q     &       &       &       &  \\
          & 2     & 5729.351 & 0.017 & 5732.028 & 7.0   & Q     &       &       &       &       &       & Q     & Q     & Q     &       &       &       &  \\
          & 3     & 5319.355 & 0.024 & 5319.011 & 7.0   & A     & F     & Q     & Q     & Q     &       & Q     & Q     & Q     &       &       &       &  \\
          & 4     & 6123.294 & 0.017 & 6124.028 & 7.0   & F     &       &       &       &       &       & A     & A     & A     &       &       &       &  \\
    1127-145 & 1     & 5193.997 & 0.214 & 5184.542 & 30.0  & Q     & T     & T     & T     & T     & -     & Q     & -     &       &       &       &       &  \\
          & 2     & 5926.442 & 0.353 & 5933.528 & 30.0  & Q     & Q     & T     &       & T     & -     & Q     &       & A     &       & A     &       &  \\
    1156+295 & 1     & 5674.348 & 0.030 & 5676.028 & 7.0   & Q     &       &       &       &       &       & Q     & Q     & Q     & Q     &       &       &  \\
          & 2     & 6038.659 & 0.785 & 6040.028 & 7.0   & Q     & A     &       & -     &       &       & T     & A     & A     & T     &       &       &  \\
    1219+285 & 1     & 5272.809 & 0.038 & 5270.011 & 7.0   & Q     & A     & T     & T     & Q     & Q     & Q     & Q     &       &       &       &       &  \\
          & 2     & 5988.436 & 0.151 & 5991.028 & 7.0   & Q     & Q     & Q     & Q     & Q     & Q     & Q     & Q     & Q     &       &       &       &  \\
          & 3     & 4877.816 & 0.010 & 4875.655 & 7.0   & A     & T     & A     & A     & A     & A     & A     & A     &       &       &       &       &  \\
          & 4     & 4884.913 & 0.014 & 4882.655 & 7.0   & A     & F     & A     & A     & A     & A     & A     & A     &       &       &       &       &  \\
    1222+216 & 1     & 5672.522 & 0.345 & 5675.992 & 7.0   & Q     & T     & Q     & Q     & Q     & Q     & Q     & Q     & Q     &       &       &       &  \\
          & 2     & 6025.625 & 0.011 & 6026.048 & 7.0   & Q     & Q     & Q     & Q     & Q     & Q     & Q     & Q     &       &       &       &       &  \\
          & 3     & 5317.278 & 0.161 & 5319.009 & 7.0   & F     & Q     &       &       &       & T     &       &       & T     &       &       &       &  \\
          & 4     & 5369.145 & 0.223 & 5368.009 & 7.0   & F     & Q     & A     & A     & A     & A     & A     & A     & A     &       &       &       &  \\
\\
    3C273 & 1     & 5295.660 & 0.896 & 5298.000 & 7.0   & Q     & T     &       &       &       &       & Q     & Q     & Q     & Q     & Q     &       &  \\
          & 2     & 6045.359 & 0.491 & 6044.000 & 7.0   & Q     &       &       &       &       &       & F     & T     & T     & Q     & Q     & Q     & Q \\
          & 3     & 5207.592 & 0.264 & 5206.000 & 7.0   & A     & A     &       &       &       &       & T     & Q     &       & Q     & T     &       &  \\
          & 4     & 5272.526 & 0.459 & 5276.000 & 7.0   & F     & T     &       &       &       &       & T     & Q     & Q     & Q     & A     &       &  \\
    3C279 & 1     & 4966.580 & 0.940 & 4969.042 & 7.0   & Q     & Q     & Q     & Q     & Q     & Q     & Q     & Q     & Q     &       & Q     &       & Q \\
          & 2     & 6011.463 & 0.398 & 6009.504 & 7.0   & Q     & T     & Q     & Q     & Q     & Q     & T     & Q     & A     &       &       &       &  \\
          & 3     & 4898.744 & 0.798 & 4899.057 & 7.0   & A     & Q     & Q     & Q     & Q     & Q     & Q     & Q     & T     &       & T     & A     & T \\
          & 4     & 5665.659 & 0.600 & 5669.028 & 7.0   & A     & A     & Q     & Q     & Q     & Q     & T     & Q     & A     & T     & A     &       & T \\
    1308+326 & 1     & 5302.082 & 0.386 & 5305.011 & 7.0   & Q     & Q     & Q     &       &       &       &       &       & Q     &       &       &       &  \\
    1406-076 & 1     & 5294.737 & 0.007 & 5298.011 & 7.0   & Q     &       &       &       &       &       & T     & Q     & T     &       & Q     &       &  \\
          & 2     & 5354.102 & 0.646 & 5354.011 & 7.0   & Q     & Q     & A     & Q     & Q     & T     & Q     & T     & A     &       &       &       &  \\
    1510-089 & 1     & 5714.774 & 0.754 & 5718.028 & 7.0   & Q     & Q     & Q     & Q     & Q     & Q     & Q     & Q     & Q     &       &       &       &  \\
          & 2     & 6064.552 & 0.923 & 6068.028 & 7.0   & Q     & T     & Q     & T     & T     &       & Q     & Q     & T     & Q     & A     &       &  \\
          & 3     & 4918.487 & 0.476 & 4917.655 & 7.0   & F     & T     & T     & T     & T     & T     & A     & A     & A     & A     & A     &       &  \\
          & 4     & 5747.276 & 0.093 & 5746.028 & 7.0   & F     & Q     & Q     & Q     & Q     & Q     & Q     & Q     & T     &       &       &       &  \\
    1611+343 & 1     & 5252.722 & 0.341 & 5256.011 & 7.0   & Q     & -     & -     & -     & -     & -     & Q     & T     &       &       &       &       &  \\
          & 2     & 5832.201 & 0.419 & 5830.028 & 7.0   & Q     &       &       &       &       &       & Q     & T     & T     & Q     &       &       &  \\
    1622-297 & 1     & 4745.490 & 0.005 & 4758.155 & 30.0  & Q     &       &       &       &       &       & Q     & Q     & T     &       & Q     &       & Q \\
          & 2     & 5350.951 & 0.728 & 5354.011 & 7.0   & Q     & T     & T     & A     & T     & A     & T     & T     &       &       &       &       &  \\
          & 3     & 5295.473 & 0.251 & 5298.011 & 7.0   & A     & A     & T     & Q     & Q     & A     & T     & Q     & T     &       & Q     &       & Q \\
    1633+382 & 1     & 5400.397 & 0.762 & 5402.000 & 7.0   & Q     & Q     & Q     & Q     & Q     & Q     & Q     & Q     &       &       & Q     & Q     & Q \\
          & 2     & 6135.430 & 0.911 & 6137.000 & 7.0   & Q     & Q     & Q     & Q     & Q     & Q     & Q     & T     & Q     & Q     &       &       &  \\
          & 3     & 5744.449 & 0.387 & 5745.000 & 7.0   & F     & F     & A     & A     & A     & A     & A     & F     & F     & F     &       &       &  \\
          & 4     & 5034.519 & 0.212 & 5038.000 & 7.0   & A     & T     & A     & A     & A     & T     & T     & T     & A     &       &       &       &  \\
    3C345 & 1     & 5826.277 & 0.420 & 5820.528 & 18.5  & Q     &       &       &       &       &       & Q     & Q     & Q     & Q     &       &       &  \\
          & 2     & 6036.932 & 0.587 & 6044.528 & 44.0  & Q     &       &       &       &       &       & Q     & Q     & Q     & Q     &       &       &  \\
          & 3     & 5067.109 & 0.269 & 5067.042 & 7.0   & A     & Q     & Q     & Q     & Q     & Q     & Q     & T     & A     & T     &       &       &  \\
          & 4     & 5110.554 & 0.669 & 5109.042 & 7.0   & A     & Q     & T     & T     & T     & T     & T     & T     & A     & T     &       &       &  \\
    1730-130 & 1     & 4980.711 & 0.156 & 4983.022 & 7.0   & Q     &       &       &       &       &       & Q     & T     & Q     &       & Q     &       & Q \\
          & 2     & 5376.704 & 0.005 & 5375.046 & 7.0   & Q     &       &       &       &       &       & A     & Q     & Q     &       & T     &       & T \\
          & 3     & 5433.603 & 0.734 & 5431.046 & 7.0   & A     &       &       &       &       &       & F     & F     & F     &       &       &       &  \\
          & 4     & 5494.502 & 0.002 & 5494.046 & 7.0   & A     &       &       &       &       &       & F     & F     & A     &       & A     &       & A \\
    1749+096 & 1     & 6070.825 & 0.092 & 6072.155 & 30.0  & Q     &       &       &       &       &       &       &       & A     &       & -     &       &  \\
          & 2     & 6135.326 & 0.125 & 6132.155 & 30.0  & Q     &       &       &       &       &       & Q     & Q     & A     & T     &       &       &  \\
          & 3     & 5427.240 & 0.185 & 5428.655 & 7.0   & A     & T     &       & Q     &       &       & Q     & Q     & F     & A     &       &       &  \\
          & 4     & 5502.790 & 0.803 & 5505.655 & 7.0   & A     & A     & Q     & Q     & Q     & Q     & Q     & Q     & A     &       &       &       &  \\
    BL Lacertae & 1     & 5033.523 & 0.419 & 5036.655 & 7.0   & Q     & T     & Q     & Q     & Q     & Q     & Q     & Q     & Q     &       &       &       &  \\
          & 2     & 5503.691 & 0.191 & 5505.655 & 7.0   & Q     & T     & Q     & Q     & Q     & Q     & Q     & Q     & Q     &       &       &       &  \\
          & 3     & 5707.822 & 0.078 & 5704.048 & 7.0   & F     & T     & A     & A     & A     & A     & A     & A     &       &       &       &       &  \\
          & 4     & 6029.555 & 0.163 & 6030.506 & 7.0   & A     & T     & A     & A     & A     & A     & A     & A     & A     &       &       &       &  \\
    3C446 & 1     & 5341.606 & 0.073 & 5351.155 & 30.0  & Q     & -     & -     &       &       & -     & A     & T     &       &       &       &       &  \\
          & 2     & 5825.797 & 0.775 & 5817.155 & 30.0  & Q     &       &       &       &       &       & A     & A     &       & -     &       &       &  \\
    CTA102 & 1     & 5126.709 & 0.047 & 5127.655 & 7.0   & Q     &       &       &       &       &       & Q     & Q     & Q     & Q     &       &       &  \\
          & 2     & 5828.375 & 0.696 & 5827.655 & 7.0   & Q     & Q     & Q     & Q     & Q     & Q     & Q     & Q     & Q     & Q     &       &       &  \\
          & 3     & 6191.256 & 0.248 & 6194.030 & 7.0   & F     & A     & T     & T     & T     & T     & T     & T     & T     & A     &       &       &  \\
          & 4     & 6245.244 & 0.594 & 6243.030 & 7.0   & F     & F     & A     &       &       &       & A     & A     & A     & F     &       &       &  \\
    3C454.3 & 1     & 5729.702 & 0.840 & 5729.655 & 1.0   & Q     &       &       &       &       &       &       &       & Q     &       &       &       &  \\
          & 2     & 6180.576 & 0.836 & 6181.655 & 7.0   & Q     & Q     & Q     & Q     & Q     & Q     & Q     & Q     & Q     & Q     &       &       &  \\
          & 3     & 5167.194 & 0.340 & 5165.042 & 7.0   & F     & F     & A     & A     & A     & A     & A     & A     & A     &       & F     &       & A \\
          & 4     & 5522.276 & 0.604 & 5522.042 & 7.0   & F     & A     & A     & A     &       &       & A     & A     & F     &       & F     &       & F \\

\enddata
\tablecomments{Activity States: Q - Quiescent; T - Transitory; A - Active; F - Flaring; Blank - No data; ``--'' - Insufficient number of observations to calculate a mean flux value.} 
\label{tab:selecteddata}
\end{deluxetable}

\clearpage

\section{Computation of Spectral Indices}

\noindent \textit{Optical Spectral Index}
\label{sec:opticalsi}

In the optical bands, we fit the blazar spectrum by a power law of the form 
\begin{equation}
\label{eqn:alpha}
S_{\nu} \propto \nu^{\alpha_{o}},
\end{equation}
where $S_{\nu}$ is the radiative flux density at frequency $\nu$ and \osi\, is the spectral index at optical wavelengths. We note that the optical spectrum that we fit with a single power law can include multiple components (emission lines, BBB, synchrotron radiation), the implication of which will be discussed in Section \ref{Results}. To compute \osi, we perform a weighted linear least-square fit using the IDL routine \texttt{LINFIT}, combining all data available in the UV$-$NIR range unless there is an obvious break in the power law in either the NIR or UV bands.   We retrieve  the slope and its error  and report these as \osi\, and $\sigma_{\alpha_{o}}$, respectively. Examples of the fit are shown  in Figure \ref{fig:SI} for two objects.  

Because we assume the model to be linear, testing the goodness of the fit to the model in the usual sense is not very meaningful in this case. The weighted $\chi^{2}$ statistic would be quite large given the small value of many of our uncertainties.   To provide some measure of the ``goodness of fit," we compute the standard deviation of the data, $\sigma$, using 
\begin{equation}
\label{eqn:stddev}
\sigma^{2} = \frac{1}{N-2}\Sigma(y_{i}-\bar{y})^2,
\end{equation}
where \textit{N} is the number of data points, $y_{i} = \log\;S_{\nu}$, and  $\bar{y}$ is the computed best-fit value \citep{Bevington}, with two parameters determined from the fit. 

\noindent \textit{X-Ray and Gamma-Ray Photon Indices}
\label{sec:xandgammasi}

Both the X-ray and $\gamma$-ray spectral indices are computed from the power-law photon index, $\Gamma$, as $\alpha = \Gamma + 1$. For $\gamma$-ray observations, $\Gamma_{\gamma}$ is derived differently depending upon whether the epoch corresponds to a quiescent or an active state. For quiescent epochs,  we extract from the 2FGL catalog \citep{Nolan2FGL} the photon index  and its uncertainty. (Note: the spectra of some sources were also fit with a log parabolic model, in which case the uncertainty in $\alpha_{\gamma}$ is not given in the 2FGL catalog and, therefore, is not listed in the table.) For active states, we calculate $\Gamma_{\gamma}$ values from the photon and spacecraft data (see Section \ref{sec:gammareduction}).   

\noindent \textit{Broadband Spectral Slopes}

Two additional spectral indices are of interest to our study: the  slope between  optical and X-ray frequencies, $\alpha_{ox}$, and the slope between  X-ray and $\gamma$-ray energies, $\alpha_{xg}$. We use the weighted mean of the fluxes in  \textit{V} band for the optical emission. If no  \textit{V}-band observations are available, preference is given to measurements in the \textit{R, J, B, UVM2}, or \textit{UVW1} bands, in that order. We use X-ray and $\gamma$-ray emission at 1 keV and 0.5 GeV, respectively, to represent the high energies. 

The computed spectral indices for all objects are summarized in Table \ref{tab:seddata}: column 1 is the object name, column 2 is the identifying epoch number (corresponding to the number displayed on the light curve plot), column 3 is the date of the earliest observation (among X-ray - NIR measurements) within the epoch, columns 4$-$9 are \gsi, \xsi, and \osi, and their respective 1-$\sigma$ uncertainties, column 10 provides the number of UV$-$optical$-$NIR observations included in the computation of \osi, and column 11 lists the standard deviation of the data relative to the best-fit line (the measurement of the ``goodness of fit" of the spectral slope for \osi).  Columns 12$-$15 are \oxsi\, and \xgsi\, and their respective 1-$\sigma$ uncertainties. Column 16 indicates the frequency band  used in the computation of \oxsi\, if no \textit{V}-band observation is available. 
\clearpage

\begin{deluxetable}{lcrrrrrrrrrrrrrc}

\rotate
\tablewidth{0pt}
\tabletypesize{\tiny}
\tablecolumns{16}
\tablecaption{Computed Spectral Indices}
\tablehead{
			\multicolumn{1}{l}{Object}&
			\multicolumn{1}{c}{Epoch}&
			\multicolumn{1}{r}{Earliest Non-$\gamma$}&
			\multicolumn{1}{r}{}&
			\multicolumn{1}{r}{}&
			\multicolumn{1}{r}{}&
			\multicolumn{1}{r}{}&
			\multicolumn{1}{r}{}&
			\multicolumn{1}{r}{}&
			\multicolumn{1}{r}{\# UV-opt-}&
			\multicolumn{1}{r}{Std. Dev.}&
			\multicolumn{1}{r}{}&
			\multicolumn{1}{r}{}&
			\multicolumn{1}{r}{}&
			\multicolumn{1}{r}{}&
			\multicolumn{1}{r}{\osi}

			\cr
			\multicolumn{1}{l}{Name}&
			\multicolumn{1}{c}{Number}&
			\multicolumn{1}{r}{Obsv.in Epoch}&
			\multicolumn{1}{r}{\gsi}&
			\multicolumn{1}{r}{$\sigma_{\alpha_{\gamma}}$}&
			\multicolumn{1}{r}{\xsi}&
			\multicolumn{1}{r}{$\sigma_{\alpha_{x}}$}&
			\multicolumn{1}{r}{\osi}&
			\multicolumn{1}{r}{$\sigma_{\alpha_{o}}$}&
			\multicolumn{1}{r}{NIR pts}&
			\multicolumn{1}{r}{of Data}&
			\multicolumn{1}{r}{\oxsi}&
			\multicolumn{1}{r}{$\sigma_{\alpha_{ox}}$}&
			\multicolumn{1}{r}{\xgsi}&
			\multicolumn{1}{r}{$\sigma_{\alpha_{xg}}$}&
			\multicolumn{1}{r}{Band}
					  \cr
		  	\multicolumn{1}{l}{(1)}&
		  	\multicolumn{1}{r}{(2)}&
		 	\multicolumn{1}{r}{(3)}&
		 	\multicolumn{1}{r}{(4)}&
		 	\multicolumn{1}{r}{(5)}&
		  	\multicolumn{1}{r}{(6)}&
		  	\multicolumn{1}{r}{(7)}&
		  	\multicolumn{1}{r}{(8)}&
		  	\multicolumn{1}{r}{(9)}&
		  	\multicolumn{1}{r}{(10)}&
		  	\multicolumn{1}{r}{(11)}&
		  	\multicolumn{1}{r}{(12)}&
		  	\multicolumn{1}{r}{(13)}&
		  	\multicolumn{1}{r}{(14)}&
		  	\multicolumn{1}{r}{(15)}&
		  	\multicolumn{1}{r}{(16)}
		}
		  	
\startdata			  	

    3C66A & 1     & 5784.410 & -0.912 &       & -1.559 & 0.253 & -1.292 & 0.002 & 14    & 0.007 & -1.503 & 0.026 & -0.780 & 0.012 &  \\

          & 2     & 6185.907 & -0.912 &       &       &       & -1.044 & 0.019 & 6     & 0.022 &       &       &       &       &  \\
          & 3     & 4744.658 & -0.893 & 0.085 & -1.971 & 0.129 & -0.872 & 0.001 & 10    & 0.057 & -1.443 & 0.010 & -0.757 & 0.005 &  \\
          & 4     & 5390.561 & -0.746 & 0.098 &       &       & -0.811 & 0.004 & 8     & 0.059 &       &       &       &       &  \\
    0235+164 & 1     & 5087.774 & -1.124 &       & -1.120 & 0.338 & -1.586 & 0.013 & 6     & 0.017 & -1.130 & 0.047 & -0.769 & 0.022 &  \\
          & 2     & 5128.683 & -1.124 &       &       &       & -1.791 & 0.009 & 10    & 0.122 &       &       &       &       &  \\
          & 3     & 4729.762 & -0.990 & 0.078 & -1.237 & 0.460 & -1.726 & 0.003 & 8     & 0.084 & -1.409 & 0.051 & -0.674 & 0.024 &  \\
          & 4     & 4758.502 & -1.056 & 0.087 & -1.679 & 0.174 & -1.627 & 0.002 & 10    & 0.081 & -1.115 & 0.018 & -0.812 & 0.008 &  \\
    0336-019 & 1     & 4711.555 & -1.475 & 0.072 &       &       & -0.260 & 0.051 & 5     & 0.001 &       &       &       &       &  \\
          & 2     & 4917.231 & -1.475 & 0.072 & -0.924 & 0.677 & -0.366 & 0.030 & 4     & 0.022 & -1.154 & 0.063 & -0.860 & 0.029 &  \\
          & 3     & 5832.461 & -1.104 & 0.217 &       &       & -1.009 & 0.023 & 9     & 0.001 &       &       &       &       &  \\
          & 4     & 5858.888 & -1.225 & 0.130 &       &       & -0.948 & 0.023 & 4     & 0.007 &       &       &       &       &  \\
    0420-014 & 1     & 5124.447 & -1.298 & 0.028 &       &       & -1.330 & 0.025 & 11    & 0.003 &       &       &       &       &  \\
          & 2     & 5508.897 & -1.298 & 0.028 & -0.984 & 0.367 & -0.807 & 0.031 & 7     & 0.049 & -1.038 & 0.050 & -0.811 & 0.023 &  \\
          & 3     & 5217.245 & -0.870 & 0.132 &       &       & -1.146 & 0.073 & 6     & 0.007 &       &       &       &       &  \\
          & 4     & 5899.334 & -1.369 & 0.196 &       &       & -1.616 & 0.065 & 4     & 0.030 &       &       &       &       &  \\
    0528+134 & 1     & 5120.762 & -1.545 &       &       &       & -0.719 & 0.030 & 9     & 0.025 &       &       &       &       &  \\
          & 3     & 5825.600 & -1.545 &       &       &       & -0.446 & 0.029 & 7     & 0.039 &       &       &       &       &  \\
    0716+714 & 1     & 4882.182 & -1.077 &       & -1.159 & 0.136 & -1.224 & 0.012 & 11    & 0.039 & -1.517 & 0.020 & -0.873 & 0.009 &  \\
          & 2     & 5587.747 & -1.077 &       & -1.436 & 0.309 & -1.200 & 0.001 & 11    & 0.035 & -1.591 & 0.031 & -0.771 & 0.014 &  \\
          & 3     & 5859.502 & -0.962 & 0.086 & -1.450 & 0.119 & -1.231 & 0.001 & 11    & 0.082 & -1.489 & 0.013 & -0.737 & 0.006 &  \\
          & 4     & 6122.279 & -1.013 & 0.135 &       &       & -1.223 & 0.006 & 5     & 0.005 &       &       &       &       &  \\
    0735+178 & 1     & 5503.001 & -1.047 & 0.035 & -1.294 & 0.481 & -1.519 & 0.007 & 9     & 0.035 & -1.502 & 0.065 & -0.754 & 0.031 & R \\
          & 2     & 6011.277 & -1.047 & 0.035 &       &       & -1.477 & 0.017 & 5     & 0.011 &       &       &       &       &  \\
          & 3     & 6070.398 & -1.374 & 0.248 & -1.223 & 0.379 & -0.975 & 0.006 & 7     & 0.033 & -1.246 & 0.037 & -0.808 & 0.017 &  \\
    0827+243 & 1     & 4767.528 & -1.674 & 0.070 &       &       & -0.548 & 0.070 & 6     & 0.010 &       &       &       &       &  \\
          & 2     & 5503.630 & -1.674 & 0.070 & -0.697 & 0.108 & -0.482 & 0.013 & 7     & 0.038 & -1.001 & 0.018 & -0.937 & 0.008 &  \\
          & 3     & 6198.701 & -1.268 & 0.229 & -0.577 & 0.095 & -0.890 & 0.011 & 3     & 0.000 & -1.080 & 0.014 & -0.801 & 0.007 &  \\
          & 4     & 6284.264 & -1.304 & 0.097 & -0.703 & 0.178 & -0.974 & 0.031 & 4     & 0.083 & -1.046 & 0.039 & -0.770 & 0.015 & UVM2 \\
    0829+046 & 1     & 5663.736 & -1.181 &       &       &       & -1.729 & 0.011 & 4     & 0.036 &       &       &       &       &  \\
          & 2     & 6089.713 & -1.181 &       & -0.430 & 0.454 & -1.501 & 0.015 & 3     & 0.038 & -1.444 & 0.081 & -0.804 & 0.037 &  \\
          & 3     & 5234.339 & -1.217 & 0.273 &       &       & -1.596 & 0.011 & 5     & 0.006 &       &       &       &       &  \\
    0836+710 & 1     & 5624.396 & -1.948 & 0.073 &       &       & -0.629 & 0.039 & 4     & 0.001 &       &       &       &       &  \\
          & 2     & 6020.937 & -1.948 & 0.073 & -0.468 & 0.102 & -0.282 & 0.022 & 3     & 0.006 & -0.797 & 0.021 & -1.016 & 0.010 &  \\
          & 3     & 5869.336 & -1.607 & 0.081 & -0.438 & 0.095 & -0.904 & 0.114 & 3     & 0.003 & -0.766 & 0.014 & -0.790 & 0.007 & J \\
          & 4     & 5923.485 & -1.609 & 0.172 & -0.451 & 0.085 & -0.453 & 0.023 & 5     & 0.067 & -0.714 & 0.013 & -0.928 & 0.006 &  \\
    OJ287 & 1     & 5296.542 & -1.232 & 0.043 & -1.259 & 0.163 & -1.338 & 0.002 & 19    & 0.075 & -1.426 & 0.017 & -0.863 & 0.008 &  \\
          & 2     & 5340.494 & -1.232 & 0.043 & -1.279 & 0.173 & -1.528 & 0.003 & 11    & 0.032 & -1.347 & 0.020 & -0.848 & 0.009 &  \\
          & 3     & 5129.847 & -1.392 & 0.176 & -0.885 & 0.069 & -1.582 & 0.002 & 12    & 0.038 & -1.339 & 0.010 & -0.821 & 0.005 &  \\
          & 4     & 6038.508 & -1.229 & 0.164 &       &       & -1.425 & 0.002 & 20    & 0.016 &       &       &       &       &  \\
    0954+658 & 1     & 4766.685 & -1.415 & 0.067 &       &       & -1.329 & 0.022 & 7     & 0.017 &       &       &       &       &  \\
          & 2     & 4781.697 & -1.415 & 0.067 &       &       & -1.242 & 0.062 & 3     & 0.022 &       &       &       &       & J \\
          & 3     & 5636.232 & -1.076 & 0.218 &       &       & -1.805 & 0.020 & 16    & 0.116 &       &       &       &       &  \\
          & 4     & 5667.827 & -1.292 & 0.253 &       &       & -1.769 & 0.071 & 17    & 0.037 &       &       &       &       &  \\
    1055+018 & 1     & 5305.329 & -1.217 & 0.039 & -0.998 & 0.408 & -1.509 & 0.019 & 8     & 0.074 & -1.164 & 0.050 & -0.860 & 0.023 &  \\
          & 2     & 6046.729 & -1.217 & 0.039 &       &       & -1.438 & 0.015 & 5     & 0.001 &       &       &       &       &  \\
          & 3     & 5664.801 & -1.243 & 0.190 &       &       & -1.418 & 0.008 & 4     & 0.007 &       &       &       &       &  \\
          & 4     & 5709.314 & -1.434 & 0.252 & -0.718 & 0.179 & -1.581 & 0.009 & 6     & 0.023 & -1.248 & 0.023 & -0.799 & 0.011 &  \\
    Mkn421 & 1     & 5306.337 & -0.771 & 0.012 &       &       & -0.521 & 0.046 & 5     & 0.001 &       &       &       &       &  \\
          & 2     & 5729.351 & -0.771 & 0.012 &       &       & -0.575 & 0.055 & 4     & 0.000 &       &       &       &       &  \\
          & 3     & 5319.355 & -0.770 & 0.089 & -1.061 & 0.012 & -0.419 & 0.063 & 4     & 0.000 & -0.717 & 0.004 & -1.166 & 0.002 &  \\
          & 4     & 6123.294 & -0.747 & 0.046 &       &       & -0.587 & 0.024 & 4     & 0.000 &       &       &       &       &  \\
    1127-145 & 1     & 5193.997 & -1.697 & 0.051 & -0.388 & 0.111 & -0.646 & 0.012 & 4     & 0.049 & -1.120 & 0.021 & -0.953 & 0.010 &  \\
          & 2     & 5926.442 & -1.697 & 0.051 & -0.665 & 0.559 & -0.331 & 0.018 & 4     & 0.031 & -1.001 & 0.048 & -0.957 & 0.025 & J \\
    1156+295 & 1     & 5674.348 & -1.295 & 0.027 &       &       & -1.112 & 0.043 & 5     & 0.018 &       &       &       &       &  \\
          & 2     & 6038.659 & -1.295 & 0.027 & -0.584 & 0.528 & -1.216 & 0.024 & 17    & 0.008 & -1.345 & 0.102 & -0.775 & 0.048 &  \\
    1219+285 & 1     & 5272.809 & -1.019 & 0.034 & -1.704 & 0.252 & -0.911 & 0.003 & 7     & 0.019 & -1.348 & 0.028 & -0.908 & 0.013 &  \\
          & 2     & 5988.436 & -1.019 & 0.034 & -1.622 & 0.440 & -1.264 & 0.004 & 7     & 0.009 & -1.554 & 0.043 & -0.837 & 0.020 &  \\
          & 3     & 4877.816 & -0.965 & 0.156 & -1.686 & 0.441 & -1.022 & 0.002 & 6     & 0.055 & -1.550 & 0.040 & -0.786 & 0.019 &  \\
          & 4     & 4884.913 & -1.470 & 0.205 & -1.776 & 0.170 & -0.973 & 0.002 & 6     & 0.049 & -1.402 & 0.016 & -0.848 & 0.008 &  \\
    1222+216 & 1     & 5672.522 & -1.231 &       & -0.592 & 0.254 & -0.146 & 0.004 & 6     & 0.013 & -1.359 & 0.035 & -0.787 & 0.016 &  \\
          & 2     & 6025.625 & -1.231 &       & -0.814 & 0.484 & -0.040 & 0.005 & 4     & 0.017 & -1.364 & 0.086 & -0.748 & 0.040 &  \\
          & 3     & 5317.278 & -0.982 & 0.035 & -0.828 & 0.237 & -0.305 & 0.009 & 3     &       & -1.377 & 0.036 & -0.521 & 0.017 & R \\
          & 4     & 5369.145 & -1.078 & 0.024 & -0.669 & 0.258 & -0.363 & 0.004 & 6     & 0.094 & -1.473 & 0.036 & -0.482 & 0.017 &  \\
          \\
    3C273 & 1     & 5295.660 & -1.616 &       & -0.658 & 0.043 & -0.458 & 0.004 & 8     & 0.024 & -1.236 & 0.006 & -1.035 & 0.003 &  \\
          & 2     & 6045.359 & -1.616 &       &       &       & -0.564 & 0.070 & 4     & 0.041 &       &       &       &       &  \\
          & 3     & 5207.592 & -1.431 & 0.074 & -0.664 & 0.026 & -0.422 & 0.003 & 4     & 0.019 & -1.165 & 0.004 & -0.915 & 0.002 &  \\
          & 4     & 5272.526 & -1.492 & 0.088 & -0.693 & 0.044 & -0.494 & 0.003 & 9     & 0.010 & -1.223 & 0.005 & -0.898 & 0.003 &  \\
    3C279 & 1     & 4966.580 & -1.340 &       & -0.797 & 0.131 & -1.696 & 0.007 & 12    & 0.040 & -0.924 & 0.017 & -0.862 & 0.008 &  \\
          & 2     & 6011.463 & -1.340 &       & -0.550 & 0.178 & -1.578 & 0.008 & 12    & 0.102 & -1.115 & 0.028 & -0.895 & 0.013 &  \\
          & 3     & 4898.744 & -1.419 & 0.083 & -0.875 & 0.215 & -1.770 & 0.006 & 13    & 0.036 & -1.096 & 0.028 & -0.742 & 0.013 &  \\
          & 4     & 5665.659 & -1.908 & 0.152 & -0.665 & 0.092 & -1.747 & 0.005 & 16    & 0.054 & -1.081 & 0.014 & -0.763 & 0.006 &  \\
    1308+326 & 1     & 5302.082 & -1.222 &       & -0.250 & 0.718 & -1.517 & 0.206 & 3     & 0.002 & -1.159 & 0.171 & -0.757 & 0.081 & R \\
    1406-076 & 1     & 5294.737 & -1.429 & 0.064 &       &       & -0.851 & 0.042 & 4     & 0.045 &       &       &       &       &  \\
          & 2     & 5354.102 & -1.429 & 0.064 & -0.742 & 1.843 & -1.401 & 0.104 & 4     & 0.012 & -1.134 & 0.180 & -0.779 & 0.084 &  \\
    1510-089 & 1     & 5714.774 & -1.388 &       & -0.489 & 0.587 & -0.628 & 0.008 & 8     & 0.022 & -1.213 & 0.083 & -0.727 & 0.039 &  \\
          & 2     & 6064.552 & -1.388 &       & -0.835 & 0.422 & -0.721 & 0.006 & 14    & 0.041 & -1.150 & 0.073 & -0.770 & 0.034 &  \\
          & 3     & 4918.487 & -1.244 & 0.025 & -0.394 & 0.139 & -1.104 & 0.004 & 17    & 0.161 & -1.309 & 0.021 & -0.550 & 0.010 &  \\
          & 4     & 5747.276 & -1.268 & 0.046 & -0.585 & 0.146 & -0.710 & 0.007 & 9     & 0.026 & -1.201 & 0.021 & -0.573 & 0.010 &  \\
    1611+343 & 1     & 5252.722 & -1.307 & 0.171 & -0.443 & 0.578 & -0.461 & 0.059 & 4     & 0.003 & -1.208 & 0.092 & -0.920 & 0.042 &  \\
          & 2     & 5832.201 & -1.307 & 0.171 &       &       & -0.422 & 0.014 & 11    & 0.023 &       &       &       &       &  \\
    1622-297 & 1     & 4745.490 & -1.339 & 0.067 &       &       & -0.647 & 0.043 & 4     & 0.010 &       &       &       &       &  \\
          & 2     & 5350.951 & -1.339 & 0.067 & -0.397 & 0.557 & -0.466 & 0.026 & 5     & 0.026 & -1.283 & 0.056 & -0.791 & 0.026 &  \\
          & 3     & 5295.473 & -1.423 & 0.230 & -0.301 & 0.399 & -0.490 & 0.025 & 6     & 0.082 & -1.114 & 0.042 & -0.782 & 0.020 &  \\
    1633+382 & 1     & 5400.397 & -1.410 &       & -0.637 & 0.546 & -0.698 & 0.018 & 6     & 0.096 & -1.073 & 0.064 & -0.760 & 0.029 &  \\
          & 2     & 6135.430 & -1.410 &       & -0.859 & 0.581 & -0.885 & 0.031 & 8     & 0.084 & -1.055 & 0.090 & -0.777 & 0.042 &  \\
          & 3     & 5744.449 & -1.155 & 0.058 & -0.562 & 0.223 & -1.559 & 0.009 & 9     & 0.025 & -1.109 & 0.046 & -0.708 & 0.021 &  \\
          & 4     & 5034.519 & -1.270 & 0.084 & -1.118 & 0.446 & -1.211 & 0.020 & 4     & 0.116 & -1.052 & 0.044 & -0.721 & 0.020 &  \\
    3C345 & 1     & 5826.277 & -1.489 & 0.056 &       &       & -1.438 & 0.030 & 7     & 0.033 &       &       &       &       &  \\
          & 2     & 6036.932 & -1.489 & 0.056 &       &       & -1.493 & 0.062 & 7     & 0.005 &       &       &       &       &  \\
          & 3     & 5067.109 & -1.073 & 0.174 & -0.859 & 0.135 & -1.515 & 0.017 & 8     & 0.022 & -1.090 & 0.022 & -0.782 & 0.010 &  \\
          & 4     & 5110.554 & -1.319 & 0.334 & -0.578 & 0.246 & -1.291 & 0.024 & 5     & 0.027 & -1.193 & 0.028 & -0.770 & 0.013 &  \\
    1730-130 & 1     & 4980.711 & -1.488 &       &       &       & -0.937 & 0.019 & 6     & 0.058 &       &       &       &       &  \\
          & 2     & 5376.704 & -1.488 &       &       &       & -1.520 & 0.027 & 5     & 0.073 &       &       &       &       &  \\
          & 3     & 5433.603 & -1.440 & 0.103 &       &       & -2.385 & 0.134 & 4     & 0.006 &       &       &       &       &  \\
          & 4     & 5494.502 & -1.132 & 0.087 &       &       & -1.061 & 0.074 & 3     & 0.001 &       &       &       &       &  \\
    1749+096 & 1     & 6070.825 & -1.243 &       &       &       & -1.536 & 0.012 & 7     & 0.030 &       &       &       &       & R \\
          & 2     & 6135.326 & -1.243 &       &       &       & -1.717 & 0.028 & 8     & 0.018 &       &       &       &       &  \\
          & 3     & 5427.240 & -1.267 & 0.198 & -0.422 & 0.179 & -1.767 & 0.025 & 9     & 0.060 & -1.404 & 0.026 & -0.766 & 0.012 &  \\
          & 4     & 5502.790 & -1.394 & 0.148 & -0.579 & 0.141 & -2.066 & 0.014 & 7     & 0.007 & -1.155 & 0.024 & -0.853 & 0.011 &  \\
    BL Lacertae & 1     & 5033.523 & -1.261 &       & -0.957 & 0.182 & -1.745 & 0.005 & 9     & 0.012 & -1.350 & 0.019 & -0.854 & 0.009 &  \\
          & 2     & 5503.691 & -1.261 &       & -0.854 & 0.112 & -1.694 & 0.006 & 8     & 0.021 & -1.323 & 0.015 & -0.884 & 0.007 &  \\
          & 3     & 5707.822 & -1.240 & 0.074 & -0.790 & 0.157 & -1.640 & 0.002 & 7     & 0.022 & -1.486 & 0.019 & -0.749 & 0.009 &  \\
          & 4     & 6029.555 & -1.070 & 0.083 & -0.913 & 0.083 & -1.619 & 0.002 & 7     & 0.052 & -1.506 & 0.010 & -0.772 & 0.005 &  \\
    3C446 & 1     & 5341.606 & -1.436 & 0.053 & 0.217 & 0.825 & -0.650 & 0.087 & 3     & 0.006 & -1.030 & 0.123 & -0.857 & 0.057 &  \\
          & 2     & 5825.797 & -1.436 & 0.053 &       &       & -0.679 & 0.025 & 4     & 0.006 &       &       &       &       &  \\
    CTA102 & 1     & 5126.709 & -1.538 &       &       &       & -0.254 & 0.011 & 7     & 0.011 &       &       &       &       &  \\
          & 2     & 5828.375 & -1.538 &       & -0.377 & 0.158 & -0.413 & 0.005 & 16    & 0.008 & -1.150 & 0.024 & -0.820 & 0.011 &  \\
          & 3     & 6191.256 & -1.006 & 0.034 & -0.616 & 0.084 & -1.108 & 0.005 & 18    & 0.034 & -1.098 & 0.014 & -0.619 & 0.007 &  \\
          & 4     & 6245.244 & -1.396 & 0.080 & -0.514 & 0.120 & -1.409 & 0.008 & 6     & 0.055 & -1.177 & 0.019 & -0.760 & 0.009 &  \\
    3C454.3 & 1     & 5729.702 & -1.379 &       & -0.566 & 0.140 & -0.889 & 0.007 & 9     & 0.115 & -1.151 & 0.022 & -0.752 & 0.010 &  \\
          & 2     & 6180.576 & -1.379 &       & -0.984 & 0.357 & -1.053 & 0.010 & 9    & 0.092 & -1.163 & 0.051 & -0.842 & 0.024 &  \\
          & 3     & 5167.194 & -1.344 & 0.023 & -0.584 & 0.039 & -1.352 & 0.001 & 9     & 0.031 & -0.972 & 0.006 & -0.749 & 0.003 &  \\
          & 4     & 5522.276 & -1.259 & 0.010 & -0.602 & 0.040 & -1.548 & 0.003 & 6     & 0.041 & -1.057 & 0.007 & -0.630 & 0.003 &  \\

\enddata
\label{tab:seddata}
\end{deluxetable}

\clearpage

\section{Trends and Correlations of Spectral Indices}

\subsection{Distributions of Spectral Indices \label{Sdsi}}
\begin{figure*}
\centering
\mbox{\
\subfloat{
	\includegraphics[trim=0cm 0cm 0.1cm 0cm, clip=true,height=.28\linewidth, angle=0]{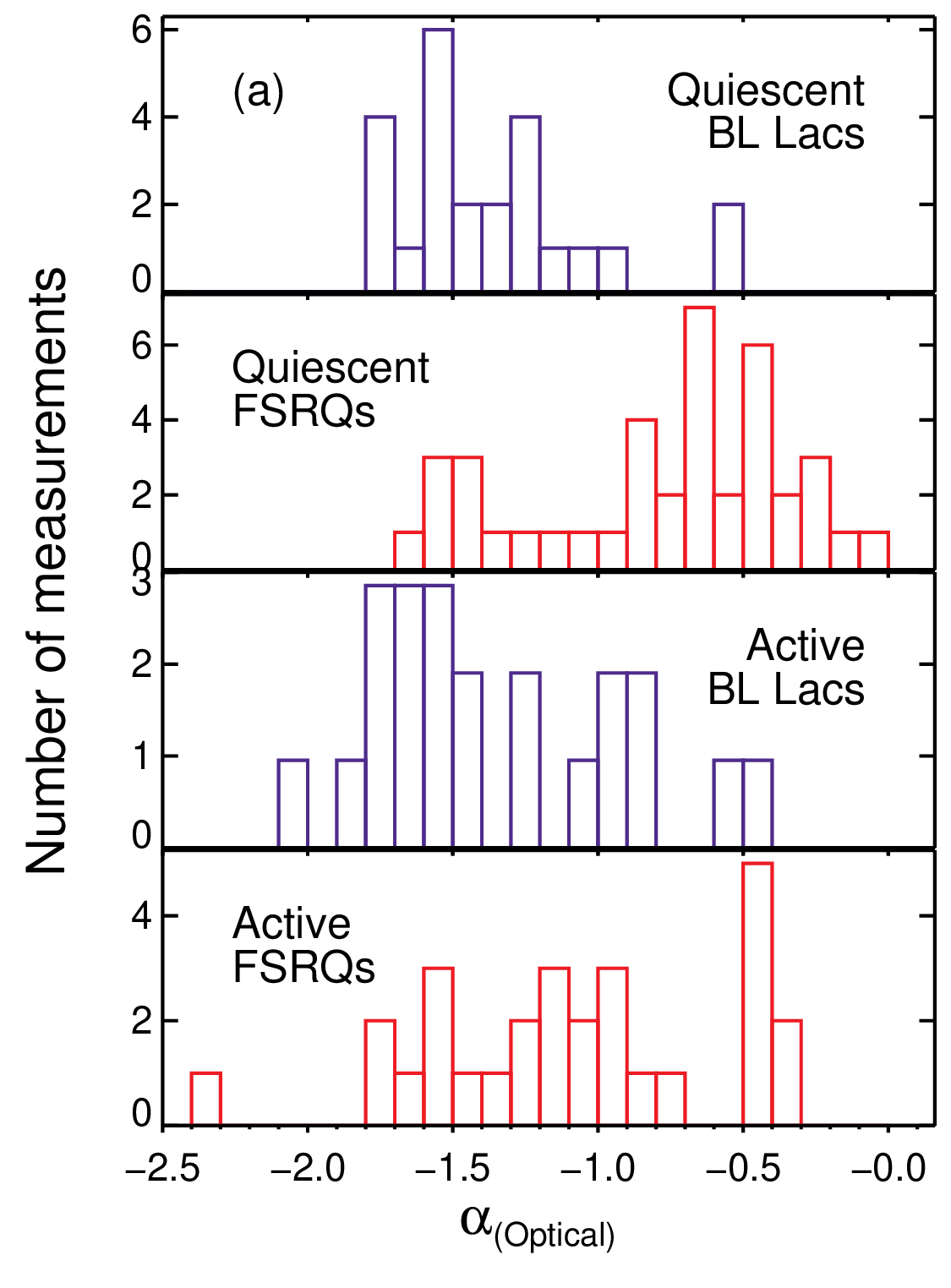}
	\label{fig:histooptical}
}
\subfloat{
	\includegraphics[trim=1.4cm 0cm 0.1cm 0cm, clip=true,height=.28\linewidth, angle=0]{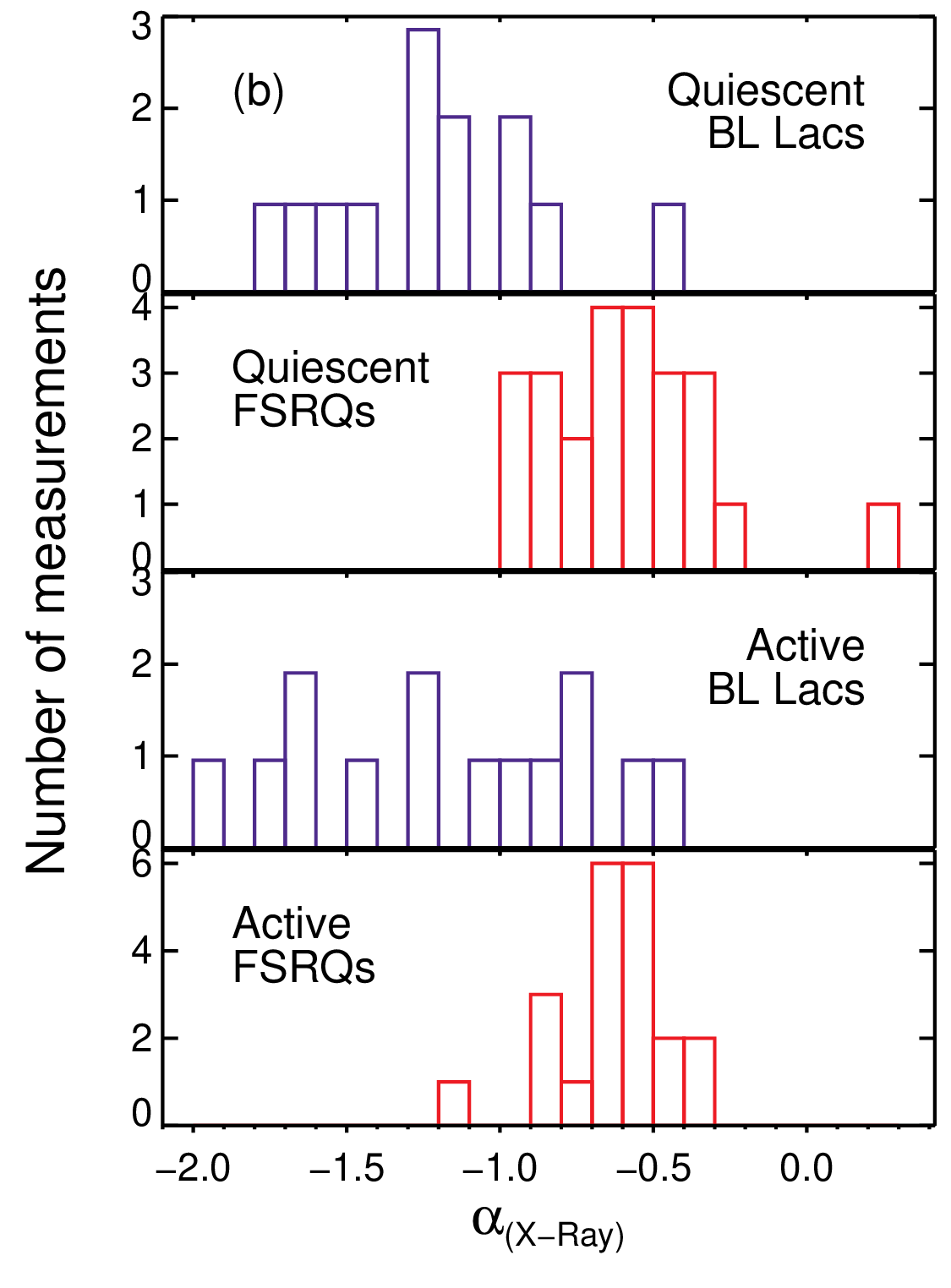}
	\label{fig:histoxray}
}
\subfloat{
	\includegraphics[trim=1.4cm 0cm 0.1cm 0cm, clip=true,height=.28\linewidth, angle=0]{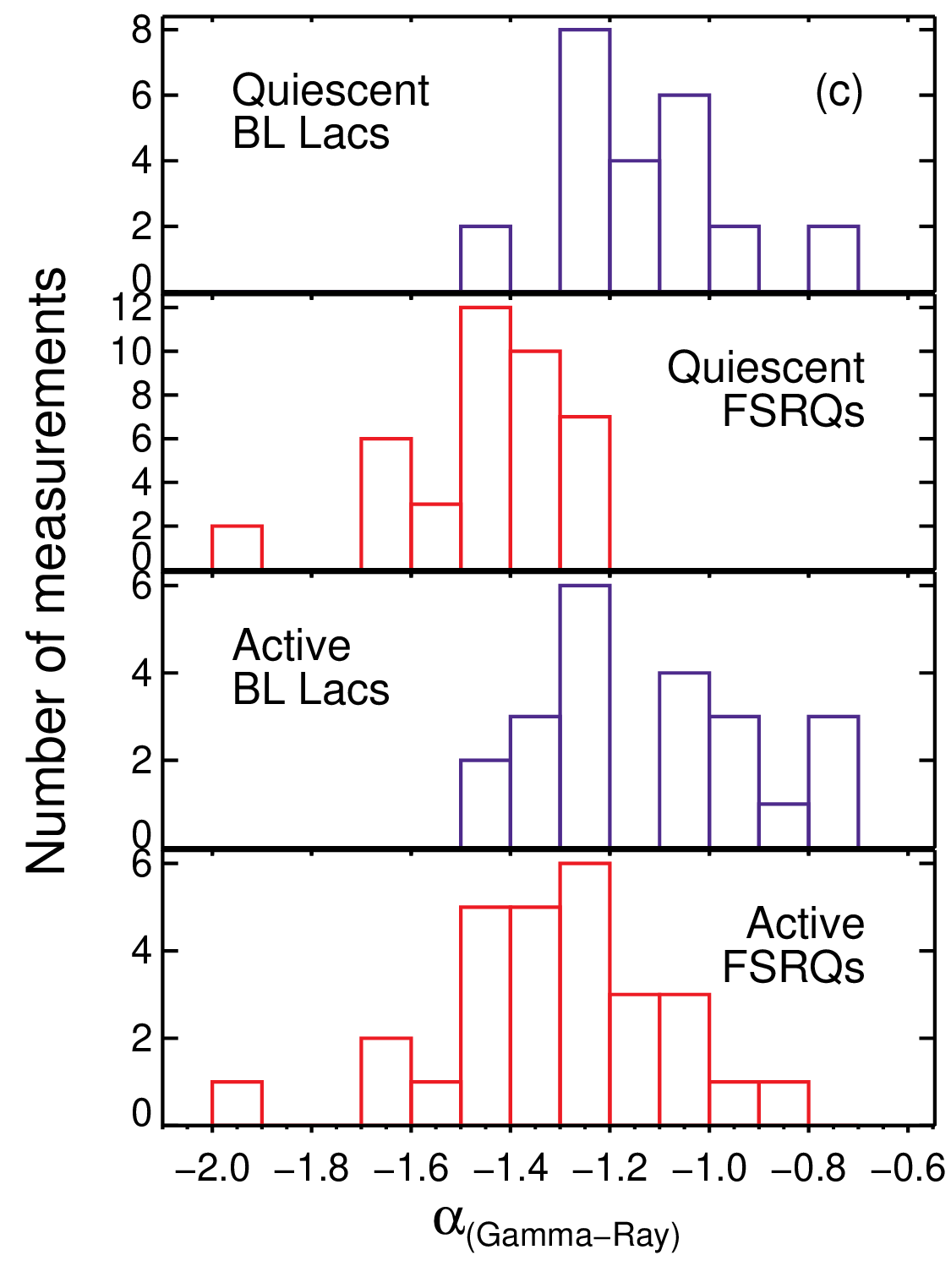}
	\label{fig:histogamma}
	}


\subfloat{
	\includegraphics[trim=1.4cm 0cm 0.1cm 0cm, clip=true,height=.28\linewidth, angle=0]{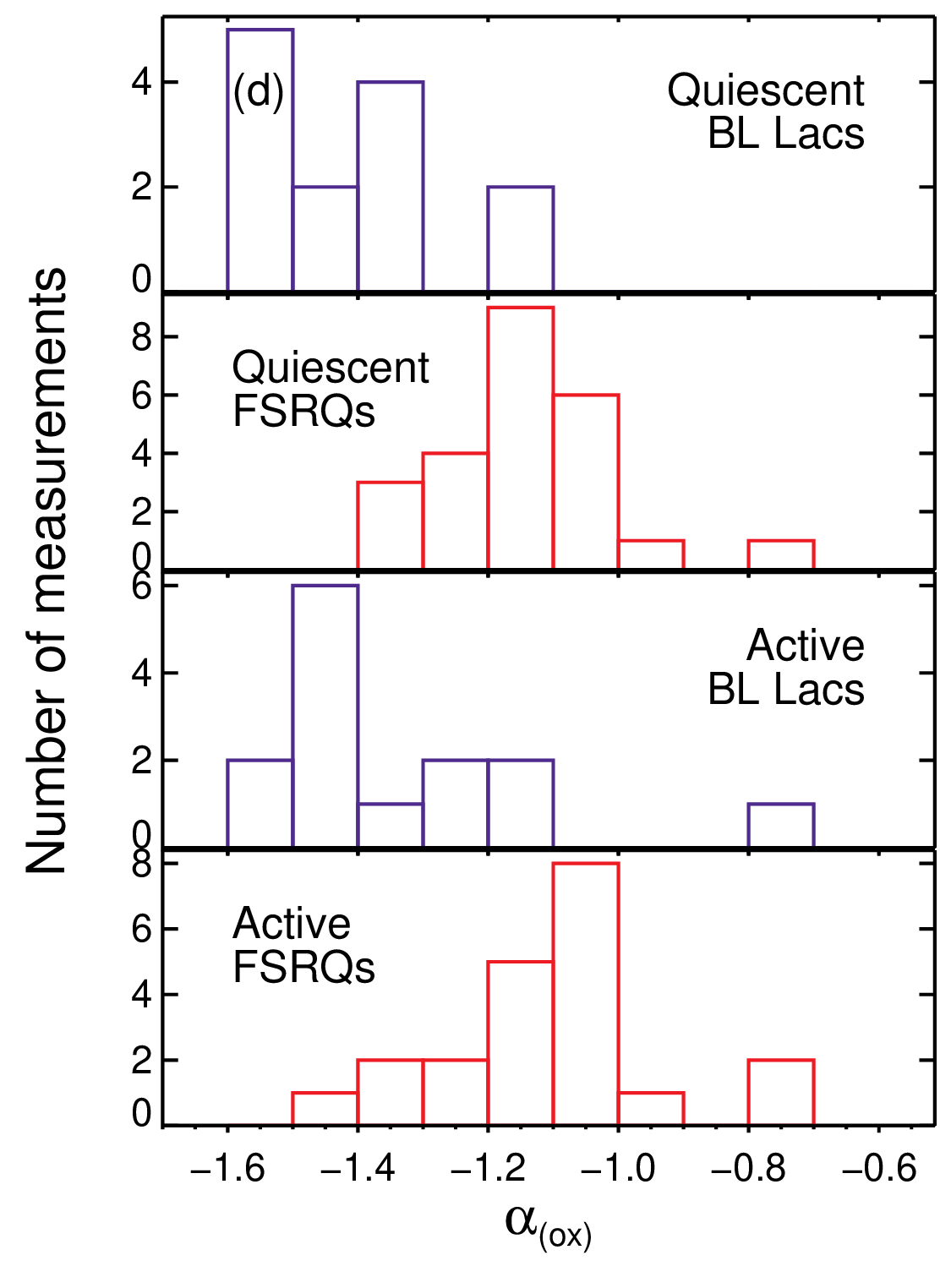}
	\label{fig:histoox}
}
\subfloat{
	\includegraphics[trim=1.4cm 0cm 0cm 0cm, clip=true,height=.28\linewidth, angle=0]{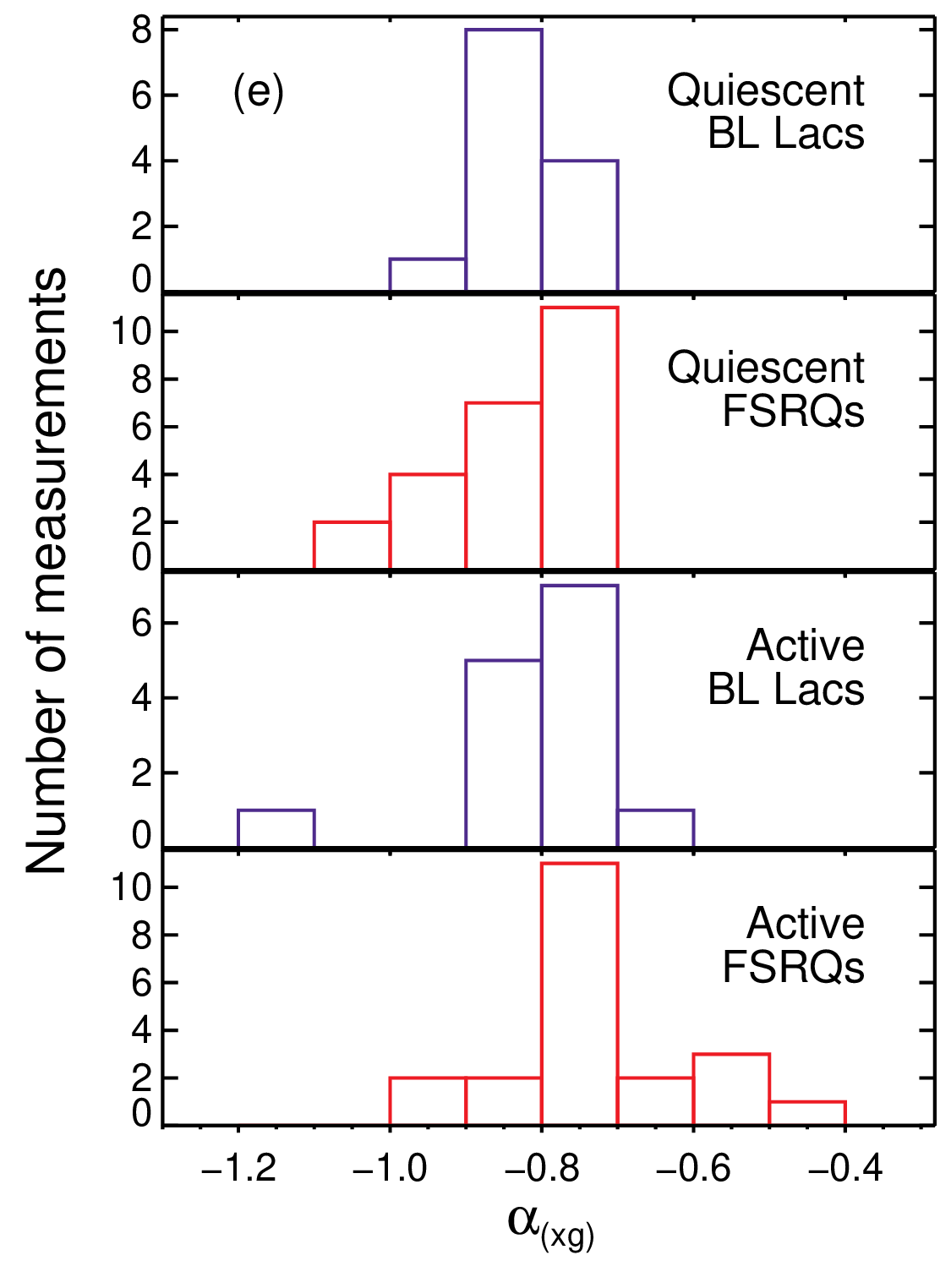}
	\label{fig:histoxg}
}
}
\caption{Distributions of spectral indices for quiescent and active states:
\protect\subref{fig:histooptical} optical, \protect\subref{fig:histoxray} X-ray,
\protect\subref{fig:histogamma} $\gamma$-ray,
\protect\subref{fig:histoox}  optical $-$ X-ray, and
\protect\subref{fig:histoxg} X-ray $-$ $\gamma$-ray.
FSRQs are plotted in red;
BL Lac objects in blue. 
}%
\label{fig:histograms}
\end{figure*}

Figure \ref{fig:histograms} presents distributions of the spectral indices \osi, \xsi, and \gsi, and the spectral index between these regions, \oxsi\, and \xgsi. We compute a mean of each spectral index from our selected epochs for each class  in each state. The results are summarized in Table \ref{tab:prefvalues}.  The standard deviation  is a good indicator of the spread of the indices. We consider a deviation within $\pm$ 0.35 ($\sim 20\%$ of the approximate spread of all indices) to be a sufficiently narrow  spread to indicate a ``preferred'' value for the index. 
	
For \osi, only BL Lacs in a quiescent state maintain a preferred value.  For \xsi, both the quiescent and the active FSRQs exhibit small deviations, with a preferred value of $\sim$ $-$0.6, as expected if the X-ray emission is produced via inverse Compton scattering by relatively low energy electrons that also emit synchrotron emission at millimeter-submillimeter (mm-submm) wavelengths. Active BL Lacs have a significant scatter in \xsi, with some values as steep as $-2$. This can be explained by a synchrotron origin of the enhanced X-ray emission in some BL Lacs. Quiescent BL Lacs exhibit a preferred value of  $-$1.2, which suggests that the quiescent X-ray emission is a mixture of IC and synchrotron emission.
 
Both quiescent and active states of both classes exhibit a preferred value of the $\gamma$-ray spectral index. The BL Lacs  show little difference in \gsi\, between quiescent and active states. \citet{ackermann2fgl} found a similar mean value for BL Lacs with a range from $-$0.90 to $-$1.17, depending upon the SED classification (LSP, ISP, HSP).  The FSRQs show a modest flattening of \gsi\, during active states.  \citet{ackermann2fgl} computed a mean value for FSRQs of $-$1.42 $\pm$ 0.17 for a much larger sample, which falls between the average values of \gsi\, during quiescent and active states. 

Both quiescent and active states of both classes exhibit preferred values of the spectral index between the optical and X-ray and between the X-ray and $\gamma$-ray regimes. The preferred values of \oxsi\, change little within each class between states, while they are different for the two classes. The preferred values of \xgsi\, are similar for the BL Lacs and FSRQs within the 1$\sigma$ uncertainty, independent of the state.
\begin{deluxetable}{lrrrr}
\tabletypesize{\small}
\tablewidth{0pt}
\tablecaption{Mean Values of Spectral Indices}
\tablecolumns{5}

\tablehead{	\multicolumn{1}{l}{}&
			\multicolumn{2}{c}{\hspace{20pt}Quiescent}&
			\multicolumn{2}{c}{\hspace{20pt}Active}
					  \cr
			\multicolumn{1}{l}{Spectral Index}&
			\multicolumn{1}{r}{\hspace{20pt}BL Lac}&
			\multicolumn{1}{r}{FSRQ}&
			\multicolumn{1}{c}{\hspace{20pt}BL Lac}&
			\multicolumn{1}{c}{FSRQ}
			\cr
		  	\multicolumn{1}{c}{(1)}&
		  	\multicolumn{1}{r}{(2)}&
		 	\multicolumn{1}{r}{(3)}&
		 	\multicolumn{1}{r}{(4)}&
		 	\multicolumn{1}{r}{(5)}
			}

\startdata
	\osi	&&&&\\
    Average Value 		& $-1.4$ 	& $-0.8$ 	& $-1.4$ 	& $-1.1$ \\
    Standard Deviation 	& 0.3  	& 0.4  	& 0.4  	& 0.5 \\
	\xsi	&&&&\\    
    Average Value 		& $-1.2$ 	& $-0.60$ 	& $-1.2$ 	& $-0.63$ \\
    Standard Deviation 	& 0.3  	& 0.27 	 	& 0.5  	& 0.18 \\
	\gsi	&&&&\\
    Average Value 		& $-1.12$ 	& $-1.46$ 	& $-1.13$ 	& $-1.31$ \\
    Standard Deviation 	&  0.17  	& 0.17  	& 0.23  	& 0.22 \\
	\oxsi	&&&&\\
    Average Value 		& $-1.40$ 	& $-1.13$ 	& $-1.32$ 	& $-1.11$ \\
    Standard Deviation 	&  0.14  	& 0.13  	& 0.22 		& 0.17 \\    
	\xgsi	&&&&\\
    Average Value 		& $-0.83$ 	& $-0.84$ 	& $-0.81$ 	& $-0.73$ \\
    Standard Deviation 	& 0.05   	& 0.09  	& 0.11 		& 0.12     
\enddata  
\label{tab:prefvalues}%
\end{deluxetable}

\subsection{Change of Spectral Indices between States}
\begin{figure}[t]
\centering
\includegraphics[height=.65\linewidth, angle=0]{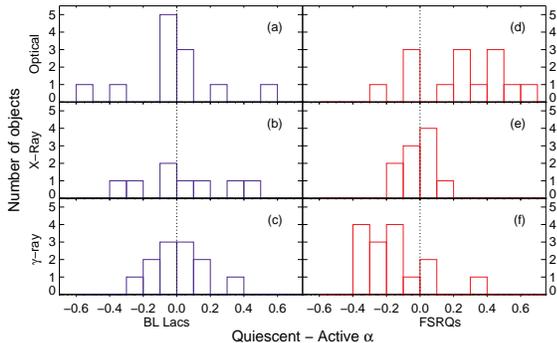}
\caption{Distribution of difference of spectral indices between quiescent and active  states for BL Lac objects ({\it left}, blue) and FSRQs ({\it right}, red), panels (a)  and (d) for \osi, (b) and (e) for \xsi, and (c) and (f) for \gsi.}
\label{fig:diff}
\end{figure}

To study the change of spectral indices between states, we compute the difference between the spectral indices of quiescent and active states for each object (between the means of $\alpha$ in the cases of two quiescent and two active states identified). Histograms of these differences are presented in Figure \ref{fig:diff}. The FSRQs tend to have a separation between quiescent and active states in both optical and $\gamma$-ray spectra, while the differences between states for the BL Lacs tend to be equally distributed.  Of the active FSRQs, 80\% have a flatter average $\gamma$-ray spectrum, with a weighted mean difference from the average quiescent spectrum of 0.16. (Some caution must be applied in this case, however, because  $\Gamma_{\gamma}$ is allowed to vary for the active states, while we use a fixed value taken from the 2FGL catalog for each object in quiescent states.) \citet{2010ApJ...710.1271A} found a weak ``harder when brighter'' effect for all FSRQs and BL Lacs except the HSP subclass, as had been previously suggested by \citet{2009MNRAS.396L.105G} for both classes when comparing some measurements from \textit{Fermi} and EGRET. For our sample of blazars, a significant ``harder when brighter'' effect is seen in the $\gamma$-ray spectral index for FSRQs, but the BL Lacs show no propensity towards a flatter or steeper spectrum, nor is there any obvious trend with SED class.    

Of the quiescent FSRQs, 73\% tend to have flatter optical spectra than during active states, while there is no statistical difference for \osi\, of BL Lacs between the two states. The difference in
behavior of \osi\, for FSRQs implies an important contribution of the emission from the accretion
disk (BBB) to the optical quiescent radiation, while accretion disk emission in BL Lacs seems to be too weak to play a significant role in the SED. In support of this latter point, the average value of \osi\, of $\sim -$1.4 in  active and quiescent BL Lacs indicates dominance of synchrotron emission during all states. This conforms with the prediction of \citet{2012MNRAS.420.2899G}, who simulated SEDs of blazars with a varying mix of 
Doppler-boosted radiation from the jet with emission from the accretion disk, broad-line region, and 
light from the host galaxy, and found strong dominance of the jet emission in BL Lacs. 

The differences of the X-ray spectral indices of FSRQs between states are equally distributed with a negligible mean of 0.001, as is evident in Figure \ref{fig:diff}e. This suggests that the same mechanism(s) is (are) employed for the X-ray production in  FSRQs, independent of the state. 
 In BL Lacs, the IC X-ray spectrum generally has a slope flatter than $-1$, whereas the slope is generally steeper for X-ray synchrotron radiation \citep[e.g.][]{Bregman1990}. 
 
 The very broad scatter of \xsi\,(quiescent) - \xsi\,(active) for BL Lacs indicates: (i) an increase in the contribution of synchrotron emission during active states for some BL Lacs (e.g., 3C66A, the largest 
positive difference); (ii) flattening of \xsi\, at active states for another group of BL Lacs (e.g., OJ287, the largest negative difference) that corresponds to an increase of the contribution of IC emission; and (iii) no change of \xsi\, for the rest of BL Lacs.  Although we cannot correlate the behavior with the SED subclasses of BL Lacs due to an insufficient amount of statistical data, the BL Lacs of the LSP type tend to have flatter X-ray spectra during active states.

\subsection{Relationships Between Spectral Indices}
\begin{figure*}[t] 
\centering
\subfloat{%
\includegraphics[trim=0cm 0cm 0cm 0cm, clip=true,height=.45\textwidth, angle=0]{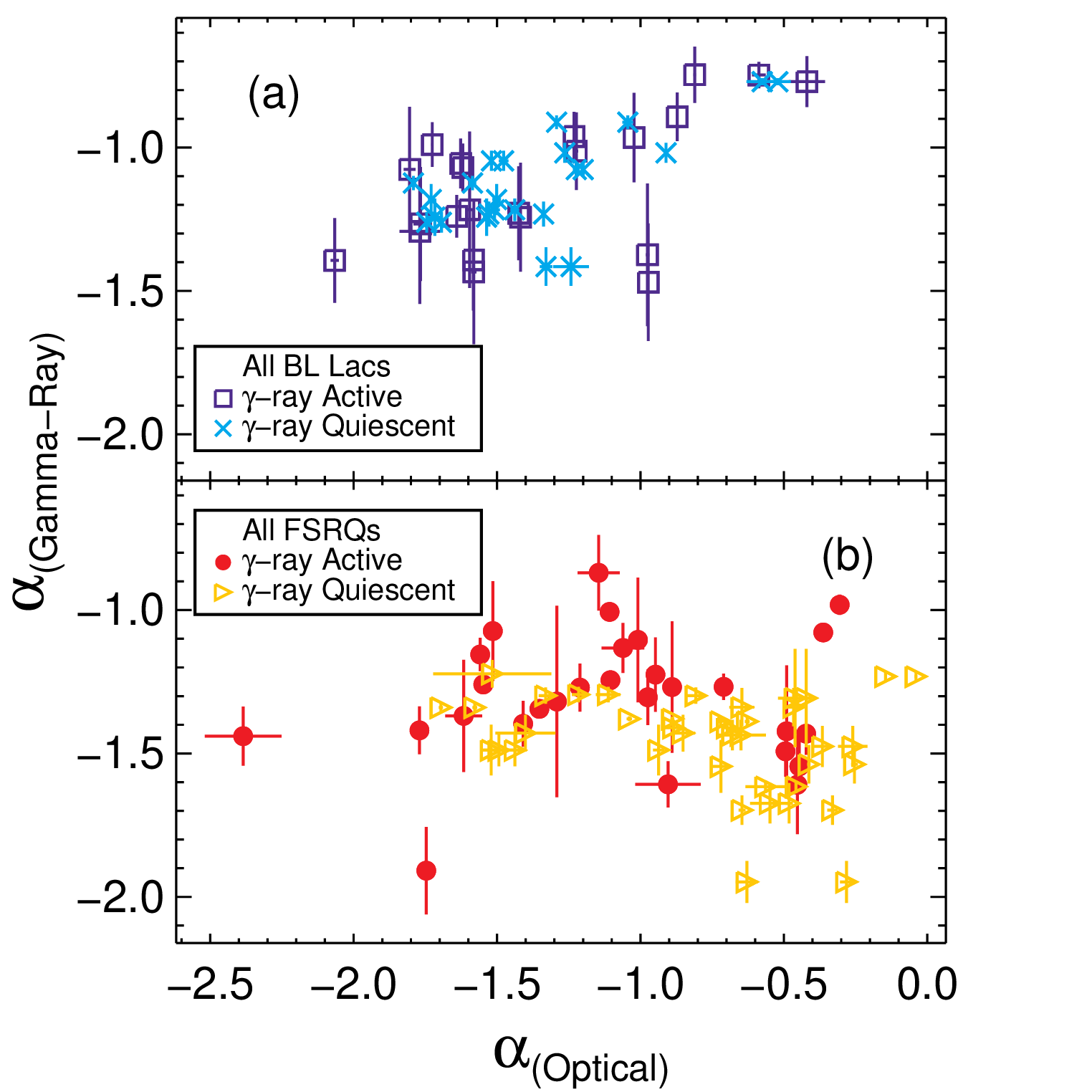}
}
\subfloat{%
\includegraphics[trim=0cm 0cm 0cm 0cm, clip=true,height=.45\textwidth, angle=0]{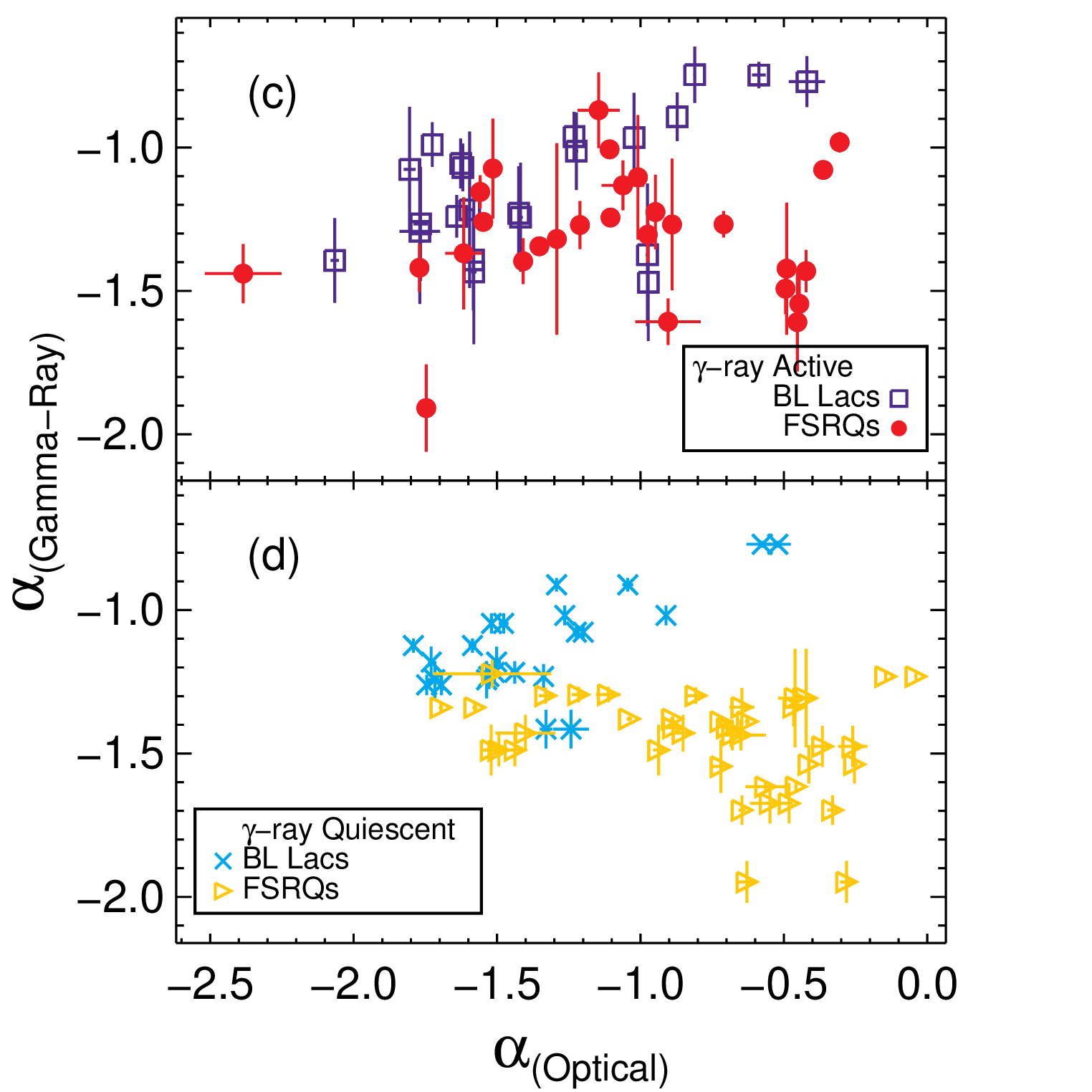}
}
\caption{Spectral indices \gsi\,vs.\ \osi\, at selected epochs (Section \ref{sec:composition}) for all blazars in the sample:  FSRQs are red-filled circles  in $\gamma$-ray  active states, yellow triangles  if quiescent, while BL Lacs are dark blue squares if $\gamma$-ray active, light blue  if quiescent. Panels are: (a) all BL Lacs, (b)  all FSRQs, (c) active BL Lacs and  FSRQs, and (d) quiescent BL Lacs and  FSRQs. [A combined plot with each data point labeled with object and epoch numbers is printed at the end of this manuscript. A complete set of individual plots, with each data point labeled with object and epoch numbers can be found in an expanded version of this paper at \url{www.bu.edu/blazars/VLBAproject.html}.]}
\label{fig:gammavsoptical}
\end{figure*}

\begin{figure*}
\centering
\subfloat{%
\includegraphics[trim=0cm 0cm 0cm 0cm, clip=true,height=.45\textwidth, angle=0]{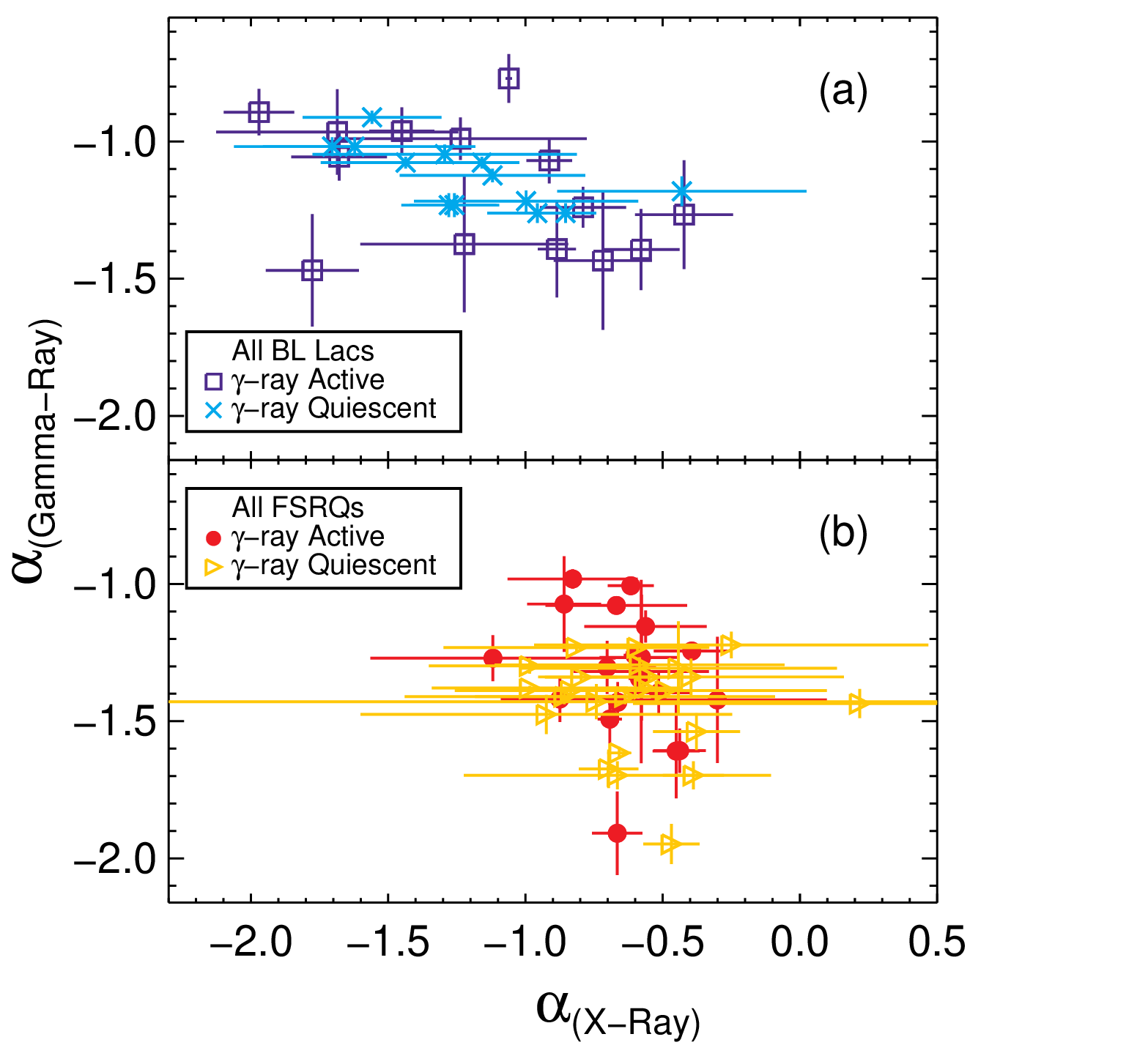}
\label{fig:gammavsxraybbff}
}
\subfloat{%
\includegraphics[trim=0cm 0cm 0cm 0cm, clip=true,height=.45\textwidth, angle=0]{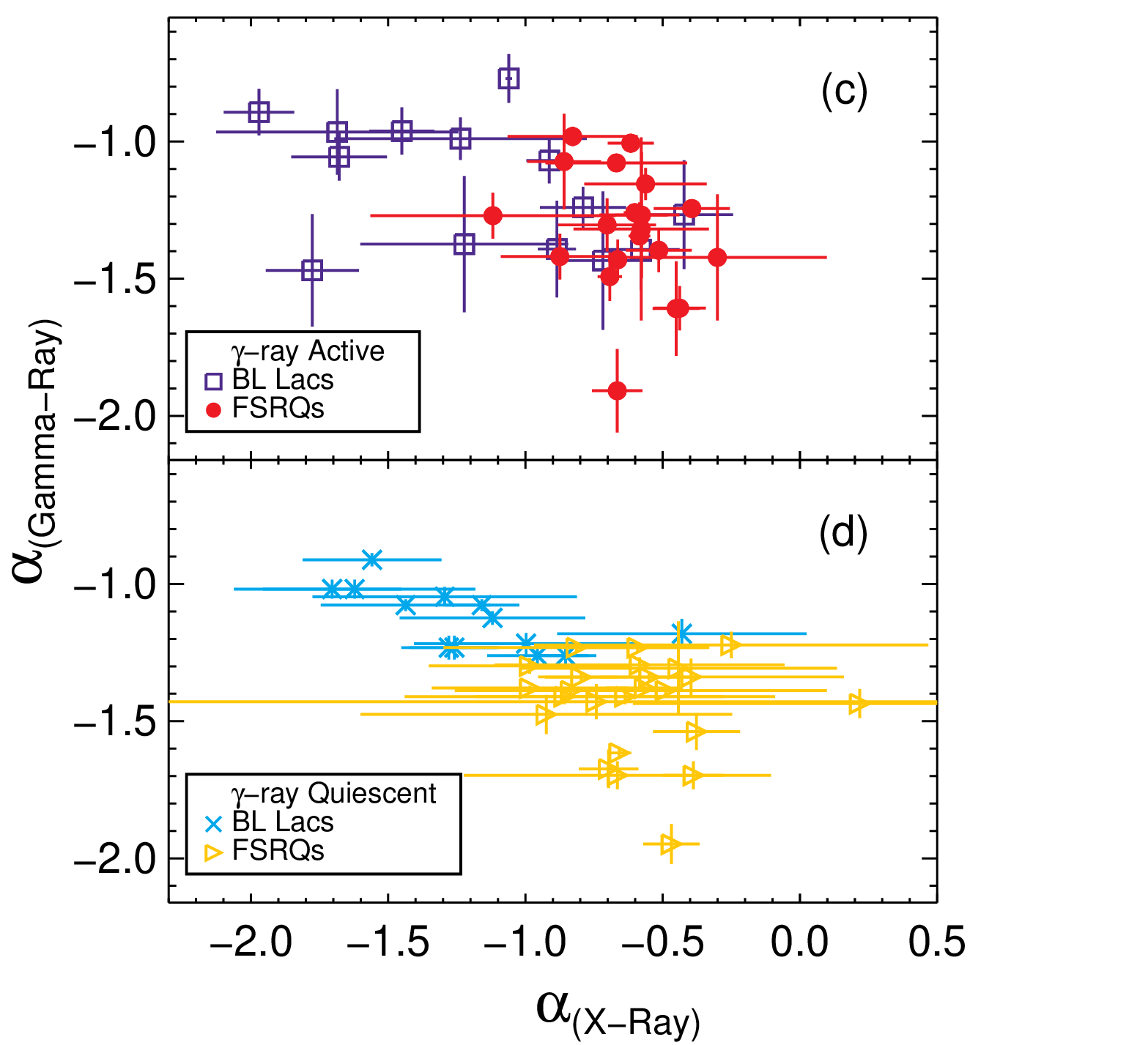}
\label{fig:gammavsxrayaaqq}
}
\caption{Spectral indices \gsi\,vs.\ \xsi. Designations are the same as in Fig.  \ref{fig:gammavsoptical}. }
\label{fig:gammavsxray}
\end{figure*}

\begin{figure*}
\centering
\subfloat{%
\includegraphics[trim=0cm 0cm 0cm 0cm, clip=true,height=.48\textwidth, angle=0]{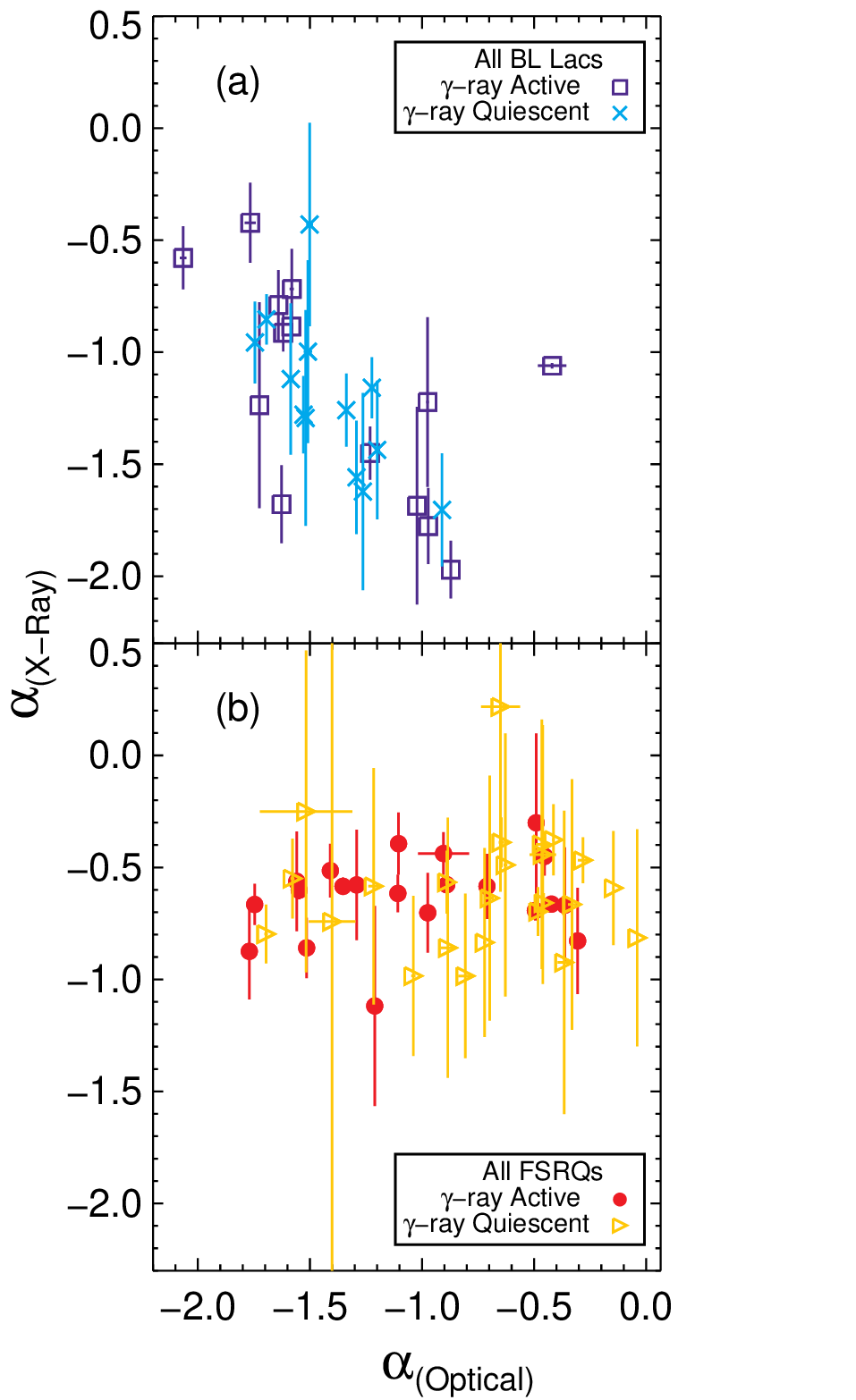}
\label{fig:xrayvsopticalbbff}
}
\subfloat{%
\includegraphics[trim=0cm 0cm 0cm 0cm, clip=true,height=.48\textwidth, angle=0]{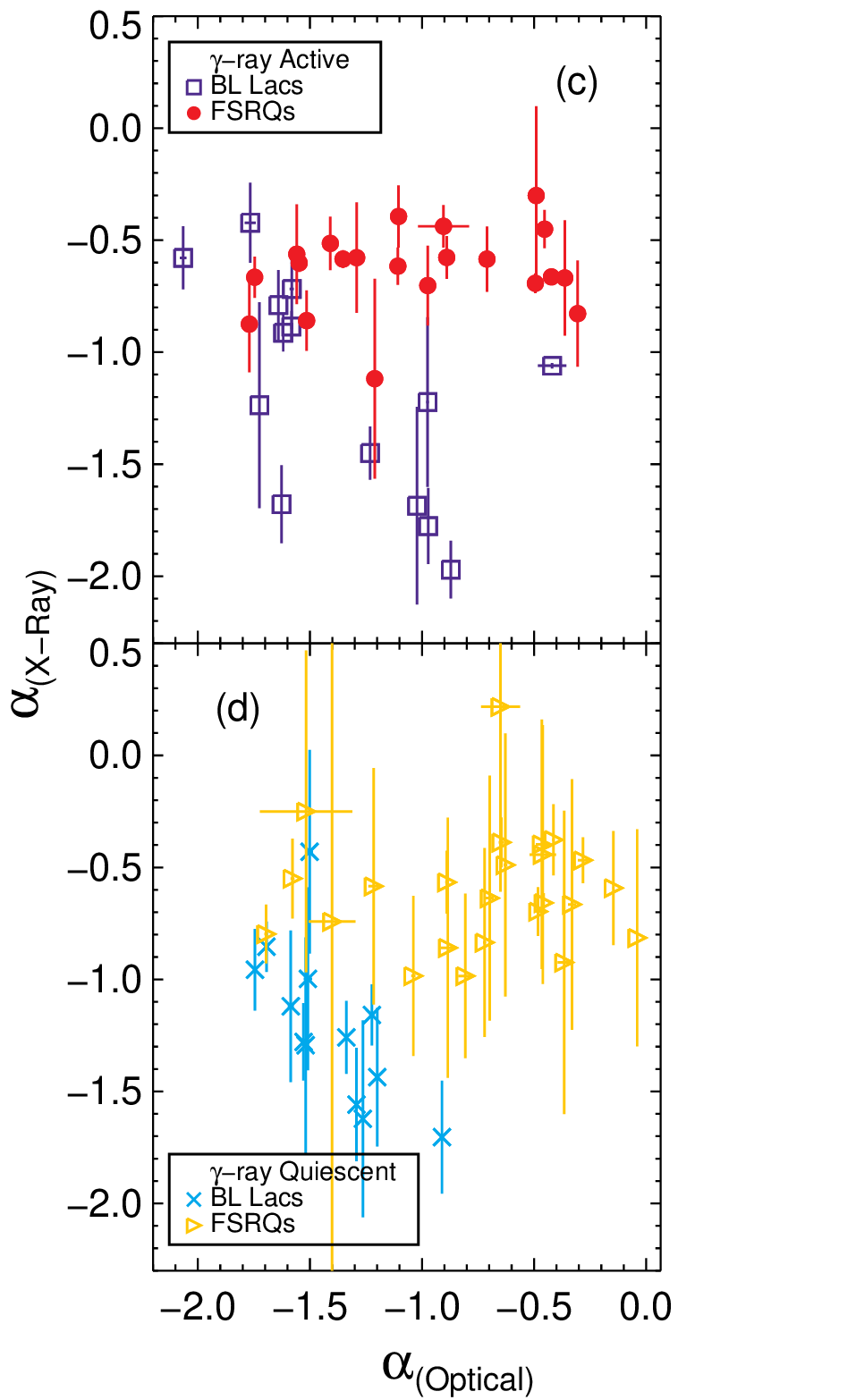}
\label{fig:xrayvsopticalaaqq}
}
\caption{Spectral indices \xsi\,vs \osi. Designations are the same as in Fig. \ref{fig:gammavsoptical}. }
\label{fig:xrayvsoptical}
\end{figure*}

\begin{figure*}
\centering
\subfloat{%
\includegraphics[trim=0cm 0cm 0cm 0cm, clip=true,height=.5\textwidth, angle=0]{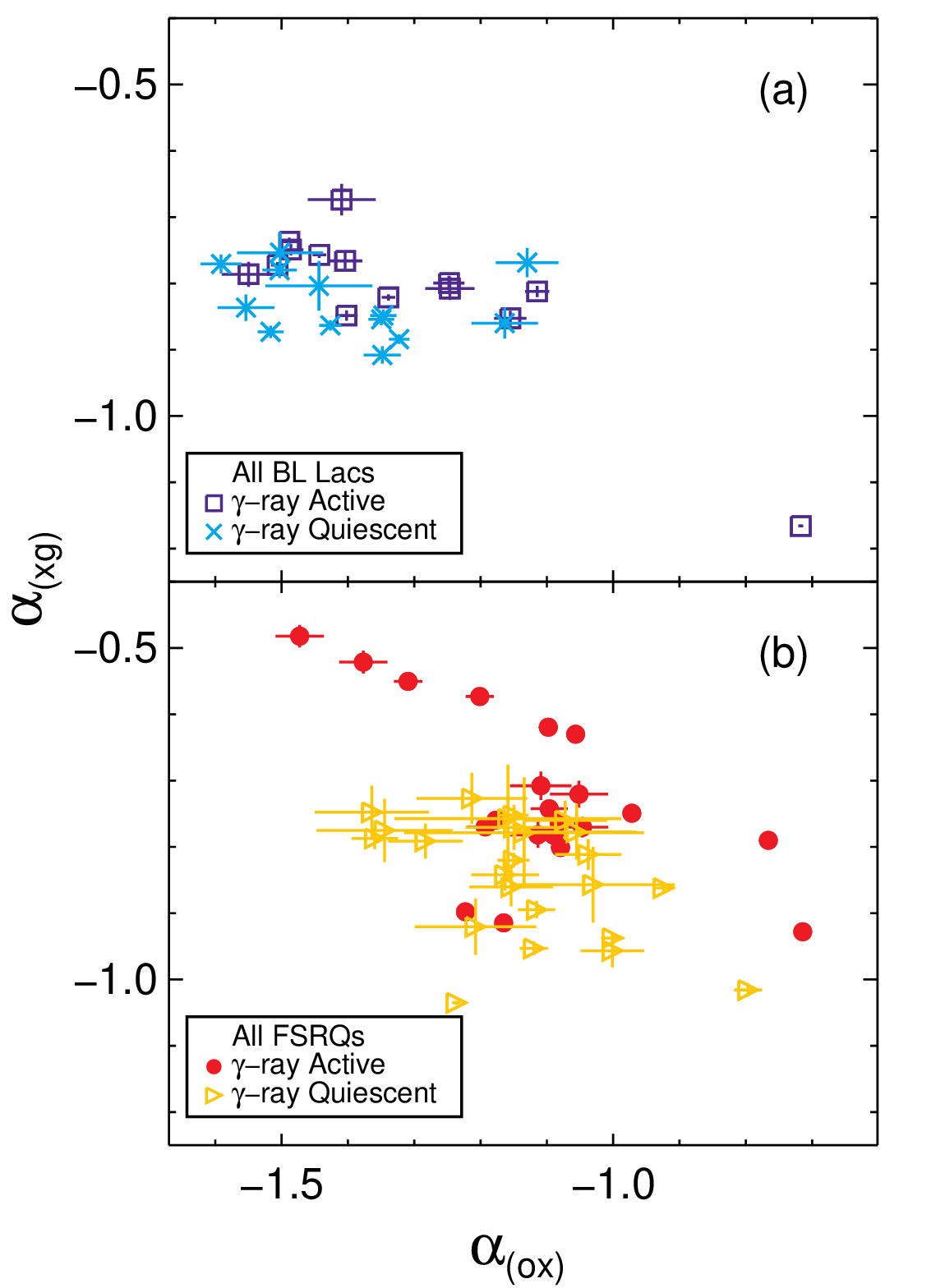}
\label{fig:xgvsoxaaqq}
}
\subfloat{%
\includegraphics[trim=0cm 0cm 0cm 0cm, clip=true,height=.5\textwidth, angle=0]{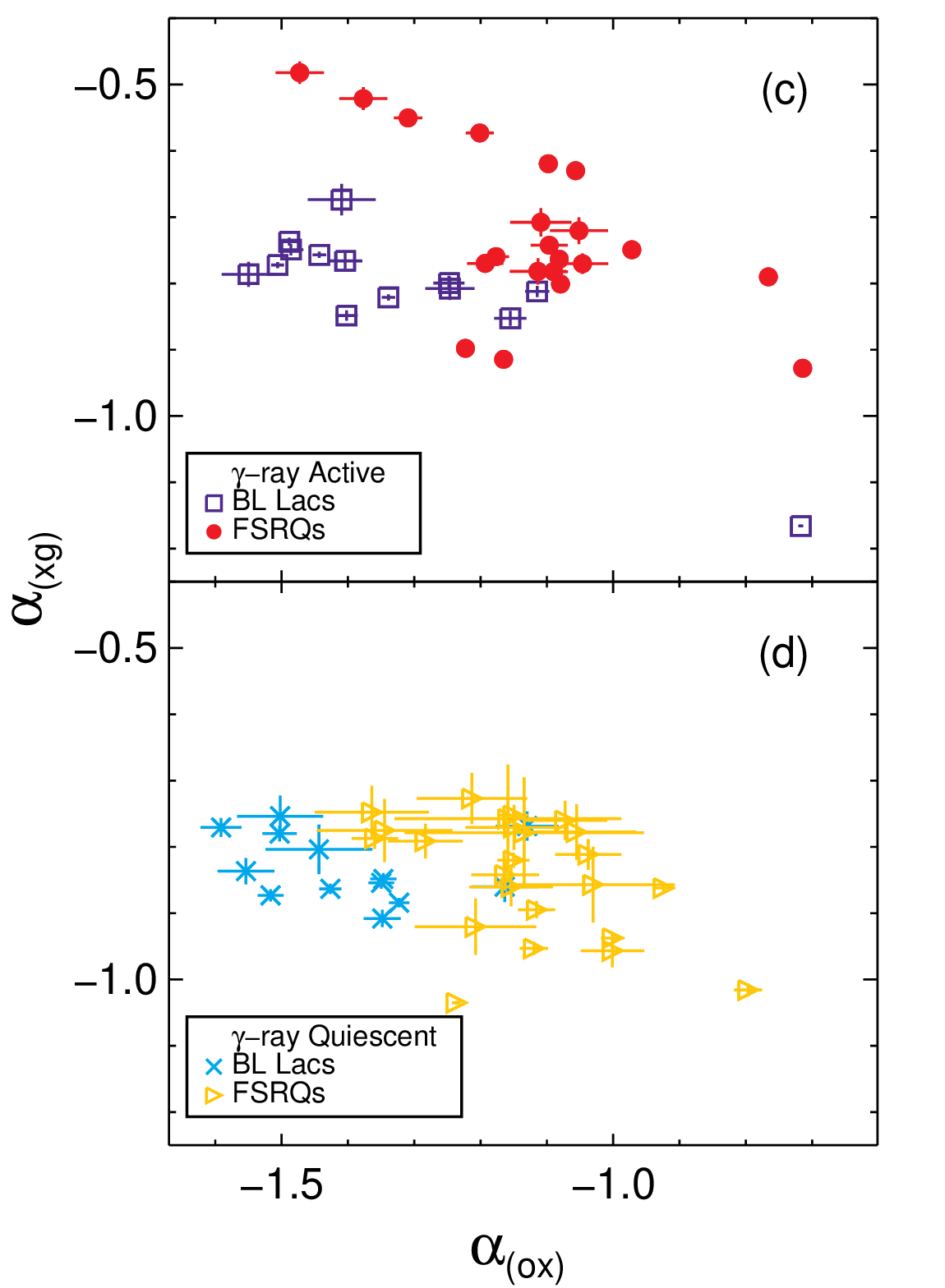}
\label{fig:xgvsoxbbff}
}
\caption{Spectral indices \xgsi\,vs \oxsi. Designations are the same as in Fig. \ref{fig:gammavsoptical}. }
\label{fig:xgvsox}
\end{figure*}

We examine relationships between the spectral indices at the different wavebands. Figures \ref{fig:gammavsoptical}, \ref{fig:gammavsxray}, \ref{fig:xrayvsoptical}, \& \ref{fig:xgvsox} show dependences between \gsi\, and \osi, \gsi\, and \xsi, \xsi\, and \osi, and between  \oxsi\, and \xgsi, respectively, for all blazars in the sample. The complete set of all plots in color and labeled with object and epoch numbers can be found in an expanded version of this paper at \url{www.bu.edu/blazars/VLBAproject.html}. We have computed Spearman's rank correlation coefficients between different spectral indices for the entire sample, as well as for different classes and states. We have used the IDL routine \texttt{R\_Correlate} to test the significance of the correlation coefficients. The results are presented in Table \ref{tab:spearmanga}, with the number of data points in the computation and the rank correlation coefficient and its significance given for each relationship. 
  
\noindent \textit{The  \gsi$-$\osi\, Plane}: Figure \ref{fig:gammavsoptical} reveals a striking difference between the quiescent BL Lacs and FSRQs: a BL Lac object with a flatter \osi\, has a flatter \gsi\,, while for the quasars a modest anti-correlation between the indices is observed.  The correlation analysis (Table \ref{tab:spearmanga}) confirms a highly significant positive correlation between \gsi\, and \osi\, of the BL Lacs independent of the state, and suggests a weak anti-correlation between \gsi\, and \osi\, of the quiescent FSRQs at $\sim$88.5\% confidence level. The latter effect disappears in active FSRQs. We associate flattening of \osi\, in FSRQs with increasing importance of the BBB contribution to the optical emission when the synchrotron flux decreases. If we assume that a pure synchrotron optical spectral index correlates with \gsi, as in the case of the BL Lacs, then the anti-correlated
behavior between \gsi\, and \osi\, for the quiescent FSRQs implies that quasars with a stronger BBB have a softer optical synchrotron spectrum. This is supported by the case of 3C273, in which the BBB dominates the optical-UV SED, while the synchrotron spectral index, as measured for the linearly polarized emission, is very steep, $-1.7$ to $-2.7$ \citep{Smith93}. However, the steep optical synchrotron index found for the quasar 3C454.3 during the prominent $\gamma$-ray outbursts, $\alpha_{o}^{syn}\sim -1.7$, is significantly steeper than  $\alpha_\gamma\sim -$1.3 \citep{2013ApJ...773..147J}; this implies that relativistic electrons that emit IR synchrotron radiation rather than optical emission are responsible for $\gamma$-ray production.

There are outliers in Figure \ref{fig:gammavsoptical} that are important to mention.  Quasar 1730$-$130 at epoch 3 and BL Lac object 1749+096 at epoch 4
(both active states) have extremely steep optical spectra ($-2.4$ and $-2.1$, respectively), and a follow-up study of additional active epochs of these objects could be enlightening. Active epoch 4 of 3C279 has a steep $\gamma$-ray spectrum ($-1.9$), while all epochs of 1222+216 are located in the flat optical$-$flat $\gamma$-ray region of both the active and quiescent FSRQs.

\begin{deluxetable}{lrrrrrrrrrrrr}
\tabletypesize{\small}
\rotate
\tablewidth{0pt}
\tablecaption{Spearman's Rank Correlation ($\rho$)}
\tablecolumns{13}

\tablehead{	\multicolumn{1}{l}{}&
			\multicolumn{3}{c}{$\alpha_{\gamma}$ and $\alpha_{o}$}&
			\multicolumn{3}{c}{$\alpha_{\gamma}$ and $\alpha_{x}$}&
			\multicolumn{3}{c}{$\alpha_{x}$ and $\alpha_{o}$}&
			\multicolumn{3}{c}{$\alpha_{xg}$ and $\alpha_{ox}$}
			\cr
			\multicolumn{1}{l}{}&
			\multicolumn{1}{r}{n}&
			\multicolumn{1}{c}{$\rho$}&
			\multicolumn{1}{c}{Signif.}&
			\multicolumn{1}{r}{\hspace{24pt}n}&
			\multicolumn{1}{c}{$\rho$}&
			\multicolumn{1}{c}{Signif.}&
			\multicolumn{1}{r}{\hspace{24pt}n}&
			\multicolumn{1}{c}{$\rho$}&
			\multicolumn{1}{c}{Signif.}&
			\multicolumn{1}{r}{\hspace{24pt}n}&
			\multicolumn{1}{c}{$\rho$}&
			\multicolumn{1}{c}{Signif.}
					  \cr
		  	\multicolumn{1}{l}{(1)}&
		  	\multicolumn{1}{r}{(2)}&
		 	\multicolumn{1}{c}{(3)}&
		 	\multicolumn{1}{c}{(4)}&
		  	\multicolumn{1}{r}{(5)}&
		 	\multicolumn{1}{c}{(6)}&
		 	\multicolumn{1}{c}{(7)}&
		  	\multicolumn{1}{r}{(8)}&
		 	\multicolumn{1}{c}{(9)}&
		 	\multicolumn{1}{c}{(10)}&
		  	\multicolumn{1}{r}{(11)}&
		 	\multicolumn{1}{c}{(12)}&
		 	\multicolumn{1}{c}{(13)}
			}

\startdata
    BL Lac Quiescent & 24    & 0.572 & 0.004 & 13    & -0.776 & 0.002 & 13    & -0.676 & 0.011 & 13    & -0.269 & 0.374 \\
    BL Lac Active & 22    & 0.473 & 0.026 & 14    & -0.442 & 0.114 & 14    & -0.631 & 0.016 & 14    & -0.732 & 0.003 \\
    All BL Lacs & 46    & 0.504 & 3.6E-04 & 27    & -0.556 & 0.003 & 27    & -0.648 & 2.6E-04 & 27    & -0.395 & 0.041 \\
          &       &       &       &       &       &       &       &       &       &       &       &  \\
    FSRQ Quiescent & 40    & -0.253 & 0.115 & 24    & -0.055 & 0.799 & 24    & 0.105 & 0.625 & 24    & -0.437 & 0.033 \\
    FSRQ Active & 28    & 0.029 & 0.883 & 21    & -0.239 & 0.297 & 21    & 0.113 & 0.626 & 21    & -0.458 & 0.037 \\
    All FSRQ & 68    & -0.258 & 0.034 & 45    & -0.134 & 0.379 & 45    & 0.078 & 0.609 & 45    & -0.289 & 0.054 \\
          &       &       &       &       &       &       &       &       &       &       &       &  \\
    All Quiescent & 64    & -0.445 & 2.3E-04 & 37    & -0.594 & 1.1E-04 & 37    & 0.405 & 0.013 & 37    & -0.238 & 0.156 \\
    All Active & 50    & 0.086 & 0.552 & 35    & -0.428 & 0.010 & 35    & 0.060 & 0.731 & 35    & -0.216 & 0.212 \\
    All   & 114   & -0.256 & 0.006 & 72    & -0.499 & 8.3E-06 & 72    & 0.233 & 0.049 & 72    & -0.180 & 0.129 \\

\enddata 
\tablecomments{\textit{n}: number of indices included in the computation; \textit{$\rho$}: rank correlation coefficient; \textit{Signif}: the two-sided significance level.} 
\label{tab:spearmanga}%
\end{deluxetable}

\noindent \textit{The  \gsi$-$\xsi\, Plane}: Figure \ref{fig:gammavsxray} shows a distinct separation in the \gsi\, $-$ \xsi\, plane for the two classes of blazars, with only a slight overlap. This is obviously driven by the separation of X-ray spectral index values between classes as discussed in \S\ref{Sdsi}.  Combining classes yields strong anti-correlations for both active (Fig.\ \ref{fig:gammavsxray}c) and quiescent (Fig.\ \ref{fig:gammavsxray}d) states. Quiescent BL Lacs show a strong anti-correlation between \gsi\, and \xsi, that becomes very weak for active BL Lacs (Table~\ref{tab:spearmanga}). In general, for a blazar in our sample, steeper \xsi\, pairs with flatter \gsi. Within IC mechanisms for $\gamma$-ray production, this suggests that for sources with a synchrotron origin of X-rays (fully or partly), lower-energy relativistic electrons participate in $\gamma$-ray production (those that generate IR-optical synchrotron emission), while for sources with X-rays via IC mechanisms, higher-energy relativistic electrons should be involved in 0.1$-$200 GeV $\gamma$-ray production (those that produce optical-UV synchrotron emission).
 
There are outliers in the \gsi\, $-$ \xsi\, plane that include three BL Lacs that are well known TeV sources:  
1219+285, 3C66A, and Mkn421.  Among the FSRQs, the quasars 3C279 and 0836+710 are distinguished by the steepness of their $\gamma$-ray spectra. Additionally, the first quiescent epoch of 3C446  is isolated in the region of flat X-ray spectra (\xsi\, = 0.22), although the uncertainty in the index is high. 

\noindent \textit{The \xsi$-$\osi\, Plane}: Figure \ref{fig:xrayvsoptical}a shows 
a strong anti-correlation between \xsi\, and  \osi\ for BL Lacs, independent of activity state, with a high confidence level (see Table~\ref{tab:spearmanga}). According to the discussion in \S\ref{Sdsi}, values of \osi\, of the BL Lacs should represent pure synchrotron spectra. The observed anti-correlation and steepness of \xsi, up to $-$2.0, imply that in BL Lacs with the hardest optical spectra, the X-ray emission is produced via the synchrotron mechanism. These are the TeV sources  Mkn421, 1219+285 and 3C66A mentioned above. As the optical spectrum softens, the contribution from IC mechanisms to the X-ray emission increases. In general, there is no overlap between the BL Lacs and FSRQs in Figures~\ref{fig:xrayvsoptical}(c,d), since the FSRQs have flatter values of \osi, indicating the presence of the BBB, and uniformly flat values of \xsi\, that point to IC mechanisms for X-ray production. However, some active BL Lacs with the flattest \xsi\, form a continuation of the sequence of active FSRQs into the steepest \osi\, values. These are among the brightest BL Lacs at radio wavelengths, 1749+096, BL Lacertae, 1055+018 and OJ287. Three quiescent quasars with the steepest \osi\, values form a continuation of the quiescent BL Lac sequence into the flattest \xsi\, values (3C~279, 1308+326, and 1406$-$076), which most likely have weaker BBB emission with respect to the jet emission than for the other FSRQs. 


\noindent \textit{The \oxsi$-$\xgsi\, Plane}: An anti-correlation is expected in this plane if 1) the X-ray flux varies with much higher amplitude than do the optical and $\gamma$-ray fluxes, or 2) the optical and $\gamma$-ray fluxes vary in unison while the X-ray flux is relatively stable in many of the sources. Neither case commonly occurs (see Table \ref{table:statedata}).  According to Table~\ref{tab:spearmanga} there is a statistically significant anti-correlation between \oxsi\, and  \xgsi\, for active BL Lacs.  However, the anti-correlation is driven by the spectral indices of Mkn~421,
which is the only HSP source in our sample. The rest of the BL Lacs show very small scatter in the values of \xgsi, with slightly flatter values during active states. Table~\ref{tab:prefvalues} shows that
the average values of \xgsi\, of FSRQs are similar to those of BL Lacs. The stability of \xgsi\ follows from the high ratio of $\gamma$-ray to X-ray frequencies, the logarithm of which is in the denominator of the X-ray -- $\gamma$-ray spectral index calculation. In this context, the line of active quasars in Figures~\ref{fig:xgvsox}b,c with \xgsi\, flatter than $-$0.7 is especially interesting, since these are the
quasars with the strongest amplitude of $\gamma$-ray activity: 1222+216, 1510-089, CTA102, 3C454.3, and 0836+710 (see Figure \ref{fig:pernorm}). The line shows a clear anti-correlation between \oxsi\, and  \xgsi,  which corresponds to case 2 above and implies  that the $\gamma$-ray and optical fluxes have significantly larger amplitudes of variation than that of the X-ray emission. This is not expected if the SSC mechanism 
is responsible for both the X-ray and $\gamma$-ray emission, since in this case the value of \xgsi\, should remain stable across activity states. The significant difference in the amplitude
of X-ray and $\gamma$-ray activity might be caused by different seed photons being scattered by the relativistic electrons: synchrotron from the jet for X-rays (SSC) and external for $\gamma$-rays (EC).
Alternatively, the X-ray variations could be smoothed out by longer timescales of energy losses of the relatively low-energy electrons participating in IC X-ray production.  There is a clear separation between the BL Lacs and FSRQs with respect to values of \oxsi, especially for the quiescent blazars (Figure \ref{fig:xgvsox}d): the  FSRQs possess flatter \oxsi\ values than those of BL Lacs. This supports the conclusion that different X-ray emission mechanisms operate in the BL Lacs and FSRQs, as pointed out in the analysis of the \xsi$-$\osi\ plane.


\section{Discussion: Implications for Emission Models}

\label{Results}
The analysis of spectral indices in each waveband and the relationship between these indices allow us to describe a ``typical'' BL Lac object or FSRQ  and contrast the results by activity state within each class. Table \ref{tab:typical} summarizes statistically significant results from this exercise.

\begin{deluxetable}{lcccc}
\tabletypesize{\small}
\tablewidth{0pt}
\tablecaption{``Typical" Quiescent or Active Object}
\tablecolumns{5}

\tablehead{	& 
			\multicolumn{2}{c}{``Typical'' BL Lac} & 
			\multicolumn{2}{c}{``Typical'' FSRQ} 
					\cr
			\multicolumn{1}{l}{}&
			\multicolumn{1}{c}{Quiescent}&
			\multicolumn{1}{c}{Active}&
			\multicolumn{1}{c}{Quiescent}&
			\multicolumn{1}{c}{Active}
					  \cr
					  
		  	\multicolumn{1}{c}{(1)}&
		  	\multicolumn{1}{c}{(2)}&
		 	\multicolumn{1}{c}{(3)}&
		 	\multicolumn{1}{c}{(4)}&
		  	\multicolumn{1}{c}{(5)}
			}

\startdata
	\multicolumn{5}{l}{Mean value:}\\
     \hspace{8pt}\osi & $-1.4 \pm 0.3$    & high dispersion               & high dispersion & high dispersion \\
     \hspace{8pt}\xsi & $-1.2 \pm 0.3$    & high dispersion               & $-0.60 \pm 0.27$ & $-0.63 \pm 0.18$ \\
     \hspace{8pt}\gsi & $-1.12 \pm 0.17$  & $-1.13 \pm 0.23$ & $-1.46 \pm 0.17$ & $-1.31 \pm 0.22$ \\
     \hspace{8pt}\oxsi & $-1.40 \pm 0.14$ & $-1.32 \pm 0.22$ & $-1.13 \pm 0.13$ & $-1.11 \pm 0.17$ \\
     \hspace{8pt}\xgsi & $-0.83 \pm 0.05$ & $-0.81 \pm 0.11$ & $-0.84 \pm 0.09$ & $-0.73 \pm 0.12$ \\
    \cr
%
    \cr     
	\multicolumn{5}{l}{Correlation probability:}\\    
     \hspace{8pt}\gsi\, and \osi & 99.6\%        & 97.4\%       & 88.5\% (anti)    & ns \\
     \hspace{8pt}\gsi\, and \xsi & 99.8\% (anti) & 88.6\% (anti)  & ns    & ns \\
     \hspace{8pt}\xsi\, and \osi & 98.9\% (anti) & 98.4\% (anti) & ns    & ns \\
     \hspace{8pt}\oxsi\, and \xgsi & ns          & 99.7\% (anti) & 96.7\% (anti) & 96.3\% (anti) \\
     \cr
      Percentage time in state: & $55 \pm 20\%$ & $9 \pm 4\%$ & $65 \pm 15\%$ & $10 \pm 8\%$\\
     \multicolumn{5}{l}{Longest uninterrupted period:}\\ 
     \hspace{8pt}Average number of days&216 &43 & 217 & 30\\
     \multicolumn{2}{l}{Normalized amplitude of $\gamma$-ray variations:}& $5.2 \pm 2.0$ && $10 \pm 12$     
\enddata 
\tablecomments{\textit{ns}: not significant.} 
\label{tab:typical}%
\end{deluxetable}

Our findings suggest that the optical emission of a ``typical'' BL Lac object is strongly dominated by synchrotron radiation at any state, independent of SED classification. This implies that any emission from the accretion disk is weak in BL Lacs, consistent with the polarimetry of BL Lacs showing no evidence for the wavelength-dependent polarization expected when the essentially unpolarized BBB contributes substantially to the optical-UV emission \citep[e.g.,][]{SmithSitko1991,Smith1996}.

The X-ray emission from BL Lacs is a mixture of synchrotron and IC radiation. The statistically significant correlation between $\alpha_o$ and $\alpha_X$ implies that the contribution of IC emission to the observed X-rays increases as the optical spectrum softens, especially for active BL Lacs. 
The optical and $\gamma$-ray spectral indices are correlated at $>$ 97\% confidence level. No difference in values of \gsi\, between quiescent and active states is observed, which implies that the same mechanism is responsible for quiescent and flaring $\gamma$-ray emission. The modest amplitude of $\gamma$-ray activity, with small scatter across the BL Lac sample, favors the SSC mechanism for $\gamma$-ray production, while slightly flatter values of \gsi\, relative to \osi\, imply that relativistic electrons radiating at both optical and IR wavelengths are involved.

A ``typical'' FSRQ has a flatter optical spectrum in quiescent than in active states, which can be attributed to the importance of the contribution of the BBB to the optical-UV continuum \citep[e.g.,][]{SMITH88, Giommi12a}. The wide dispersion of optical spectral indices is then due to diversity in the relative strength of the BBB among FSRQs rather than to variations in the slope of their synchrotron spectra. We anticipate that once the BBB component is subtracted, the residual synchrotron spectral index will show a smaller scatter in \osi, as in BL Lacs, and also as is the case for \gsi\, for both the BL Lacs and FSRQs. A modest
anti-correlation between \gsi\, and \osi\, for the quiescent FSRQs implies a possible connection between the properties of the BBB and jet if the anti-correlation is driven by the contribution of the BBB to the optical emission. The latter is probable, since the anti-correlation disappears during active states. In this scenario, a quasar with a stronger BBB has softer optical synchrotron and $\gamma$-ray spectra in quiescent states. The $\gamma$-ray spectrum of an FSRQ flattens during active states, which
implies more efficient acceleration of relativistic electrons if the $\gamma$-rays originate via IC mechanisms. This should cause flattening of the optical synchrotron spectra during active states as well. However, to test such an assumption and a possible connection between BBB and jet properties, pure synchrotron optical spectra of FSRQs should be extracted from the observations by subtracting the BBB spectrum from the continuum.

We find a uniform preferred value of \xsi\, $\sim-$0.6, among the FSRQs that is the same as the average spectral index of blazars measured at wavelengths of 0.8 to 4 mm \citep{Giommi12a}. This supports the hypothesis that IC scattering from relativistic electrons emitting synchrotron radiation at mm-submm wavelengths is responsible for X-ray production in a typical FSRQ, independent of the activity state. Whether the X-rays are from the SSC or EC mechanism, or a combination of the two, might depend on the blazar and its activity state.
The large dispersion in the amplitude of $\gamma$-ray activity, and the anti-correlated behavior between
\xgsi\, and \oxsi\, for the FSRQs displaying the highest amplitude of $\gamma$-ray outbursts, require different mechanisms
of $\gamma$-ray production during different activity states. There is most likely a mixture of SSC and EC emission, with a dominance of external IC during the highest $\gamma$-ray states, as has been modeled for some blazars \cite[e.g.,][]{BON11,ANN12}.

\section{Summary}

\label{summary}
We have assembled---and  de-reddened at  NIR, optical and UV wavelengths---observational measurements obtained from  2008 through 2012 of 33  blazars by ten ground- and space-based observatories.  
We have computed a mean flux value for each frequency band for each source and used these values to determine whether the object was in a quiescent or active state in each band. The state of the object in the $\gamma$-ray band was the basis for defining quiescent and active periods. The frequency and length of quiescent and active periods, and the maximum flux achieved during active periods, were compared between the BL Lacs and FSRQs. Up to four epochs per source were selected for further analysis of spectral indices at $\gamma$-ray, X-ray, and, optical wavelengths. All IR through X-ray observations selected for an epoch were obtained within a 24-hour period, with an average span of 9.0 hours. 
We find significant diversity in the properties of the BL Lacs and FSRQs in each spectral regime analyzed: 
\begin{enumerate}
\item The FSRQs exhibit the highest amplitude of $\gamma$-ray activity, while the duration of an average active period in the source frame is similar for the FSRQs and BL Lacs. On the other hand, the fraction of time when a quasar is dormant exceeds that of a BL Lac object by $\sim$10\%, with less scatter.
\item Comparison of the behavior of \osi\, between activity states suggests weak accretion disk emission in the BL Lacs, while the contribution of the BBB to the optical emission of the FSRQs dominates quiescent states.
\item The lack of significant variations in $\gamma$-ray spectral indices of the BL Lacs between activity states, the relatively low ratio of $\gamma$-ray to synchrotron luminosity, and the good correlation between
\gsi\, and \osi, implies that the same inverse Compton mechanism --- most likely SSC --- is responsible for the $\gamma$-ray production at different activity states.
\item The anti-correlation between \xgsi\, and \oxsi\, for the FSRQs during the most extreme activity at $\gamma$-ray energies suggests that the SSC mechanism is insufficient to explain the enhanced $\gamma$-ray flux in these objects. Hence, the EC mechanism for $\gamma$-ray (but not necessarily X-ray) production is favored by the data.
\item The analysis of X-ray spectral indices indicates that the X-ray emission of the BL Lacs is a mixture of synchrotron and inverse Compton radiation. IC scattering dominates during active states of the LSP BL Lacs, while IC scattering by $<$ 1 GeV electrons can explain the entire X-ray emission of the FSRQs at any state.
\end{enumerate}

The relationships among the various spectral indices therefore imply strong connections between the emission at pairs of wavebands: mm-submm and X-ray for FSRQs and LSP BL Lacs, optical and X-ray for ISP and HSP BL Lacs, and IR-optical and $\gamma$-ray for FSRQs and LSP BL Lacs. These connections should be apparent in timing studies of multi-waveband light curves of blazars. We are in the process of compiling such light curves over a sufficiently long time span ($\sim 5$ years) to test whether the predictions of such correlations are fulfilled.

\acknowledgments
We thank the anonymous referee for providing valuable comments and suggestions that improved several sections of the paper. The data acquisition and analysis for this study was supported by National Science Foundation grant AST-0907893, NASA Fermi Guest Investigator grants NNX08AV65G, NNX09AT99G, NNX10AO59G, NNX10AV15G, NNX11AO37G, NNX11AQ03G, NNX12AO90G, and NASA Swift Guest Investigator grants NNX09AR11G, NNX10AL13G, NNX10AF88G, NNX12AF09G, and NNX12AE90G. The effort at Steward Observatory was
funded in part by NASA through Fermi Guest Investigator grants NNX08AW56G, NNX09AU10G, and NNX12AO93G. The St. Petersburg State University team acknowledges support from RFBR grants 12-02-00452 and 12-02-31193. 
The research at the IAA-CSIC is supported by the Spanish Ministry of Economy and Competitiveness and the Regional Government of Andaluc\'{i}a (Spain) through grants AYA2010-14844 and P09-FQM-4784, respectively. The {\it Swift} effort at PSU is supported by NASA contract NAS5-00136. The PRISM camera at
Lowell Observatory was developed by K.\ Janes et al. at BU and Lowell Observatory, with funding from
the NSF, BU, and Lowell Observatory. The Liverpool Telescope is operated on the island of La Palma by Liverpool John Moores University in the
Spanish Observatorio del Roque de los Muchachos of the Instituto de Astrofisica de Canarias, with funding
from the UK Science and Technology Facilities Council.
The Calar Alto Observatory is jointly operated by the Max-Planck-Institut f\"ur Astronomie and the Instituto de 
Astrof\'{\i}sica de Andaluc\'{\i}a-CSIC. This study is partly based on data taken and assembled by the WEBT collaboration and stored in the WEBT archive at the Osservatorio Astronomico di Torino - INAF (http://www.oato.inaf.it/blazars/webt/). 

{\it Facilities:} \facility{Perkins}, \facility{Liverpool:2m}, \facility{CAO:2.2m}, \facility{Bok}, \facility{SO:Kuiper}, \facility{CTIO:1.3m}, \facility{Swift}, \facility{FERMI (LAT)}


\bibliographystyle{apj} 



\clearpage
\textbf{\LARGE{Supplemental Material}}
\vspace*{10pt}

The following plots combine panels a $-$ d of Figures \ref{fig:gammavsoptical} $-$ \ref{fig:xgvsox}, with each data point labeled with object and epoch numbers, included in this version for your convenience. 

An expanded version of this paper with a complete set of light curves, SEDs, and labeled spectral index relationship plots for all sources  can be found at \url{www.bu.edu/blazars/VLBAproject.html}.

\begin{figure*}[t]%

\centering
	\includegraphics[width=0.85\paperwidth, angle=90]{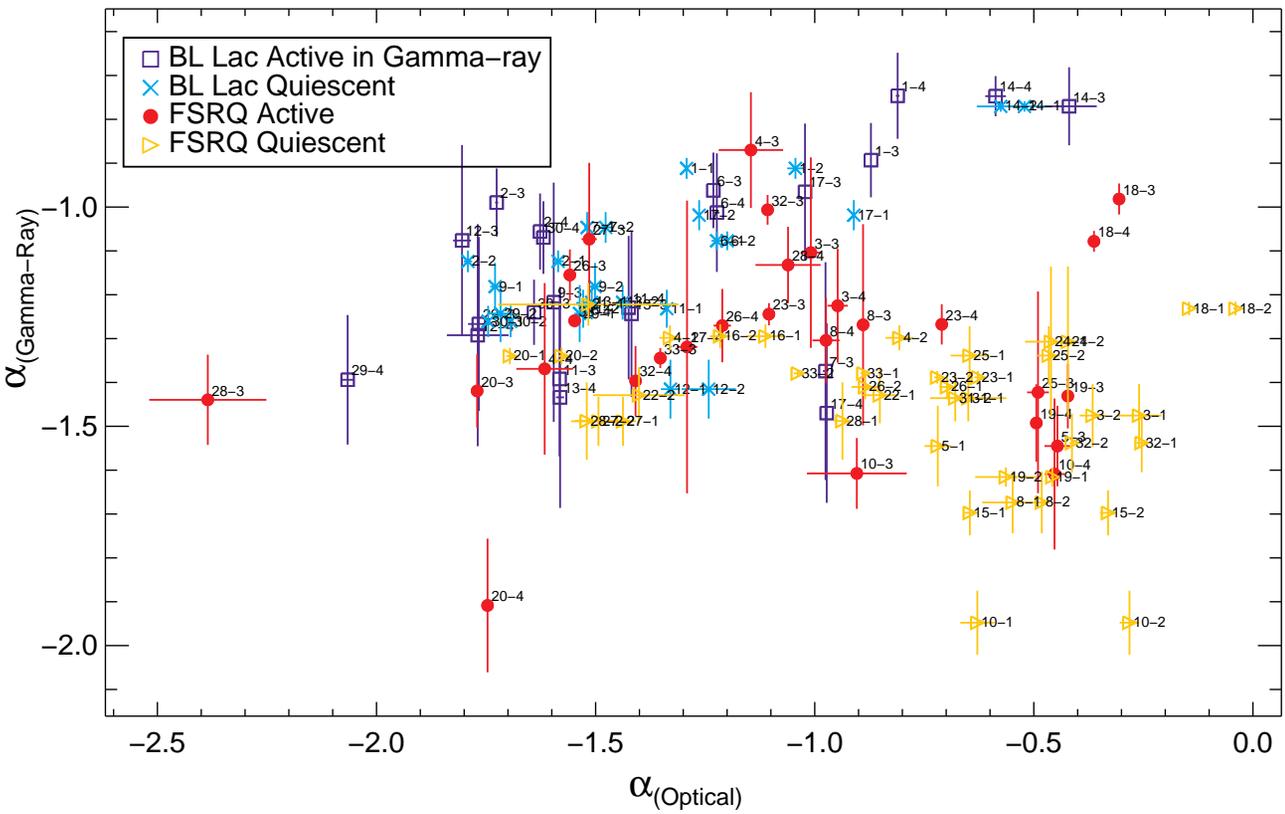}%

\caption{Spectral indices \gsi\,vs.\ \osi\, at selected epochs (Section \ref{sec:composition}) for all blazars in the sample:  FSRQs are red circles  in $\gamma$-ray  active states, yellow if quiescent, while BL Lacs are dark blue if $\gamma$-ray active, light blue if quiescent. Each data point is labeled with object (see Table \ref{tab:sources}) and epoch numbers (see Table \ref{tab:selecteddata}).}
  
\label{fig:9all}%
\end{figure*}

\begin{figure*}[t]%

\centering
	\includegraphics[trim=1.0cm 0cm 3.5cm 0cm, clip=true,width=0.85\paperwidth, angle=90]{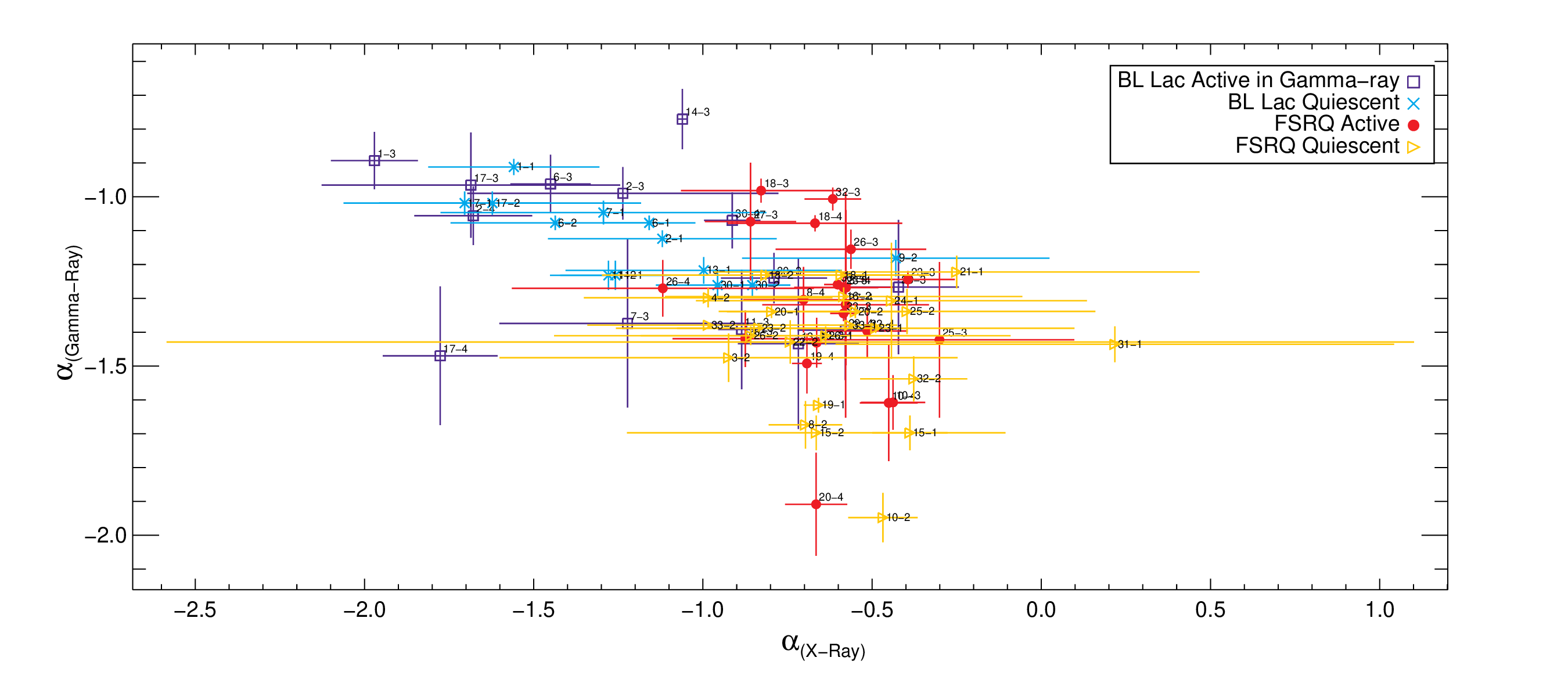}%

\caption{Spectral indices \gsi\,vs.\ \xsi. Designations are the same as in Fig.  \ref{fig:9all}. }
  
\label{fig:10all}%
\end{figure*}

\begin{figure*}[t]%

\centering
	\includegraphics[trim=0cm 2.0cm 0cm 3.0cm, clip=true,width=0.7\paperwidth, angle=0]{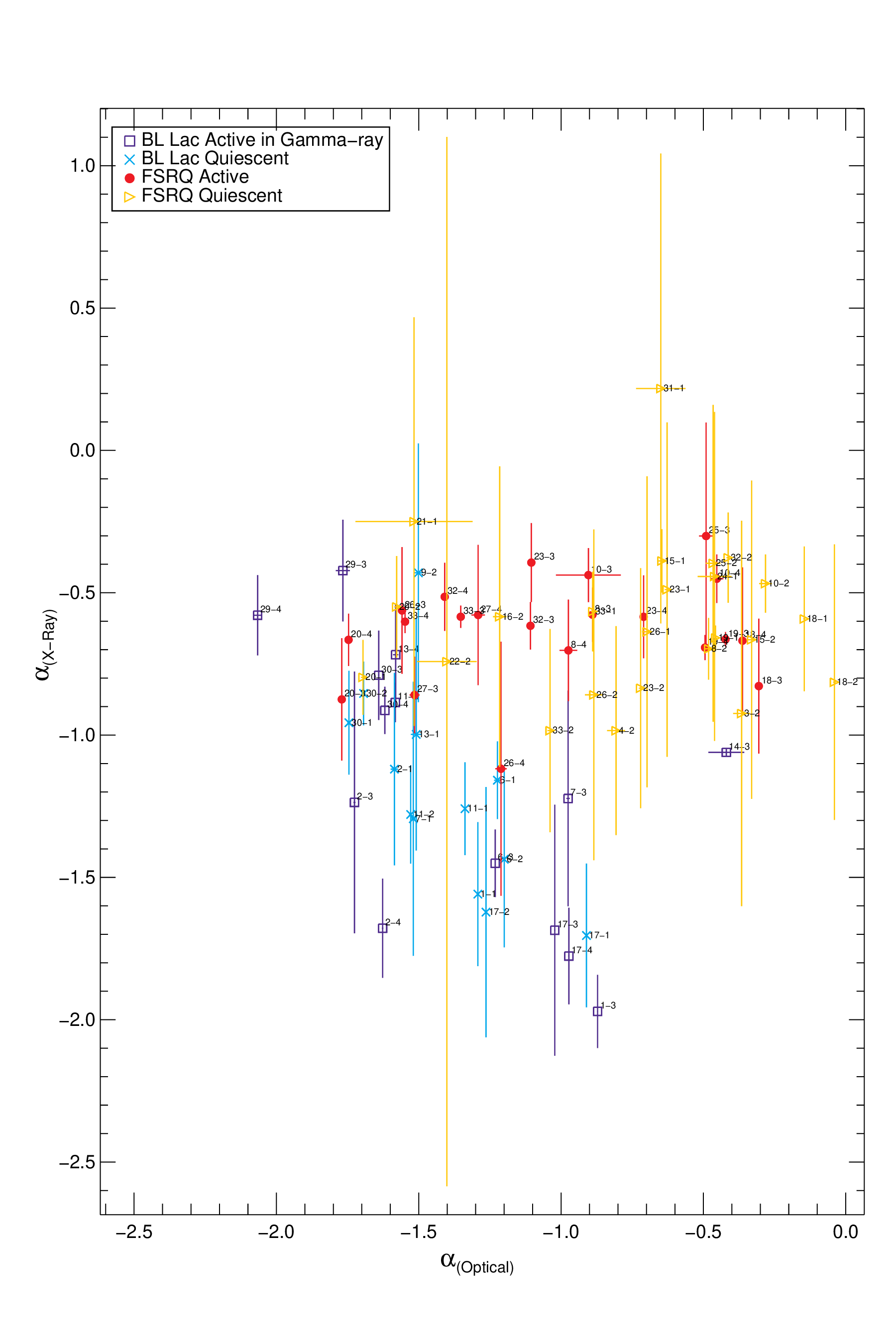}%

\caption{Spectral indices \xsi\,vs \osi. Designations are the same as in Fig. \ref{fig:9all}. }
  
\label{fig:11all}%
\end{figure*}

\begin{figure*}[t]%

\centering
	\includegraphics[width=0.85\paperwidth, angle=90]{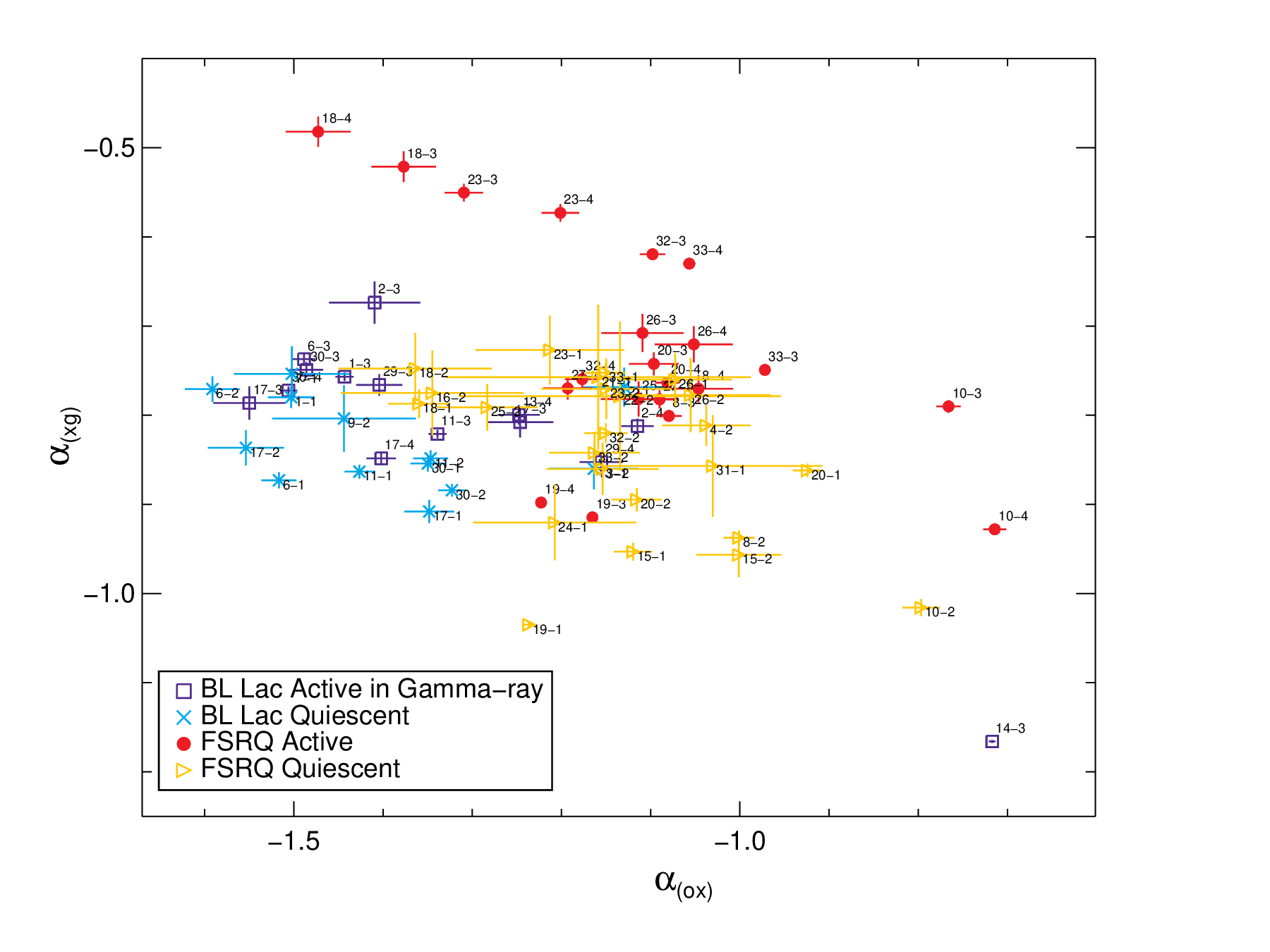}%

\caption{Spectral indices \xgsi\,vs \oxsi. Designations are the same as in Fig. \ref{fig:9all}. }
  
\label{fig:12all}%
\end{figure*}

\end{document}